\documentclass[twocolumn,twocolappendix]{aastex63}
\usepackage{hyperref}
\usepackage{multirow}
\newcommand{\free}{\textit{Free}}
\newcommand{\ndc}{\textit{Conv}}
\newcommand{\dc}{\textit{Deconv}}
\newcommand{\nsrc}{n_{S\acute{e}rsic}}

\newcommand{\Reff}{{\rm{R}_{e}}}
\newcommand{\niter}{\rm{N}_{iter}}
\newcommand{\lre}{\lambda_{R_{e}}}
\newcommand{\sratio}{\rm{{R}_{1}/{R}_{2}}}
\newcommand{\vrot}{\rm{V}_{ROT}}

\newcommand{\fpsf}{\rm{FWHM}_{\rm{PSF}}}
\newcommand{\fdc}{\rm{FWHM}_{\rm{Deconv}}}
\newcommand{\fndc}{\rm{FWHM}_{\rm{Conv}}}

\newcommand{\idl}{\texttt{IDL }}
\newcommand{\python}{\texttt{Python3}}

\newcommand{\github}{\texttt{GitHub}}
\newcommand{\ppxf}{\texttt{pPXF}}

\received{}
\revised{}
\accepted{}

\shorttitle{Stellar Kinematics Restoration}
\shortauthors{Chung et al.}

\begin{document}

\title{PSF Deconvolution of the IFU Data and Restoration of Galaxy Stellar Kinematics}

\correspondingauthor{Changbom Park}
\email{cbp@kias.re.kr}

\author[0000-0002-3043-2555]{Haeun Chung}
\affiliation{University of Arizona, Steward Observatory, 933 N Cherry Ave, Tucson, AZ 85721, USA}
\affiliation{Astronomy Program, Department of Physics and Astronomy, Seoul National University,
1 Gwanak-ro, Gwanak-gu, Seoul 08826, Republic of Korea}
\affiliation{School of Physics, Korea Institute for Advanced Study, 85 Hoegiro,
Dongdaemun-gu, Seoul 02455, Republic of Korea}

\author[0000-0001-9521-6397]{Changbom Park}
\affiliation{School of Physics, Korea Institute for Advanced Study, 85 Hoegiro,
Dongdaemun-gu, Seoul 02455, Republic of Korea}

\author{Yong-Sun Park}
\affiliation{Astronomy Program, Department of Physics and Astronomy, Seoul National University,
1 Gwanak-ro, Gwanak-gu, Seoul 08826, Republic of Korea}




\begin{abstract}
We present a performance test of the Point Spread Function deconvolution algorithm applied to astronomical Integral Field Unit (IFU) Spectroscopy data for restoration of galaxy kinematics. We deconvolve the IFU data by applying the Lucy-Richardson algorithm to the 2D image slice at each wavelength.
We demonstrate that the algorithm can effectively recover the true stellar kinematics of the galaxy, by using mock IFU data with diverse combination of surface brightness profile, S/N, line-of-sight geometry and Line-Of-Sight Velocity Distribution (LOSVD). In addition, we show that the proxy of the spin parameter $\lre$ can be accurately measured from the deconvolved IFU data. 
We apply the deconvolution algorithm to the actual SDSS-IV MaNGA IFU survey data. The 2D LOSVD, geometry and $\lre$ measured from the deconvolved MaNGA IFU data exhibit noticeable difference compared to the ones measured from the original IFU data. 
The method can be applied to any other regular-grid IFU data to extract the PSF-deconvolved spatial information.

\end{abstract}



\section{Introduction}\label{sec:ch2intro}
Integral Field Spectroscopy (IFS), or 3D spectroscopy is an observational technique
used to collect the two-dimensional spatial information 
on the spectral properties of the target object.
IFS observation can be performed by using a single or multiple Integral Field Unit(s) (IFU(s)), a module that captures one contiguous region on the sky.
Starting from SAURON IFU \citep{2001MNRAS.326...23B}, many IFS instruments (GMOS \citep{2002PASP..114..892A}, VIMOS \citep{2005A&A...439..845L}, IMACS \citep{2011PASP..123..288D}, PMAS/PPAK \citep{2006PASP..118..129K}, KMOS \citep{2013Msngr.151...21S}, MUSE \citep{2010SPIE.7735E..08B}) have been developed in the optical and near-infrared. Nowadays there are thousands of publicly available IFU data from a number of IFU surveys such as ATLAS$\rm{^{3D}}$ \citep{2011MNRAS.414..888E}, DiskMass \citep{2010ApJ...716..198B}, CALIFA \citep{2016A&A...594A..36S}, SAMI \citep{2018MNRAS.481.2299S}, and MaNGA \citep{2015ApJ...798....7B}. 
However, all of the IFU data from fore-mentioned ground-based surveys have a common limitation (unless corrected by Adaptive Optics): spatial information degradation corresponding to the Point Spread Function (PSF). PSF is a combination of the atmospheric seeing, the aberration from the telescope and instrument optics, and the sampling size/scheme. Notably, the effect becomes more severe for the data obtained by bare fiber-based IFU, because of the physical gap between sampling elements which enlarges effective PSF size. 
Due to the effects of PSF, every derived, measured, or fitted quantities from the IFU data are smoothed and becomes spatially correlated. 
To extract the spatially resolved information as much as possible from the IFU data, one must minimize the effects of PSF.
A way to correct for the PSF effects is the forward modeling; use of flux-weighted PSF convolution to the 2D model quantities \citep{2008MNRAS.390...71C, 2015AJ....150...92B}. However, this is only an approximation that does not fully reflect the PSF effects.  

Historically, there were numerous attempts that tried to mitigate the effects of PSF on 2D images in the field of signal/image processing in particular (see the summaries by \citet{2011MNRAS.418..258B}, \citet{2014ITIP...23.4322V} and references therein). However, those techniques are not directly applicable to the astronomical data since they are optimized to three channel color images or images with different characteristics compare to astronomical images.
There were studies in the field of astronomy that adopted deconvolution, such as optimal spectrum extraction from the CCD image \citep{2000ApJ...529.1136C, 2003AJ....125.2266L}, or reduction of the $Spitzer$ slit spectroscopy data \citep{2008ISTSP...2..802R}. More recently, several techniques \citep{6080853, 6080865, 2011MNRAS.418..258B, 2014ITIP...23.4322V} were proposed to restore the
3D-correlated IFU data in both spatial and spectral direction in the context of MUSE \citep{2003SPIE.4841.1096H}.
\citet{2011MNRAS.418..258B} utilized a prior knowledge on the correlation between spatial and spectral direction to deconvolve the IFU data by using a regularized $\chi^2$ method. 
This technique requires two hyper-parameters for the deconvolution, however, the parameters are determined not by quantitative criteria but by visual inspection of the results from various sets of parameters through trial and error.
\citet{2014ITIP...23.4322V} proposed to use the nonlinear deconvolution technique on the IFU data with Markov-Chain Monte-Carlo in Bayesian framework, to recover the flux, relative velocity, and the velocity dispersion distribution of the target. The technique was demonstrated on simulated IFU data from mock observation of objects with two separated emission lines.

In this work, we explore a general method to mitigate the effects of PSF that can be applied to any kind of IFU data. In particular, we study the performance of the PSF deconvolution method applied to extended sources (galaxies) to restore their true kinematics. This work was motivated for the study of stellar kinematics of 
SDSS-IV MaNGA survey galaxies. We use the Lucy-Richardson (LR) algorithm \citep{1972JOSA...62...55R, 1974AJ.....79..745L}, which is one of the simplest deconvolution techniques which requires a minimum number of parameters. We validate the algorithm using mock IFU data and show that the kinematics of galaxies can be well-restored through our deconvolution procedure. In addition, we apply the deconvolution method to measure the spin parameter $\lre$ \citep{2007MNRAS.379..401E}, which is a widely-used proxy of the galaxy angular momentum. 

The structure of this paper is as follows.  In \autoref{sec:psfdeconv}, we introduce the LR deconvolution algorithm and its implementation to the IFU data. We demonstrate the validity of the deconvolution technique using the mock IFU data in \autoref{sec:pvmock}. In \autoref{sec:appmanga} we illustrate the example of deconvolution to the MaNGA IFU data. Finally, We show how the deconvolution can be used to improve the measurement of spin parameter $\lre$ in \autoref{sec:appspin}, and present a summary in \autoref{sec:summary}.
\section{PSF Deconvolution of IFS Data}\label{sec:psfdeconv}

\subsection{Lucy-Richardson Deconvolution Algorithm}\label{sub:lrdeconv}
Lucy-Richardson (LR) deconvolution algorithm is an iterative procedure to recover an image which is blurred (convolved) by a PSF. The algorithm is introduced here in a simple form,
    \begin{equation}\label{eq:lr}
    u^{{n+1}} = u^{{n}} \cdot \Bigg(\frac{d}{u^{{n}} \otimes p } \otimes p\Bigg)
    \end{equation}
where $u^{n}$ is $n^{th}$ estimate of the two-dimensional maximum likelihood solution ($u^0 = d$), $d$ is the original PSF-convolved image, $p$ is 2D PSF, and $\otimes$ denotes 2D convolution. If $d$ follows the Poisson Statistics and $u^{n}$ converges as iteration proceeds, $u^{n}$ becomes the maximum likelihood solution \citep{4307558}. The LR deconvolution method has several advantages that 1) it is straight-forward to implement, 2) it requires only a few parameters to perform, 3) it can perform fast on an average computing machine (takes less than 4 minutes on 2.67 GHz single core CPU when applied to  72 $\times$ 72 $\times$ 4563 cube ($\rm x \times y \times wavelength$)). If the shape of the PSF is known as Gaussian, 
then only two parameters are required to the procedure:
1) Full-Width-Half-Maximum (FWHM) of Gaussian and 2) a number of iterations. The algorithm produces a non-negative solution since it assumes Poisson Statistics. However, there are well-known drawbacks of the algorithm, which are 1) the noise amplification, and 2) the ringing artifact structure around sharp feature, both happen as the number of iteration ($\niter$) increases \citep{1998ApJ...494..472M}. 
Therefore, the relation between the number of iteration and the quality of the deconvolved data should be investigated before using the deconvolved data for further scientific analysis.

\subsection{Implementation to the IFU Data}\label{sub:dcapply}
We develop a {\python} code to apply the LR deconvolution algorithm to an optical IFU data. We consider an IFU data as a combination of 2D images at multiple wavelength bins, and perform deconvolution method to the 2D image slice at each wavelength bin independently. In other words, we apply the deconvolution method only in the spatial directions, not in the spectral direction. The core part of the procedure is written to follow \autoref{eq:lr}. We implement Fast-Fourier Transform (FFT)\citep{cooley1965algorithm, press2007numerical} to increase the speed of the procedure. The algorithm requires 2D image of PSF which has identical size to the input 2D image slice. Here we use 2D Gaussian function image as a PSF but it can be any other shape in practice. 

To cope with the wavelength dependency of the $\fpsf$ size, we assume the size of $\fpsf$ as a linear function of wavelength, and deconvolve 2D image slice at each wavelength bin with corresponding $\fpsf$ size.
We apply a zero padding on the 2D slice image to increase its size to $\rm{2^N} \times \rm{2^N}$ before deconvolution to maximize the execution speed of FFT.
After the zero padding, the zero-padded pixels, bad pixels and all non-positive pixels are marked. 
The marked pixels are replaced by proper non-zero values to avoid having oscillation feature around the masked pixels or having invalid pixel values after the deconvolution.
These marked pixels are substituted by an iterative value-correction process, which alters the marked pixels to the average of the nearest positive pixel values. The value-correction process is applied multiple times until the boundary of the data is extended by three times of $\fpsf$. This process significantly reduces the artificial effect due to the sharp edge in the result of the deconvolution. 
Finally, the LR deconvolution algorithm is performed on the value-corrected $\rm{2^N}$ $\times$ $\rm{2^N}$ size image. The values which were replaced by the value-correction process are masked to zero after the deconvolution, and the padded region is cut out. We present the deconvolution code in {\python} for public use available on \github\footnote{\url{http://github.com/astrohchung/deconv}. An example code to deconvolve a MaNGA IFU data and compare the 2D kinematics measured from the original and the deconvolved MaNGA data is provided (Partially reconstruct \autoref{fig:sdssex}).} under a MIT License and version 1.0 is archived in Zenodo \citep{chung2021code}. 

\section{Deconvolution method parameter determination and performance test}\label{sec:pvmock}
In this section, we verify the reliability of the deconvolution method and also determine the proper value of the deconvolution parameter, which is the number of iteration. We also check the acceptable range of the other deconvolution parameter, $\fpsf$, when the value is different from the correct $\fpsf$ value which is originally imprinted to the PSF-convolved IFU data. We use three sets of mock IFU data: first one where no PSF is convolved, second one where a PSF is convolved to the first one, and third one where the PSF is deconvolved from the second one by our deconvolution method. The first set of mock IFU data is generated by using a model galaxy with various photometric and kinematic parameters. We use differences between the true galaxy model parameter values and the corresponding parameter values which are extracted from the PSF-deconvolved mock IFU data as metrics of the deconvolution performance. Using those metrics, we quantify the effect of the deconvolution, and determine the proper deconvolution procedure parameter values ($\niter$ and $\fpsf$).

\begin{figure}
\includegraphics[width=\linewidth]{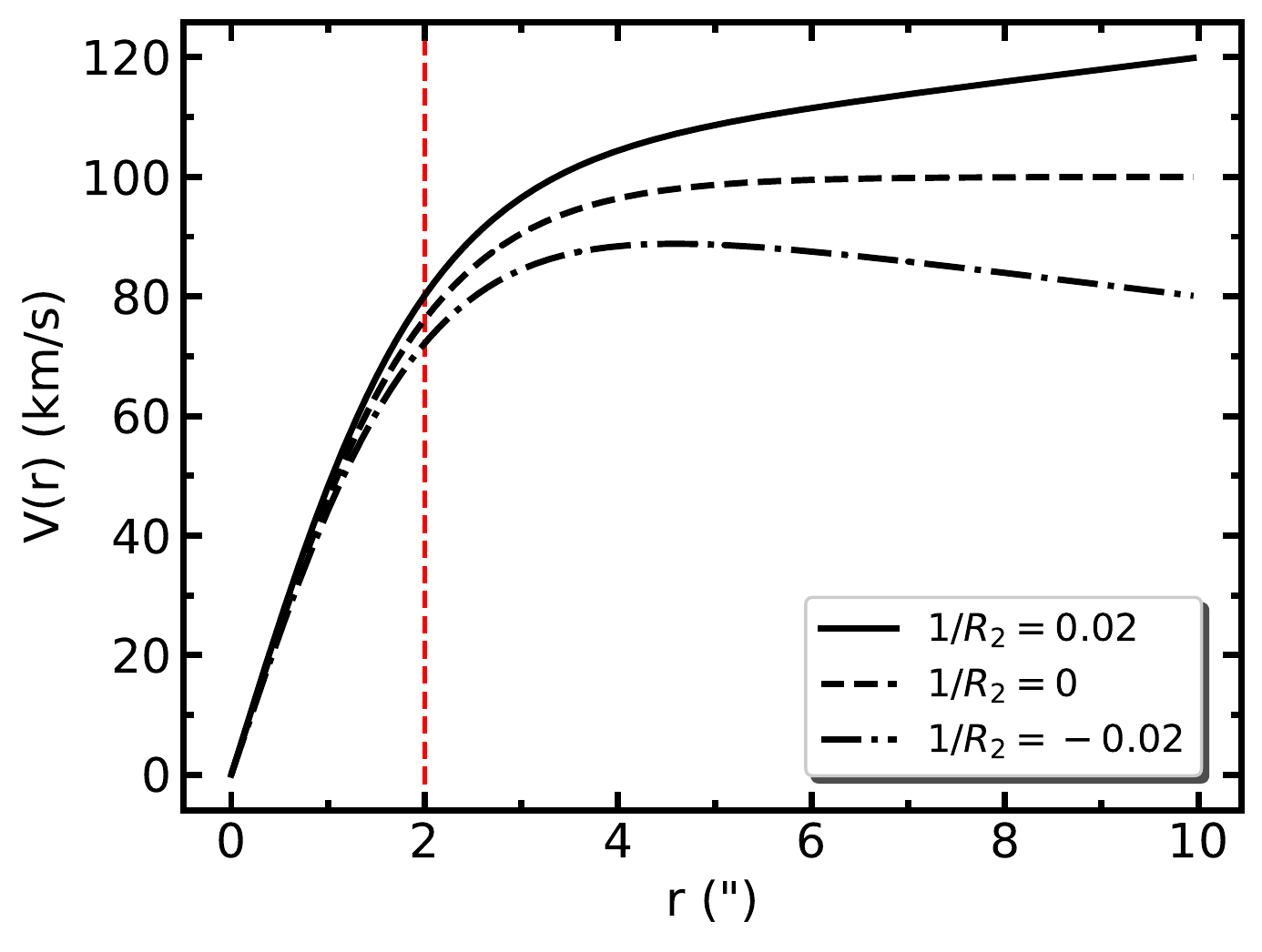}
\caption{Example of rotation curve model. Each line shows different shape at the outskirt
described by $1/R_2$ value (when $V_{\rm{ROT}}=100$ km/s and $R_1$=2${\arcsec}$) Vertical dotted line indicates $\rm R_1$.}
\label{fig:vrfunc}
\end{figure}

\subsection{Mock Galaxy Model}\label{subsec:rcmodel}
We define a mock galaxy model which resembles an actual rotating galaxy. Our mock galaxy is composed of simple photometric and kinematic models, which are flux distribution with the S\'{e}rsic profile and kinematic distribution with thin-disk approximated galaxy rotation curve (RC) function and a simple radial velocity dispersion function. 

We use a model of galaxy with infinitely thin-disk shape and ordered rotation.
There are several functional forms to describe the typical shape of the disk galaxies: an arc tangent \citep{2008A&A...484..173P}, a hyperbolic tangent \citep{2013ApJ...768...41A}, and an inverted exponential \citep{2011RAA....11.1429F}. 
All these model have a RC converging to a constant velocity at their outer radii,
namely the well-known $flat$ rotation curve. 
Although it is non-trivial to describe the complex shape of the real RC in a simple form, we try to improve the current model while maintaining its simple form.
We propose the following RC model which is a combination of the hyperbolic tangent function and a linear term,

\begin{equation} \label{eq:vr}
    V(r) = V_{\rm ROT}\;\Bigg[\rm tanh\Bigg(\frac{r}{R_{1}}\Bigg)+\frac{r}{R_{2}}\Bigg]
\end{equation}

where $V_{\rm ROT}$ is a maximum circular velocity if $1/{R_2}=0$, $R_1$ is a characteristic radius where the curve slope changes, and $1/{R_2}$ is the slope of the curve at its outer radii. \autoref{fig:vrfunc} shows a few examples of this model with different signs of $1/{R_2}$. The advantage of this model is that it can describe the inclined/flat RC at the outer radii, as well as the rigid body rotation motion at near the center of galaxy, which are typically observed in the rotation curve of real galaxies. We would like to point out that there is a degeneracy between $R_1$ and $1/{R_2}$ in term of the shape of the curve. For example, the shape of RC model with certain $R_1=a$ and $1/{R_2}=b$ is identical to the other RC model with $R_1=ca$ and $1/{R_2}=b/c$. Therefore, to compare the shape of different set of our RC model parameters, a normalized RC outer radius, $R_1/{R_2}$, should be used.

\begin{deluxetable}{cc}
\tabletypesize{\footnotesize}
\tablecolumns{2}
\tablewidth{0pt}
 \tablecaption{ Mock IFU Data Parameters (Group 1 and 2) \label{tab:grtb}}
 \tablehead{ \colhead{Parameter} &  \colhead{Value} 
}
 \startdata  
 IFU field of view ($\arcsec$)     &  32        \\
 IFU radial coverage in $\Reff$     & 2.5               \\
S/N at 1$\Reff$         &  10, 20, 30                              \\
S\'{e}rsic index        &  1, 4                                     \\
Inclination  ($\arcdeg$)        &  40, 55, 70                            \\
Position Angle ($\arcdeg$)        &  15                           \\
$V_{ROT}$ (km/s)               &  200                         \\
 & 3, -0.05 \\ 
 & 3, 0.00 \\
$R_1$ ($\arcsec$), $1/R_2$ ($1/\arcsec$) & 3, 0.05 \\ 
& 2, 0.05 \\ 
& 4, 0.05 \\
$\sigma_0$ (km/s)             &  150         \\                       
$\sigma^{\prime}$              &  1         \\                       
\multirow{2}{*}{FWHM coefficient $c_0$ ($\arcsec$)} &  2.6 (Group 1)  \\
                                                &  2.3, 2.6, 2.9 (Group 2)   \\
FWHM coefficient $c_1$ ($\times 10^{-5} \; \arcsec/$\AA) & -1.2 \\
Redshift             &  0.02         \\
 \enddata
 
 \tablecomments{S/N at 1 $\Reff$ is defined as median S/N of a spaxel around 1 $\Reff$ per spectral element. }

\vspace{-0.5cm}
\end{deluxetable}

We use a line-of-sight velocity dispersion function as follows,
\begin{equation}\label{eq:sig}
    \sigma_r=\frac{\sigma_0}{{\sigma^{\prime}}r/R_{1}+1},
\end{equation}
where $\sigma_0$ is a velocity dispersion at the center, $r$ is a circular radial distance from the center of a galaxy, and $R_1$ is a characteristic scale which is set to be identical to the one in the RC model. 
The slope of $\sigma_r$ is mainly described by the $R_1$, but $\sigma^{\prime}$ is introduced to provide an additional freedom to the slope (\autoref{sub:spinmock}). The $\sigma_r$ form is taken from \citet{2018MNRAS.477.4711G} with slight modification, and it also well describes the actual velocity dispersion distribution of galaxies (see \autoref{sub:sdssresult}).

\begin{deluxetable}{cc}

\tabletypesize{\footnotesize}
\tablecolumns{2}
\tablewidth{0pt}

 \tablecaption{ Mock IFU Data Parameters (Group 3) \label{tab:mctb}}

 \tablehead{ \colhead{Parameter} &  \colhead{Value} 
}

 \startdata  
 IFU field of view (\arcsec)     &  12, 17, 22, 27, 32        \\
 IFU radial coverage in $\Reff$         & 1.5, 2.5              \\
S/N at 1$\Reff$         &  10 - 30                              \\
S\'{e}rsic index        &  1, 4                                     \\
Inclination  (\arcdeg)        &  10 - 80                            \\
Position Angle (\arcdeg)        &  15                           \\
$V_{ROT}$ (km/s)               &  50 - 300                         \\
$R_1$  (\arcsec)                  & 1 - 4                               \\
$1/R_2$  (1/\arcsec)              &  -0.1  -  0.1                    \\
$\sigma_0$ (km/s)             &  50 - 300                         \\
$\sigma^{\prime}$              &  1         \\                       
FWHM coefficient $c_0$ (\arcsec)       &  2.3 - 2.9           \\
FWHM coefficient $c_1$ ($\times 10^{-5}$ \arcsec/\AA)      &  -3.6 - 1.2       \\
Redshift             &  0.02         \\
 \enddata
 \tablecomments{S/N at 1 $\Reff$ is defined as median S/N of a spaxel around 1 $\Reff$ per spectral element. When values of a parameter are listed with comma, one of the value is randomly selected. When values of a parameters is given in range (with hyphen), value is selected randomly within the range.}
 
\vspace{-0.5cm}
\end{deluxetable}

\subsection{Mock IFU Data}\label{subsec:mockifs}
We generate three $groups$ of mock IFS data using the fore-mentioned photometric and kinematic galaxy model. 
Each group of mock IFU data is determined by multiple $sets$ of model parameters, and each mock IFU data is generated to follow the two-dimensional velocity, velocity dispersion and flux distribution determined by a set of model parameters. The detail of mock IFU generation process is described in \autoref{app:mockgen}. Here, we only describe the composition of each mock IFU data group. 

The purpose of Group 1 is to investigate the performance of the deconvolution with respect to the number of deconvolution iterations. 
We determine sets of model parameters as in \autoref{tab:grtb} to elaborate the diverse properties of galaxies. We use the realistic model parameters which could represent the photometric and kinematic distributions of actual galaxies such as the target galaxies of SDSS-IV MaNGA IFU survey. The S/N at one half-light radius (1 $\Reff$) is defined similarly to the MaNGA data, where ranges S/N=14-35 per spatial element per spectral resolution element in $r$ band at 1 $\Reff$ \citep{2015ApJ...798....7B}. We also choose the shape and size of mock IFU field of view as same as the MaNGA IFU data, which has hexagonal shape with the field of view size of 12$\arcsec$ to 32$\arcsec$ in vertex to vertex with the size of spatial element as 0.5$\arcsec$ $\times$ 0.5$\arcsec$.
Combination of each parameter; S/N at 1 $\Reff$, S\'{e}rsic index ($\nsrc$), inclination angle, $R_1$, and $1/R_2$ yields 90 sets of mock galaxies ($3 \times 2 \times 3 \times 5$ = 90) (see \autoref{subsec:rcmodel} for the definition).
For each set of galaxy parameters we construct three types of mock IFU data. Type 1 ($\free$) is an ideal IFU data without any PSF convolution or noise (i.e. free from atmospheric seeing effects and optical aberrations). Type 2 ($\ndc$) is a realistic IFU data where Gaussian PSF is convolved and the Gaussian noise is added. Type 3 ($\dc$) is a PSF-deconvolved IFU data which is obtained by performing the deconvolution method to the Type 2 IFU data. 
We generate 25 $\ndc$ IFU data from each $\free$ mock IFU data by adding Gaussian random noise with 25 different random seeds. 
By using the distribution of the parameters measured from mock IFU data with different random noise, we obtain the statistical distribution of each extracted galaxy model parameters.
Also, we assume the wavelength dependent $\fpsf$ which corresponds to the $\rm{FWHM}_{\rm{PSF, \lambda}}=c_0 + c_1 \times \rm{\lambda}$, where $c_0$ and $c_1$ are as in the \autoref{tab:grtb}. This wavelength-dependent PSF is to represent the wavelength dependency of PSF in the real data (\autoref{subsec:mangapsf}). Lastly, 50 $\dc$ IFU data are produced per each $\ndc$ IFU data with $\niter$ = 1 to 50. In total, 90 $\free$, 2,250 $\ndc$, and 112,500 $\dc$ mock IFU data are produced as Group 1. 

Group 2 is designed to investigate the impact of the two types of $\fpsf$ value to the performance of the deconvolution method; 1) $\fndc$ value which was convolved to the PSF-Free IFU mock data, and 2) $\fdc$ value which is used for the deconvolution procedure. 
This is to verify the effect of deconvolution in practical situation where 1) each IFU data is observed with various atmospheric seeing size and 2) the $\fdc$ being different from the actual effective $\fndc$. These effects are identified to ensure that the deconvolution provides more accurate kinematics compare to the one from the non-deconvolved data even with a little inaccurate $\fdc$. We again construct three types of mock IFU data using the parameters given in \autoref{tab:grtb}. Group 2 - Type 1 data is identical to the Group 1 - Type 1 data. For each of the Group 2 - Type 1 $\free$ IFU data, we produce 75 $\ndc$ IFU data by using 3 different $c_0$ values and the 25 different random noise seed per each $c_0$ value. 13 $\dc$ IFU data are produced per each $\ndc$ IFU data with 13 different $\fdc$ values, which ranges within $\pm$ 0.3$\arcsec$ from the $c_0$ value with 0.05$\arcsec$ interval. 
\added{
We have chosen $\pm$ 0.3$\arcsec$ range considering the {$\fpsf$} distribution of actual IFU survey data (2.2$\arcsec$ to 2.7$\arcsec$ in $g$-band, \autoref{fig:fwhm}, \autoref{subsec:mangapsf}). It is known that such {$\fpsf$} could vary up to 10\% of over the field of view of a single IFU \citep{2016AJ....152...83L}. This means that an average $\fpsf$ difference over a single IFU field of view will be less then 0.3$\arcsec$. Thus, the $\pm$ 0.3$\arcsec$ range is an extreme case where the entire {$\fpsf$} used for the deconvolution is $\sim$10\% larger or smaller compared to the actual {$\fpsf$}. 
}
$\niter$ is fixed as 20 times. In total, 90 $\free$, 6,750 $\ndc$, and 87,750 $\dc$ mock IFU data are produced as Group 2. 

Lastly, we produce Group 3 data using a range of mock galaxy model parameters as in \autoref{tab:mctb}. This is to verify the performance of deconvolution in more diverse combination of galaxy photometric and kinematic distributions. 40,000 sets of galaxy model parameters are determined randomly in Monte-Carlo way, and 1 $\free$, 1 $\ndc$, 1 $\dc$ mock IFU are generated for each set. In total, 40,000 $\free$, 40,000 $\ndc$, and 40,000 $\dc$ IFU data are produced as Group 3.

\subsection{Kinematics Measurement and Rotation Curve Model Fitting}\label{sub:mockkm}
We measure the line-of-sight kinematics from the mock IFU data produced in \autoref{subsec:mockifs} and fit the RC model on the measured 2D kinematic distribution to extract the RC model parameter values.
We use an \idl version of the Penalized-Pixel Fitting ($\ppxf$)\citep{2004PASP..116..138C, 2017MNRAS.466..798C} procedure to extract the Line-Of-Sight-Velocity-Distribution (LOSVD) from the mock IFU data. To minimize the $\ppxf $ computation time, we use model SEDs identical to the ones that we used for the mock generation (see \autoref{app:mockgen}), and fit only the velocity and the velocity dispersion without any additive or multiplicative Legendre polynomials or high-order kinematic moments. Following the recipe from \citet{2017MNRAS.466..798C}, 1) we match the spectral resolution of the model SED to that of the mock IFU data, and 2) we de-redshift the mock IFU spectra to the rest frame before extracting the LOSVD. We also masked the wavelength around the known emission lines, although there is no emission line in the mock IFU spectra. Considering the wavelength coverage of the mock IFU data (3,540 to 7,410 \AA; see \autoref{app:mockgen}), we limit the fitting wavelength range as from 3700 to 7400 \AA $\;$ for the LOSVD measurement. 

We fit our RC model (\autoref{eq:vr}) to the extracted 2D velocity map of mock galaxies to quantify the shape of the rotation curve. From the fitting, we obtain the RC model parameters ($V_{ROT}, R_1, 1/R_2$) and the kinematic geometrical parameters (center x, center y, position angle and inclination angle). The fitting procedure uses the minimum $\chi^2$ method that finds a set of parameters which is minimizing the $\chi^2$ between the true 2D velocity map and the measured 2D velocity map. The following equation describes the 2D model velocity map,
    \begin{equation}\label{eq:vobs}
    V_{\rm obs}(r',\phi') = V_{\rm SYS}+V(r)\;{\rm sin}\,i\; {\rm cos}(\phi-{\phi}_{0})
    \end{equation}
where $r'$ is the distance from the kinematic center of the galaxy to each pixel on the sky, $r$ is galaxy-centric radius in the de-projected plane, $V_{\rm SYS}$ is a systematic line-of-sight velocity of the kinematic center, $i$ and $\phi_{0}$ are kinematic inclination angle and the position angle in the observed (projected) plane. Including the delta $\Delta$x and $\Delta$y from the kinematic center position in the observed plane, eight parameters are fitted simultaneously ($V_{SYS}$, $V_{ROT}$, $R_1$, $1/R_2$, $i$, $\phi_0$, $\Delta x_{cent}$, and $\Delta y_{cent}$).

The minimum $\chi^2$ method is sensitive to the initial values when there are multiple fitting parameters, in particular for the geometrical parameters ($i$, $\phi_0$, $\Delta x_{cent}$, and $\Delta y_{cent}$). To fit the 2D RC model with suitable initial parameter values, we first fit the 2D S\'{e}rsic model to the reconstructed $g$-band image of the mock IFUs before fitting the RC model. The geometrical parameters obtained from 2D S\'{e}rsic model are used as the initial value of the 2D RC model fitting.   

There are two caveats in fitting our RC model.
\begin{enumerate}
\item The velocity map should cover sufficiently large radial range along the major axis compare to the $R_1$, otherwise the $1/R_2$ parameter cannot be accurately determined. In particular, it is important to have sufficient radial coverage along the \textit{major} axis. The radial coverage along the minor axis contributes significantly less than the coverage along the major axis to the RC model fitting, because of the ${\rm cos}(\phi-{\phi}_{0})$ term in \autoref{eq:vobs}.
\item The 2D RC model is less sensitive to the galaxies with too low or too high inclination angle. Due to the ${\rm sin}\,i$ term in the \autoref{eq:vobs}, $V_{ROT}$ term is often inaccurately measured at low inclination angle (close to face on). At high inclination angle, the fitting result is not reliable because of the relatively small number of data points along the major axis, and the significant PSF convolution effects which scrambles the information between the measured quantities on and out of the major axis, even in the PSF-deconvolved mock IFU data.
\end{enumerate}
In \autoref{sec:rcmodelvalid}, we analyze the RC model fitting result of the Group 3 mock IFU data and derive analytic criteria to ensure the accuracy of the RC model fitting result. We find that when the result satisfies $R_{max, S/N >3, major} / R_1 > 2.5$ and the fitted inclination angle falls on $75 \arcdeg > i > 25 \arcdeg$, the fitting results are considered reliable. 
In addition, we find that the model parameter values measured from the mock IFU data with field of view equal to 12$\arcsec$ are not well recovered because of insufficient number of valid data points (S/N $>$ 3) in such a narrow field of view with a given spatial element size (0.5$\arcsec$ by 0.5 $\arcsec$).
In further analysis, we only consider the fitting results those are satisfying the above criteria ($R_{max, S/N >3}^{major} / R_1 > 2.5$ and $75 \arcdeg > i > 25 \arcdeg$).

\subsection{Results and Discussion}
In this subsection, we present the performance of our deconvolution method by using the mock IFU data.
We show the relation between the restored kinematics and the deconvolution parameters ($\niter$, $\fdc$) and discuss the adequate choice of the deconvolution parameters. 
Lastly, we demonstrate the feasibility of applying our deconvolution method to more generalized cases, by showing the test result of the deconvolution method to mock IFU data with various combination of the galaxy surface brightness distribution, galaxy inner and the outer kinematics, its geometry, radial coverage and S/N of data, geometry, and size of the convolved PSF size.

\begin{figure*}
\includegraphics[width=\linewidth]{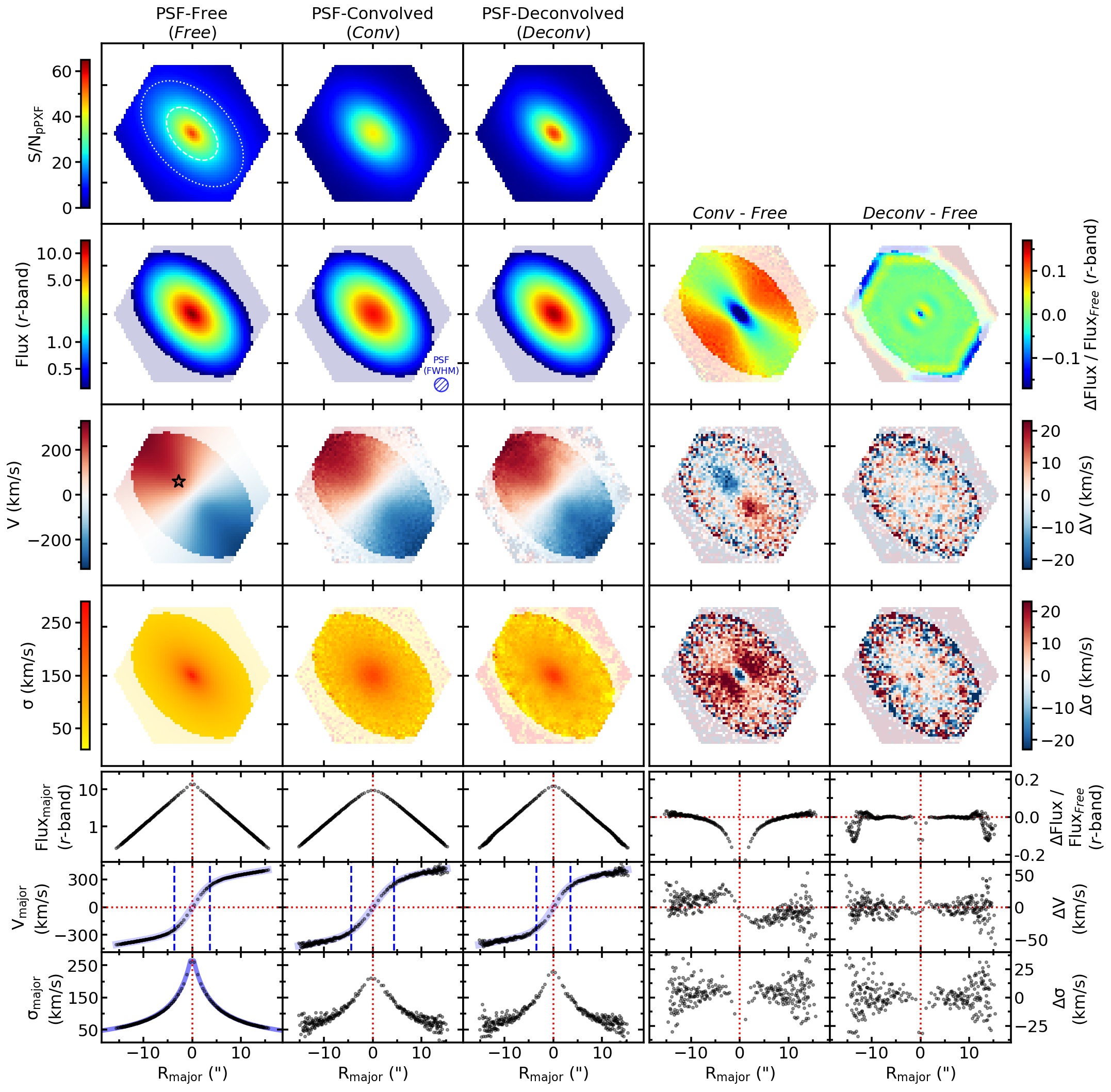}
\caption{ 
Plots demonstrating the effects of the PSF convolution and deconvolution on the 2D maps of S/N, $r$-band flux, line of sight velocity (V), and velocity dispersion ($\sigma$). 1D radial profiles of the $r$-band flux, line of sight velocity, and velocity dispersion along the major-axis are also shown.
The first, second, and third column represent the 2D or 1D distribution of the measured quantities from PSF-Free ($\free$), PSF-Convolved ($\ndc$), and PSF-Deconvolved ($\dc$) mock IFU data. The mock IFU data is selected from Group 3 Monte-Carlo mock IFU samples (see text).
The fourth (fifth) column show the difference between the quantities from the $\ndc$ ($\dc$) and the $\free$ mock IFU data, respectively.
The size of the major tick in the 2D maps is 10$\arcsec$. A dashed (dotted) ellipse is over-plotted on the top left corner panel to represent the size of 1$R_e$ (2$R_e$). FWHM of the convolved PSF is shown as a blue hatched circle in the Flux - $\ndc$ panel. Black open star on V - $\free$ panel is the location of the example spectrum in \autoref{specex}. Spaxels with S/N $<$ 3 are paled out in the 2D maps except for the S/N map. Only data points within $\pm 5 \arcdeg$ of major axis are shown in the radial profiles for clarity. Blue paled-out lines are under-plotted on the $V_{major}$ profiles ($\free, \ndc, \dc$) and the $\sigma_{major}$ profile ($\free$ only) to represent the fitted RC (and $\sigma$) model functions. Blue vertical dashed lines in the $V_{major}$ profiles denote $R_1$ of the corresponding fitted RC model function. }
\label{fig:dcex}
\end{figure*}

\subsubsection{Effects of PSF Convolution and Deconvolution}\label{subsub:dcex}
\autoref{fig:dcex} shows the effects of PSF convolution and deconvolution by using test result from
one of the Group 3 (Monte-Carlo) mock IFU data ($\nsrc=1$, ${S/N}_{1 \Reff}=25$, $i=48\arcdeg $,  $V_{ROT}$=212 km/s, $R_1$=3.7$\arcsec$, $1/R_2$=0.02 (1/$\arcsec$), $\sigma$=74 km/s,  $\rm FWHM_{PSF}$=2.88$\arcsec$, field of view = 32$\arcsec$).
Panels on the leftmost column show the 2D or 1D quantities measured or extracted from the $\free$ IFU data. The quantities match very well with the model 2D photometric and kinematic distributions which we put into, meaning that the mock IFU data is constructed accurately in accordance with the model parameters.  The second left column presents distributions from $\ndc$ IFU data, and the second right column displays the difference between the leftmost and the second left column.

As expected, the panels clearly exhibit noticeable changes in all three quantities (flux, velocity, and the velocity dispersion) caused by the PSF convolution.
The difference in $\rm{Flux}_{major}$ and $\rm{V}_{major}$ 1D profiles (the second right column) also show evident deviation between the $\ndc$ and the $\free$. In particular, the characteristic radius of the rotation curve ($R_1$, represents the size of the inner linear part) is increased by the PSF convolution. The overall velocity dispersion around the center is also increased, but at the very center the dispersion is decreased. This is caused by the combination of the PSF convolution effects on the line-of-sight (LOS) velocity, and the velocity dispersion distribution. 
The convolved PSF \textit{increases} the velocity dispersion, in particular along the minor axis because of the the opposite direction of LOS velocity around the minor axis. On the other hand, the convolved PSF smooths the velocity dispersion distribution so that it 
\textit{decreases} the velocity dispersion at the center but \textit{increases} the dispersion around the center because the center has both the brightest point and the highest velocity dispersion. 

The central column presents the distributions from $\dc$ IFU data, and the rightmost column shows the difference between the $\dc$ and $\free$. It is clear that the difference between the $\dc$ and the $\free$ is significantly less than the same between the $\ndc$ and the $\free$. 
Compare to the $\ndc$ column, the apparent b/a ratio is decreased, the flux at the center is increased, and the $\rm R_1$ of the rotation curve is now much closer to the one from the $\free$ column. The difference in both velocity and velocity dispersion distribution is also much diminished. This result clearly exhibits that the flux, velocity, and the velocity dispersion distribution from the PSF-deconvolved IFU data are indeed well-recovered toward the true distributions.
However, the distribution near the edge of the galaxy becomes fuzzier and shows some systematic feature, in particular in the flux distribution. This is partially due to the low S/N near the edge of mock IFU data, and partially due to the edge effect of the deconvolution. We would like to point out that the edge effect in this example is already significantly reduced by the iterative value-correction process (see \autoref{sub:dcapply}). Without the iterative value-correction process, the edge effect makes distinctive artificial hexagonal shape oscillating pattern on the entire image.  We put additional examples of Group 1 mock IFU data in \autoref{app:dcex} to show the result with different input distributions. The examples in \autoref{app:dcex} demonstrate that the deconvolution method on IFU data is working effectively well and the method restores the distributions of photometric and kinematic quantities close to the true distributions.

\begin{figure}
\includegraphics[width=0.49\textwidth]{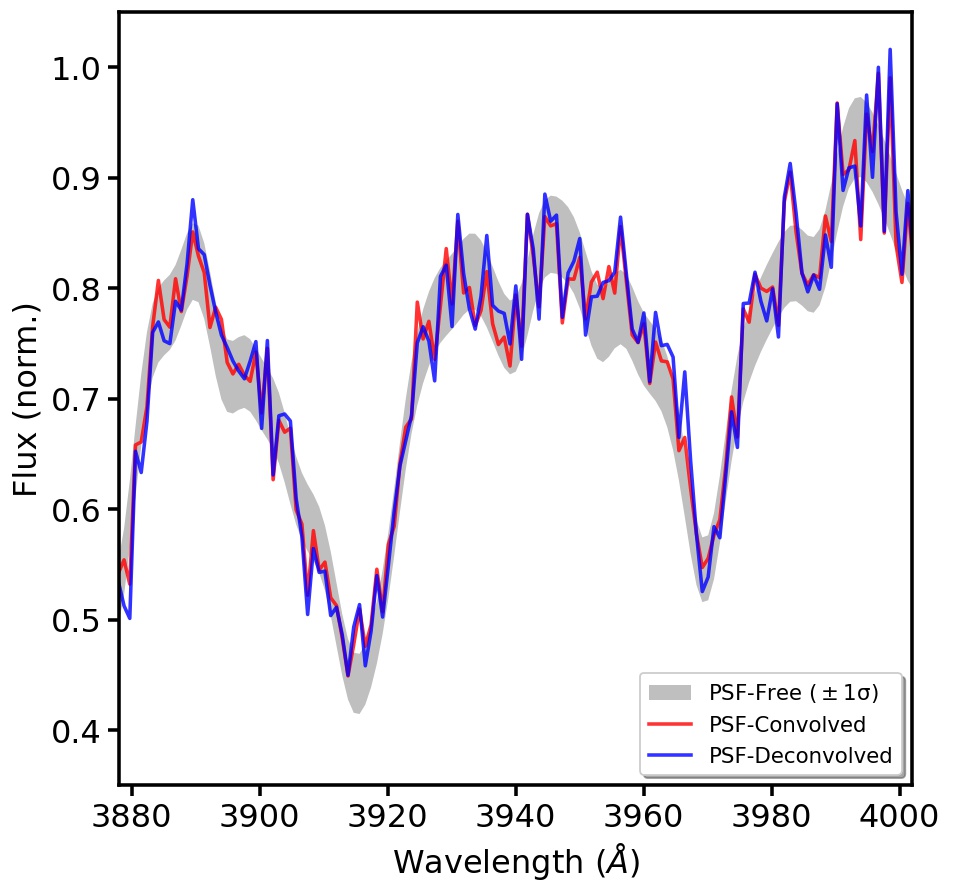}
\caption{Spectrum of a spaxel whose $\Delta V$ between the $\ndc$ and the $\free$ is -20 km/s. Only the spectra around the Ca H\&K lines are shown. The location of this spaxel is marked as a black open star in \autoref{fig:dcex}. Median S/N of this spaxel is 36. Each spectrum is normalized by the median of each to show only the difference between spectra in their shape. 
The thickness of the spectrum from $\free$ (grey) represents $\pm 1 \sigma$ error at each wavelength bin. 
}\label{specex}
\end{figure}

\begin{figure*}
\includegraphics[width=\textwidth]{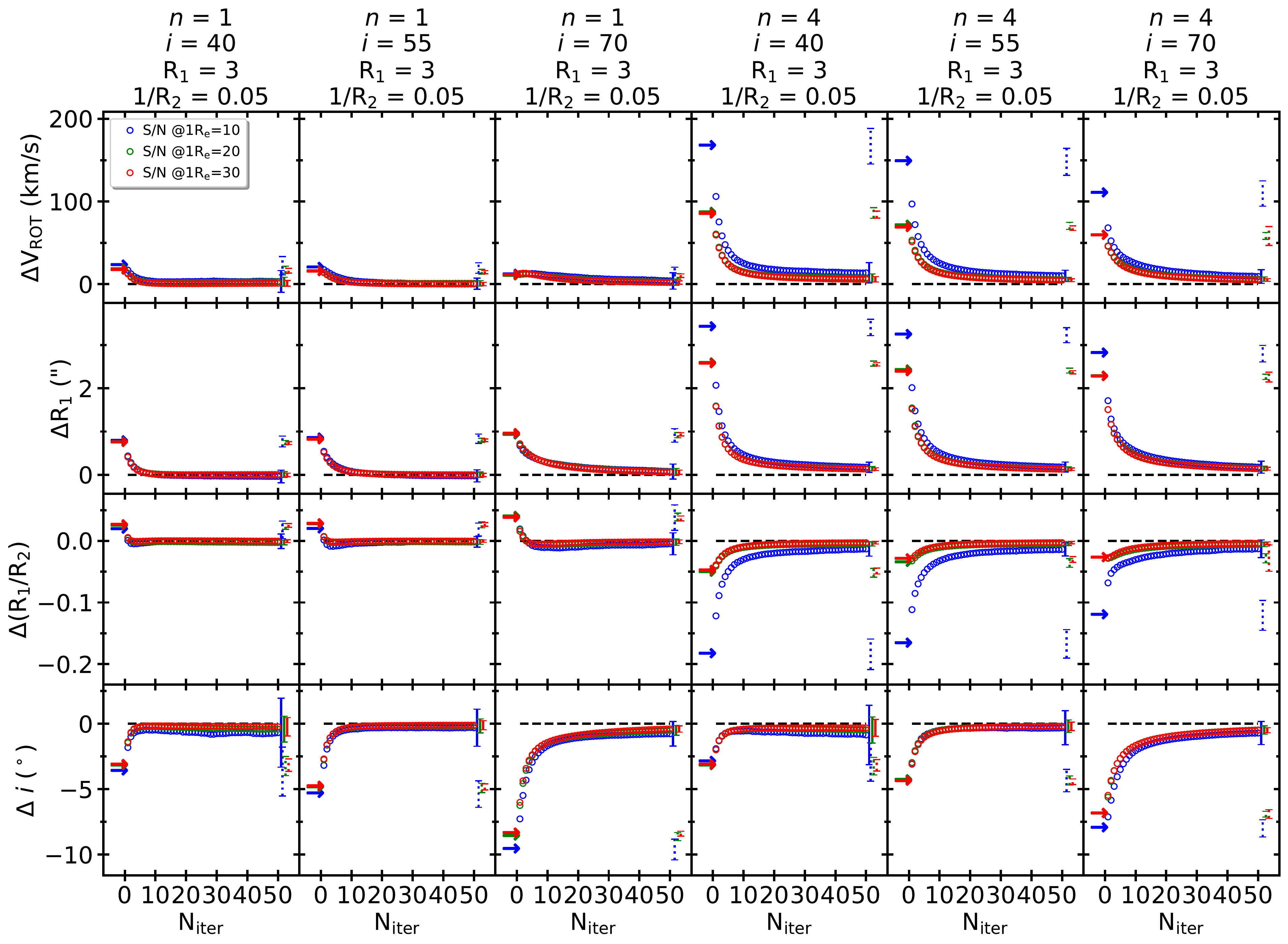}
\caption{
Difference between the true RC model parameters and the fitted model RC parameters from the PSF-deconvolved mock data ($\dc$)  with respect to the number of LR deconvolution iteration ($\niter$ = 1 to 50). 
Each column shows the results from the mock IFU data with different S\'{e}rsic index ($\nsrc$) and kinematic inclination angle ($i$). 
In each panel, $\rm\Delta$ of one of the four RC model parameters ($\rm \Delta{V}_{ROT}, \Delta{R}_{1}, \Delta({R}_{1}/{R}_{2})$, and $\rm\Delta{i}$) with different number of deconvolution iteration ($\niter$) are plotted as open circles. Color represents the S/N at 1 $\Reff$ value that is used to generate the corresponding PSF-free mock IFU data ($\free$) of each open circle.
Black dashed lines are plotted at the difference of 0 as a guidance.
Arrow points out the value of the fitted parameters from the PSF-convolved mock IFU data ($\ndc$).
For clarification, we put only one solid error bar per S/N at 1 $\Reff$ value in each panel instead of putting error bars on every open circles. The error bar represents the standard deviation of each $\rm\Delta$ parameter values from 25 different random seeds (There is almost no dependency of the standard deviation of the $\rm\Delta$ parameters with respect to $\niter$). The dotted error bar is corresponding standard deviation from the difference between the true RC model parameter value and the fitted model RC parameter value of the PSF-convolved ($\ndc$) mock IFU data.
\label{fig:niter_diffincl}}
\end{figure*}

\begin{figure*}
\includegraphics[width=\textwidth]{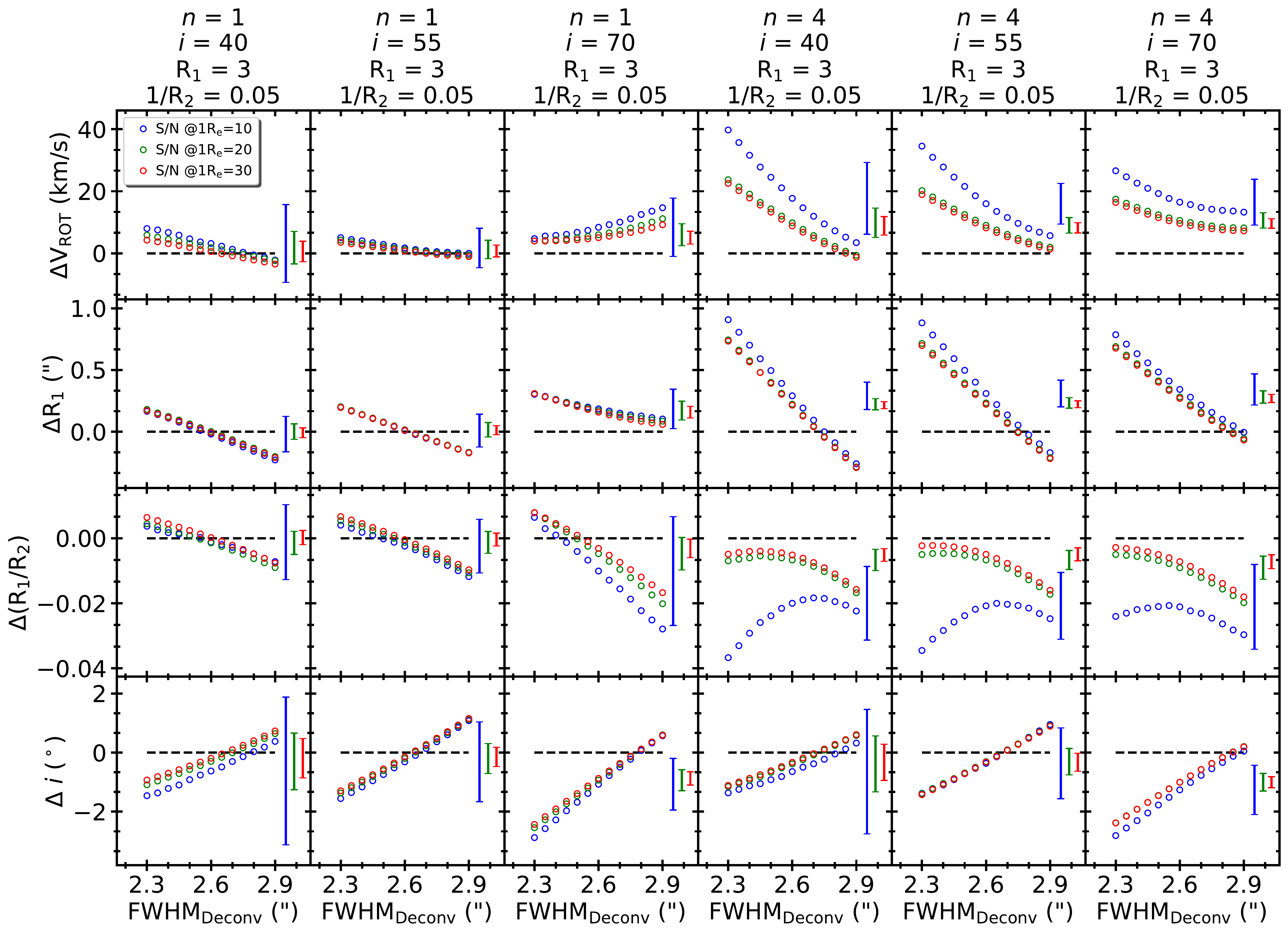}
\caption{
Difference between the true RC model parameters and the fitted model RC parameters from the PSF-deconvolved mock data ($\dc$)  with respect to the FWHM of PSF used for the deconvolution (with fixed $\niter=20$). 
Each column shows the results from the mock IFU data with different S\'{e}rsic index ($\nsrc$) and kinematic inclination angle ($i$). 
In each panel, $\rm \Delta$ of one of the four RC model parameters ($\rm \Delta{V}_{ROT}, \Delta{R}_{1}, \Delta({R}_{1}/{R}_{2})$, and $\rm \Delta{i}$) with different PSF FWHM size used for the deconvolution ($\fdc$) are plotted as open circles. Color represents the S/N at 1 $\Reff$ value that is used to generate the corresponding PSF-free mock IFU data ($\free$) of each open circle. Again, black dashed lines are plotted at the difference of 0 as a guidance. 
For clarification, we put only one solid error bar per S/N at 1 $\Reff$ value in each panel instead of putting error bars on every open circles. The solid error bar represents the standard deviation of each $\rm \Delta$ parameter values from 25 different random seeds, as in \autoref{fig:niter_diffincl} (There is almost no dependency of the standard deviation of the $\rm\Delta$ parameters with respect to $\fdc$). Note that y-axis scale of in this figure is smaller than that of \autoref{fig:niter_diffincl}.
\label{fig:fwhm_diffincl}}
\end{figure*}

We visualize the effect of the deconvolution method in the wavelength dimension in \autoref{specex}. \autoref{specex} shows an example of spectra at the spaxel where $\Delta V$ between the $\ndc$ and the $\free$ is about -20 km/s. Because the size of one wavelength bin of the spectrum corresponds to 69 km/s, $\Delta V$ of -20 km/s ($\sim$0.29 pixel) is hardly recognized between the spectra by eyes, even around the strong absorption lines. It is also noticed that the $\dc$ spectrum is slightly noisier than the $\ndc$ spectrum. 
The mean difference between the $\ndc$ and the $\free$ spectrum at this spaxel is 4.0\%, but the corresponding difference between the $\dc$ and the $\free$ spectrum is 4.6\%.
In fact, the noisier $\dc$ spectrum is expected by the effect of LR deconvolution algorithm (noise amplification). Although the $\dc$ spectrum is noisier than the $\ndc$ spectrum, the overall shape of the $\ndc$ spectrum has changed and shifted through the deconvolution process, and the line-of-sight velocity and the velocity dispersion of the $\dc$ spectrum are better recovered to the true value.

\subsubsection{Deconvolution Parameters}\label{subsub:dcparams}
\autoref{fig:niter_diffincl} represents the difference between the fitted and the true RC model parameter value as $\niter$ increases from 1 to 50. Error bar is calculated from the 25 mock IFU data with different random seeds which we implemented for the noise realization. Since the deviation from the true value depends on the galaxy model parameters, we show the result from multiple model galaxies at each column from the Group 1 mock IFU data. The figure shows the case of the mock data with $\nsrc$ = 1, 4 and $i$ = 40, 55, 70$\arcdeg$. 
Here we present the difference in $R_1/R_2$ rather than $1/R_2$, because $R_1/R_2$ value better describes the overall shape of rotation curve without degeneracy (see \autoref{subsec:rcmodel}). 

It is evident that the difference between fitted RC model parameter values measured from the $\dc$ and the true RC model parameter decreases $\niter$ increases. Although the difference does not \deleted{at $\niter$ = 20,} \added{always} converge to zero \added{as $\niter$ increases,}  it is clear that the difference is significantly reduced by the deconvolution method. Note that the size of $\rm 1\sigma$ error of the fitted RC model parameter values from $\dc$ is smaller than the $\rm \Delta$ between the parameter values measured from $\ndc$ and the true values. This result clearly exhibits that the kinematic parameters are reasonably well-restored closely to the true values, even considering the measurement error.

To visualize the effect of $\niter$, we show varying 2D $r$-band flux, line of sight velocity, and velocity dispersion map as $\niter$ changes in \autoref{niter:img}, using the same mock IFU data as in the \autoref{fig:dcex}. Note that the variation is shown for selected $\niter$ = 0 (Conv; No deconvolution), 1, 2, 3, 10, 20, and 30. Similar to the trend of difference between the true RC model parameters and the fitted model RC parameters to $\niter$ (\autoref{fig:niter_diffincl}), the amount of difference between 2D map from true and deconvolved IFU data is larger for the small $\niter$.

\begin{figure*}
\includegraphics[width=\textwidth]{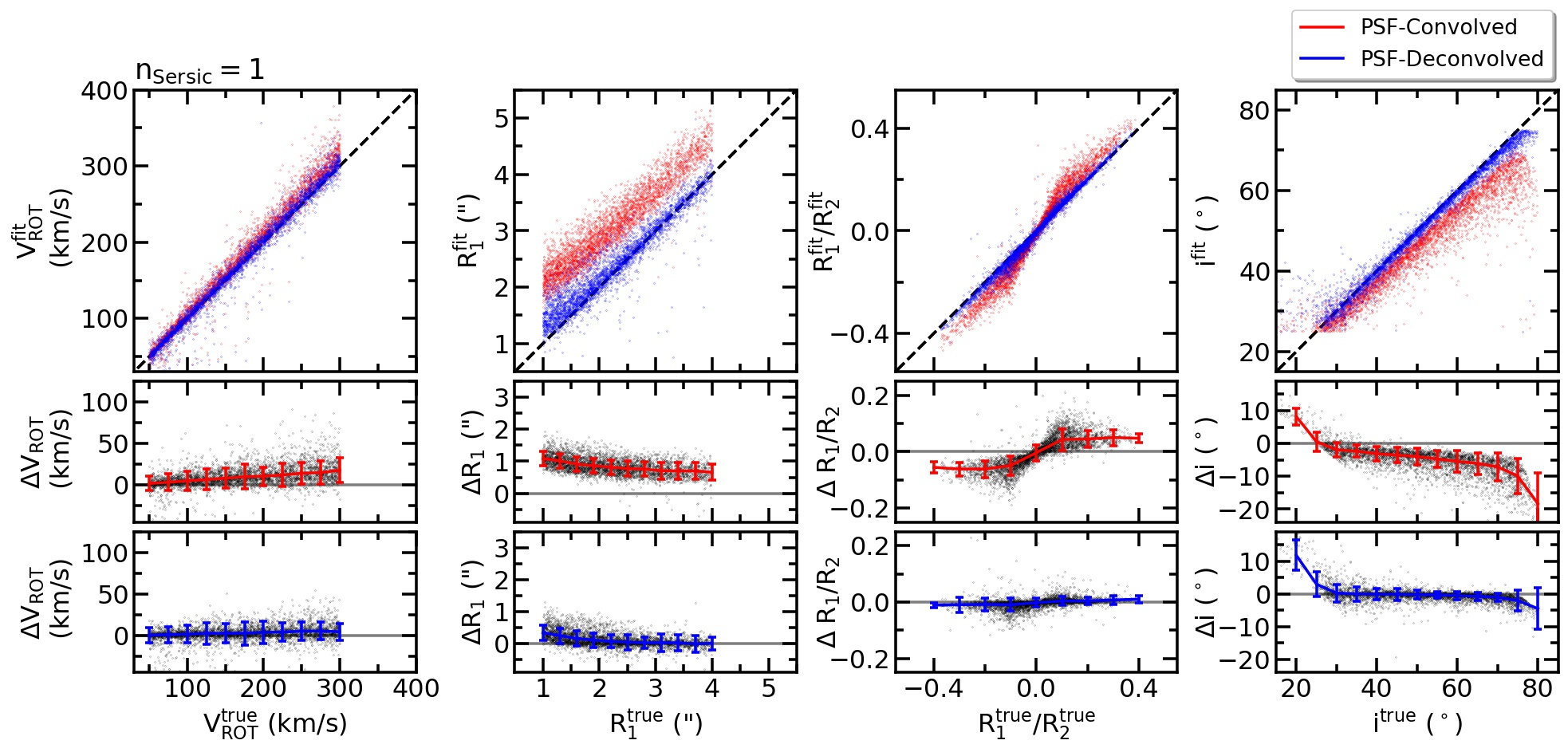}
\caption{Top panel of each column shows 1:1 relation between the fitted RC model parameter values from $\ndc$ IFU data and the true parameter value (red), and the relation between the fitted RC model parameter values from $\dc$ IFU data and the true value (blue), when $\nsrc = 1$. Middle (bottom) panel of each column presents the difference between the fitted RC model parameter values from $\ndc$ ($\dc$) IFU data and the true value with respect to the true parameter values. The error bar in the middle and the bottom panel shows the 1-$\sigma$ range of the data points within each arbitrary bin size.}
 \label{fig:mcn1}
\end{figure*}

\deleted{
Considering the overall trend of the $\rm \Delta$ parameter values with respect to $\niter$, no significant improvement on the $\rm \Delta$ parameter values are expected at beyond $\niter$ = 20. In addition, additional artifact in the flux distribution can be arisen at beyond $\niter$ = 20. Therefore, considering the overall dependency of the measured RC model parameter values with respect $\niter$, we conclude that $\niter$ = 20 as the adequate number of iteration of our deconvolution method to obtain a reasonably good result.
}

\added{
Considering the overall trend of the $\rm \Delta$ parameter values with respect to $\niter$, it is not obvious to determine the optimal $\niter$ value. Most of the parameters are rapidly converged to the true values during the first $\niter \sim 5$. At $\niter > 5$, the slope of $\Delta$ parameter is reduced although the slope is generally stiffer when $\nsrc=4$ compared to $\nsrc=1$ case. In general, the change of the parameter value depending on $\niter$ is smaller than the error bar when $\niter > 10$ (see also the figures in \autoref{app:niter}). For example, when S/N of data is low (S/N at 1 $\rm R_e$ $\lesssim$ 10), in practice, $\niter$ $>$ 10 does not improve the $\rm \Delta$ parameter for the cases in \autoref{fig:niter_diffincl}. 
Moreover, in some cases the $\rm \Delta$ parameter is even increased as $\niter$ increses (e.g. the first column of the forth row and the fourth column of the forth row in \autoref{fig:niter_diffincl} when S/N at 1 $\rm R_e$ = 10). 
This is because the measured parameter values are dominated by the intrinsic error (low S/N) in the data, not by $\niter$. On the other hand, since the error bar is decreasing when the data has high S/N, higher $\niter$ may provide more accurate result if S/N is high. However, as can be seen in \autoref{fig:niter_diffincl}, the difference between the parameters measured by $\niter = 20$ and $\niter > 20$ is considerably smaller than the difference between the parameters measured by $\niter = 0$ and $\niter = 20$. In addition, the trend of $\Delta$ parameter with medium S/N (S/N at 1 $\rm R_e$ = 20) and high S/N (S/N at 1 $\rm R_e$ = 30) are nearly overlapped in all $\niter$ vs. $\Delta$ parameters, compared to the difference between trend of the low S/N (S/N at 1 $\rm R_e$ = 10) and medium S/N cases. (see also \autoref{niter:r1ih0}, \autoref{niter:r1ih1}, \autoref{niter:r0ih2}, \autoref{niter:r2ih2}). 
This implies that the trend with even higher S/N (S/N at 1 $\rm R_e$ $>$ 30) will also follow the similar trend as for the medium S/N and high S/N cases. In this work, considering the S/N at 1 $\rm R_e$ of MaNGA data (14-35, \citet{2015ApJ...798....7B}), we use $\niter$ = 20 as the number of deconvolution iteration. 
}
In \autoref{app:niter}, we present additional similar figures with various mock galaxy model parameters to support the validity of our deconvolution method.

\autoref{fig:fwhm_diffincl} presents the difference between the fitted RC model parameter and the respective true value as a function of the Gaussian PSF FWHM used for the deconvolution ($\fdc$). $\fdc$ is varied from 2.3$\arcsec$ to 2.9$\arcsec$ with 0.05$\arcsec$ increment when the FWHM of the convolved Gaussian PSF ($\fndc$) is $2.6\arcsec$. $\niter$ is fixed as 20. Again the error is calculated from the result with 25 mock IFU data generated with different random seeds. 
The figure shows the case of the mock IFU data with the combination of $\nsrc$ = 1, 4 and $i$ = 40, 55, 70$\arcdeg$ with fixed $R_1$ and $1/{R_2}$ as 3 and 0.05. Indeed there is a dependency of the fitted parameters to the $\fdc$ value, but variation of the value is not significant when the $|\fdc-\fndc| < 0.3 \arcsec$, considering the error bar. As the $\fdc$ is varied, the difference between the parameters from the $\dc$ (open circles) to the true value changes but not always linearly. In all cases, the measured parameter values from the deconvolved IFU data are clearly getting closer to the true value, compare to the values without deconvolution (values measured from $\ndc$ mock IFU data). Considering all four kinds of fitted parameters, the best result is obtained when $\fdc = \fndc$, although the difference between the fitted and the true model parameters from the $\dc$ and the $\free$ are not always minimum at $\fdc = \fndc$. From this test result, we conclude that in most cases, the deconvolved IFU data produces fairly consistent result when the $\fdc-\fndc$ is less than $0.3 \arcsec$ (i.e. when the measurement error of the size of $\fndc$ is less than $0.3 \arcsec$). In \autoref{app:fwhm}, we present supplementary figures with different $\fndc$ values ($2.3 \arcsec$ and $2.9 \arcsec$) and different mock galaxy model parameters.

\begin{figure*}
\includegraphics[width=\textwidth]{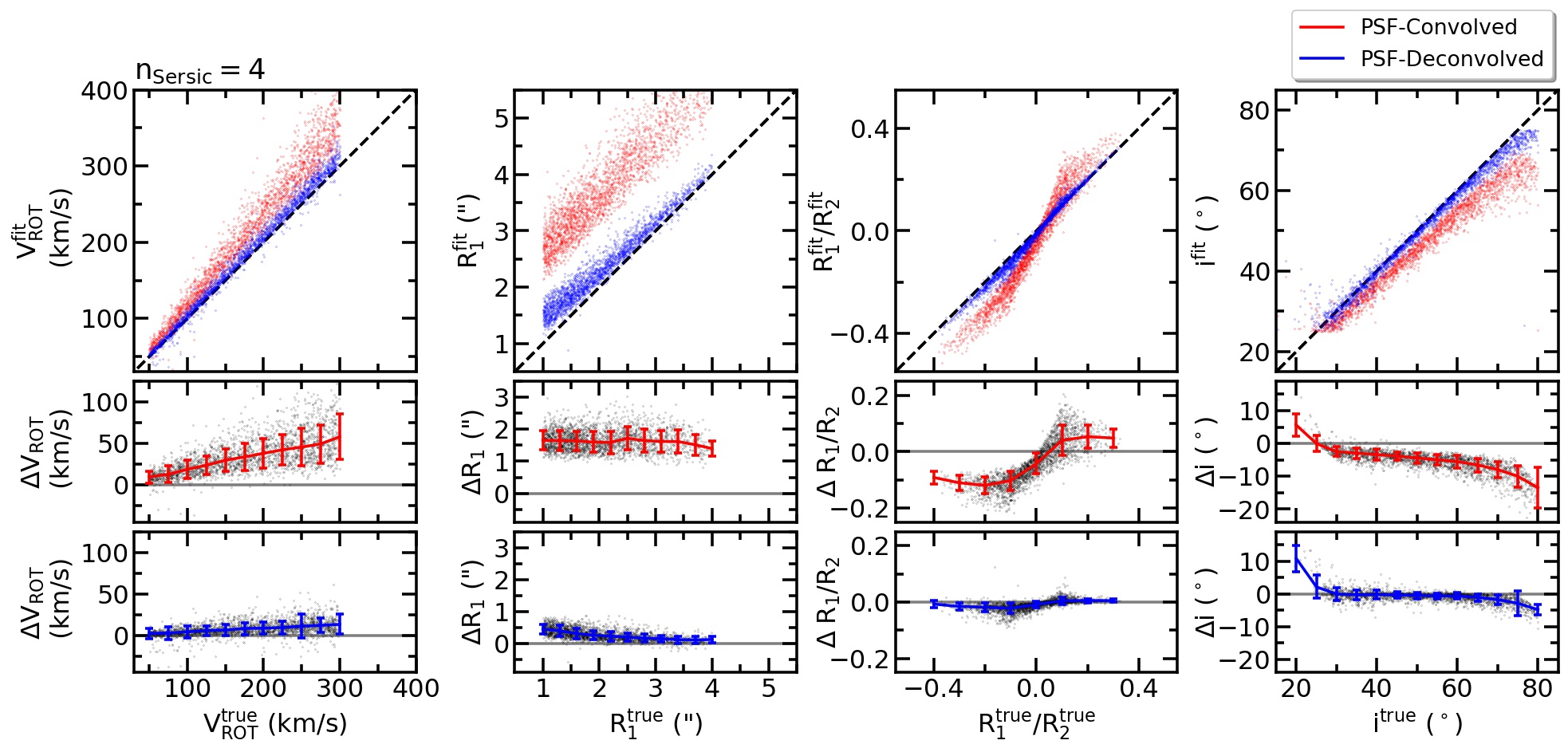}
\caption{Same as \autoref{fig:mcn1} but with mock IFU data of $\nsrc = 4$}
 \label{fig:mcn4}
\end{figure*}

\subsubsection{Results from the Monte-Carlo Mock IFU data}
Here we present the result of the deconvolution method performance verification test with Group 3 Monte-Carlo mock IFU data.
This is to validate that the deconvolution method works well not only with the mock galaxy model with certain combination of model parameter values, but also with diverse combination of the galaxy model parameters.
We divide the results according to $\nsrc$ value because the results are highly correlated with $\nsrc$. \autoref{fig:mcn1} and \autoref{fig:mcn4} shows the results with $\nsrc = 1$ and $\nsrc = 4$, respectively. Note that we only include the results when the $\dc$ mock IFU data satisfies the fitting qualification criteria, which are IFU field of view equal or wider than 17$\arcsec$, $R_{major, S/N > 3}$/$R_1 > 2.5$, and $ 75\arcdeg> i > 25\arcdeg$. The number of mock IFU data used for \autoref{fig:mcn1} is 2,354, and for \autoref{fig:mcn4} is 3,820.
In \autoref{fig:mcn1}, all $V_{ROT}$, $R_1$, ${R}_{1}/{R}_{2}$ and $i$ model parameters measured from the $\dc$ mock IFU data show good agreement with the true value. 
On the contrary, the model parameter values measured from $\ndc$ mock IFU data show considerable deviations from the true value. 
In \autoref{fig:mcn4}, again all parameters measured from the $\dc$ mock IFU data show good agreement with the true value. 
The model parameter values measured from $\ndc$ mock IFU data show larger discrepancy in the case of $\nsrc=4$.

\begin{figure*}
\includegraphics[width=\textwidth]{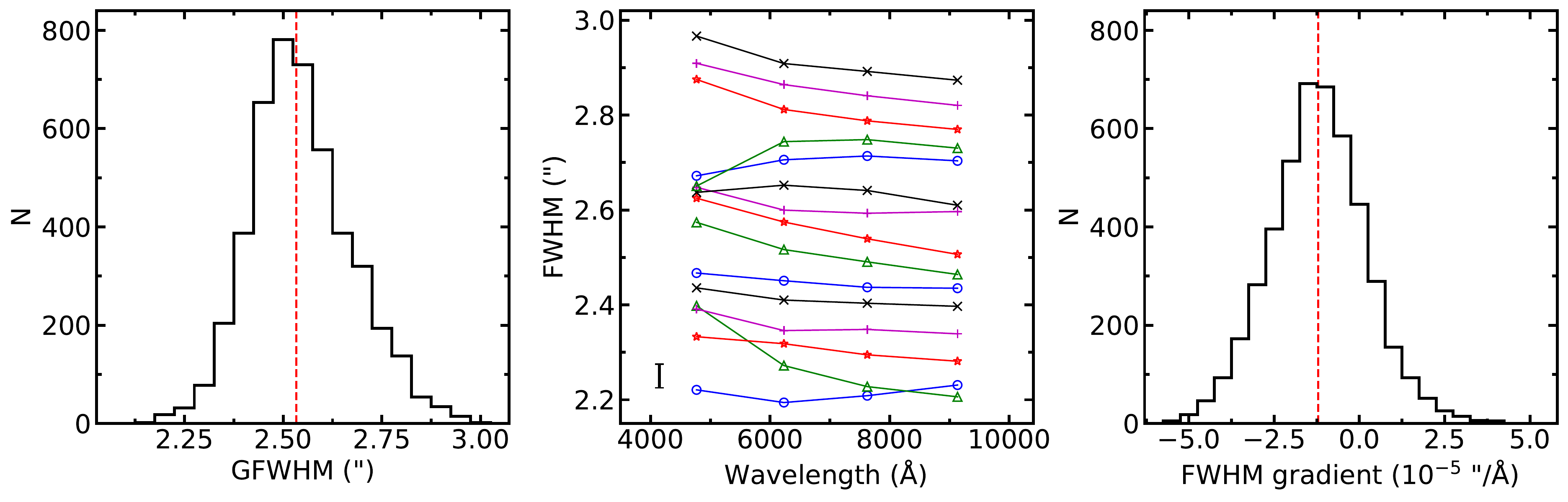}
\caption{(left) Distribution of the reconstructed FWHM in $g$-band. Median value is 2.53$\arcsec$. (middle) Wavelength-dependent $\fpsf$ of the selected MaNGA galaxies. Each connected line represents $griz$ FWHM of PSF from a particular sample. Samples are randomly selected for illustrative purpose. Error bar shows $\rm \pm$0.025$\arcsec$ range (1\% of 2.5$\arcsec$). (right) Distribution of wavelength-dependent FWHM gradient. The gradient is obtained by fitting a linear function to the reconstructed FWHM at $griz$-bands. Median value is $-1.21 \times 10^{-5} \arcsec/$\AA.}
 \label{fig:fwhm}
\end{figure*}

Results from the figures show that our deconvolution method successfully restores the kinematic properties of galaxies. It also shows that the measured parameter values from the $\ndc$ mock IFU data have a noticeable deviation from the true value, especially when $\nsrc=4$. It can be interpreted as the PSF convolution effect becomes more significant when there is a steeper relative flux slope between the adjacent spaxels. This effect is most evident for $R_1$ parameter.
$R_{1}^{fit}-R_{1}^{true}$ of the $\ndc$ mock IFU data show a median offset of 1.8 $\arcsec$ in \autoref{fig:mcn4}. This large offset also affects $1/R_{2}$, where many $1/R_{2}$ values from $\ndc$ are measured in the condition where they did not meet the fitting qualification criteria.

\section{Application to SDSS-IV MaNGA IFU Data}\label{sec:appmanga}

\subsection{MaNGA Point Spread Function}\label{subsec:mangapsf}
We use IFU data from the third public release of the MaNGA \citep{2015ApJ...798....7B}, that is a part of SDSS DR15 \citep{2019ApJS..240...23A}. Among the 4,824 DR15 MaNGA cube data, we select 4,426 unique galaxies from the MaNGA main galaxy sample (Primary, Color-enhanced primary, and secondary; \citet{2017AJ....154...86W}) by removing repeated observations, duplicated galaxies with different MaNGA-ID \footnote{https://www.sdss.org/dr15/manga/manga-caveats/}, and special targets (IC342, Coma, and M31). For the repeated observations and duplicated galaxies, we choose data observed by a bigger IFU. If both are observed by IFU with the same size, then we use the data with higher blue channel S/N as recorded in the FITS header of the data.  
In the context of deconvolution, it is important to know the accurate information about the shape and size of PSF that is convolved to each MaNGA IFU data. According to \citet{2015AJ....150...19L, 2016AJ....152...83L, 2016AJ....152..197Y},
it is known that 1) size of PSF FWHM ranges between 2.2$\arcsec$ and 2.7$\arcsec$ in $g$-band, 2) shape of PSF is well-described by a single 2D circular Gaussian function, 3) FWHM of the fitted model Gaussian function agrees with the measured FWHM within 1 - 2\%, 4) PSF FWHM varies less than 10\% across the field of view within a single MaNGA IFU. MaNGA IFU data provides the reconstructed MaNGA PSF image in $griz$ band as well as $griz$ PSF FWHM values in its header. The $g$-band PSF FWHM distribution of the entire SDSS DR15 MaNGA data is shown in the left panel of \autoref{fig:fwhm}. To account for the wavelength dependency of MaNGA PSF FWHM (\autoref{fig:fwhm}, middle panel), we fit a simple linear function (first order polynomial) to the PSF FWHM of $griz$-bands to interpolate/extrapolate the PSF FWHM value at other wavelengths. The average absolute difference between the reconstructed PSF FWHM values recorded in the IFU data header and the PSF FWHM values from the fitted linear function is 0.007$\arcsec$ with standard deviation of 0.006$\arcsec$, calculated from the entire MaNGA IFU data.

Considering the error of the reconstructed PSF FWHM of MaNGA IFU data (1-2\% or 0.025-0.05 in arcsec)\citep{2016AJ....152...83L}, we conclude that the PSF FWHM from the fitted linear function gives reasonable PSF FWHM at each wavelength bin. The distribution of the slope of the fitted linear functions is shown in the right panel of \autoref{fig:fwhm}.

\begin{figure*}
\includegraphics[width=\textwidth]{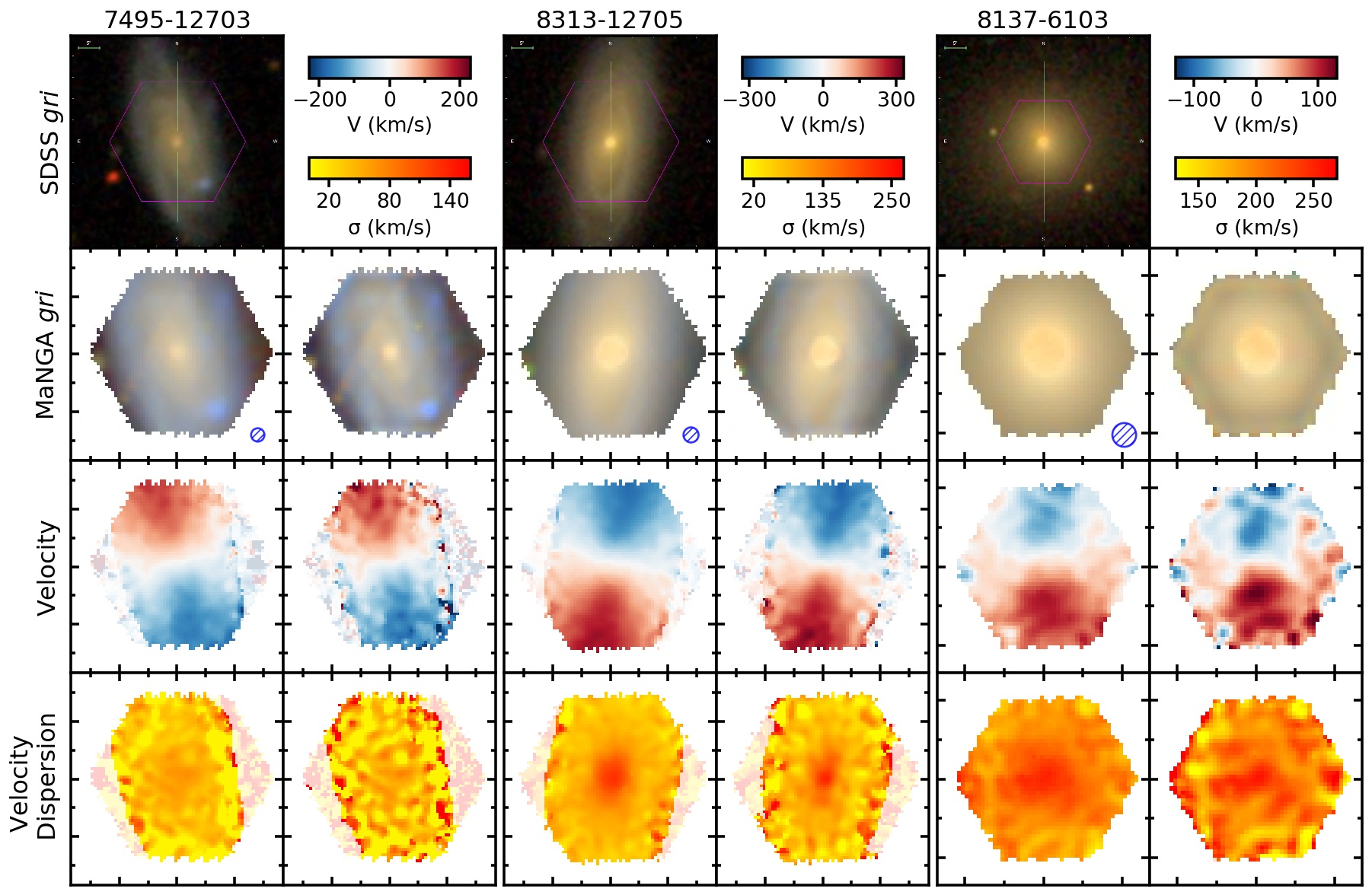}
\caption{The results of the PSF deconvolution on three MaNGA galaxies. The number at the top of each SDSS $gri$ image is the PLATE-IFU designation of a given galaxy. For each galaxy, images in the left column show reconstructed MaNGA \textit{gri} image, velocity and velocity dispersion distribution obtained from the $original$ MaNGA data. Images in the right column are those from the PSF-\textit{deconvolved} MaNGA data. Hatched blue circle represents PSF FWHM size of each galaxy. Spaxels with median ${\rm S/N}_{\rm pPXF} <$ 3 are paled out in the velocity and velocity dispersion distributions. Each major tick interval corresponds to 10$\arcsec$.}
\label{fig:sdssex}
\end{figure*}

\subsection{Measurements of Kinematic Parameters}\label{sub:mangakin}
We measure the line-of-sight velocity and the velocity dispersion from 4,425 unique MaNGA galaxies. The measurement procedure is similar to the procedure that is described in \autoref{sub:mockkm}, but with several differences.
Instead of using one single-stellar population model template, we use 156 single-stellar population model SED templates from MILES stellar library \citep{2006MNRAS.371..703S, 2011A&A...532A..95F, 2010MNRAS.404.1639V} generated by using unimodal initial mass function \citep{1996ApJS..106..307V} and Padova+00 isochrones \citep{2000A&AS..141..371G}, age from 1 to 17.78 Gyr, and metallicity (Z) from -2.32 to 0.22 (26 ages $\times$ 6 metallicites = 156). We use an option to use 6$^{\rm{th}}$ order additive and multiplicative Legendre polynomials during the fitting to account for the low-order difference and offset between the MILES model and data. We mask the spectrum pixels around the known emission lines.
Model SED templates are convolved with a Gaussian function to match the spectrum resolution of MaNGA data as provided in the SPECRES HDU.

\begin{figure*}
\includegraphics[width=\textwidth]{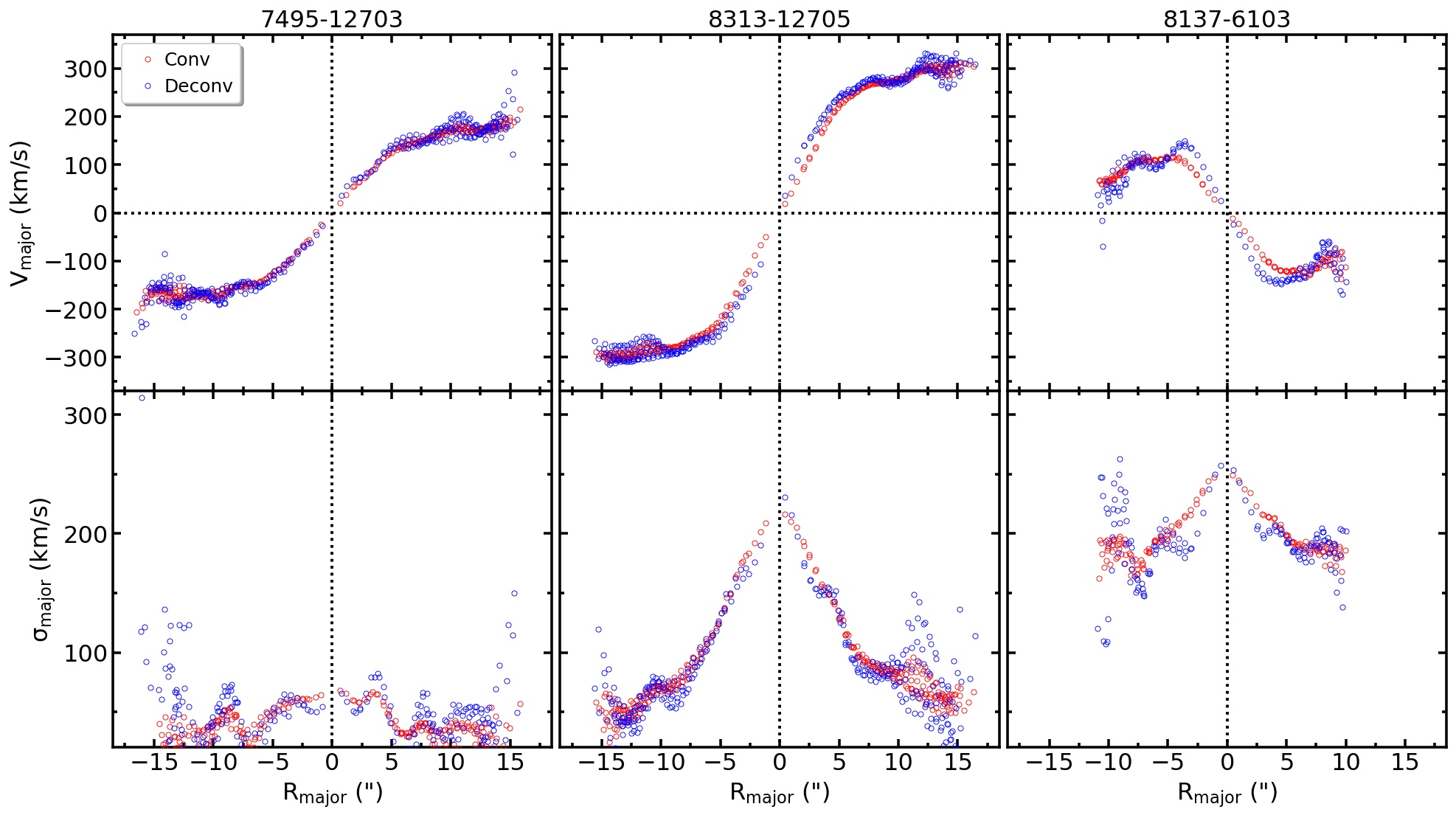}
\caption{The velocity and velocity dispersion profiles from the original and deconvolved MaNGA data. Only the data points from $\pm 10 \arcdeg$ of the major axis with median ${\rm S/N}_{\rm pPXF} > 3$ are shown for clarity. }
\label{fig:sdssrc}
\end{figure*}

\subsection{Results}\label{sub:sdssresult}
\autoref{fig:sdssex} shows the result of deconvolution applied to the three of the MaNGA galaxies as an example (PLATE-IFU: 7495-12703, 8313-12705, 8137-6103). These galaxies are chosen based on their shape of the rotation curve and the velocity dispersion profile.
Each reconstructed $gri$ image obtained from the deconvolved MaNGA ($Deconv$) data shows a noticeable difference compare to the reconstructed $gri$ image from the original MaNGA ($Ori$) data. The MaNGA-$Deconv$ data shows more sharpened substructures. The restored substructures are not artifacts created by the deconvolution method but are actual substructures which can be seen in the SDSS $gri$ image that has higher spatial resolution. \deleted{Size of one tick is 10$\arcsec$.}
The velocity distribution also shows the apparent change, especially around the center of galaxies (i.e. the velocity gradient becomes steeper). The velocity dispersion exhibits some changes as well, and shows narrower dispersion distribution near the center and sharper substructures. The restored substructures can be understood intuitively as a result of deconvolution. The difference between MaNGA-$Ori$ and MaNGA-$\dc$ data can be seen more prominently in \autoref{fig:sdssrc}. The figure clearly exhibits the changes in the velocity and the velocity dispersion distribution along the galaxy major axis.

\section{Measurement of the Spin Parameter}\label{sec:appspin}

\begin{figure*}
\includegraphics[width=\textwidth]{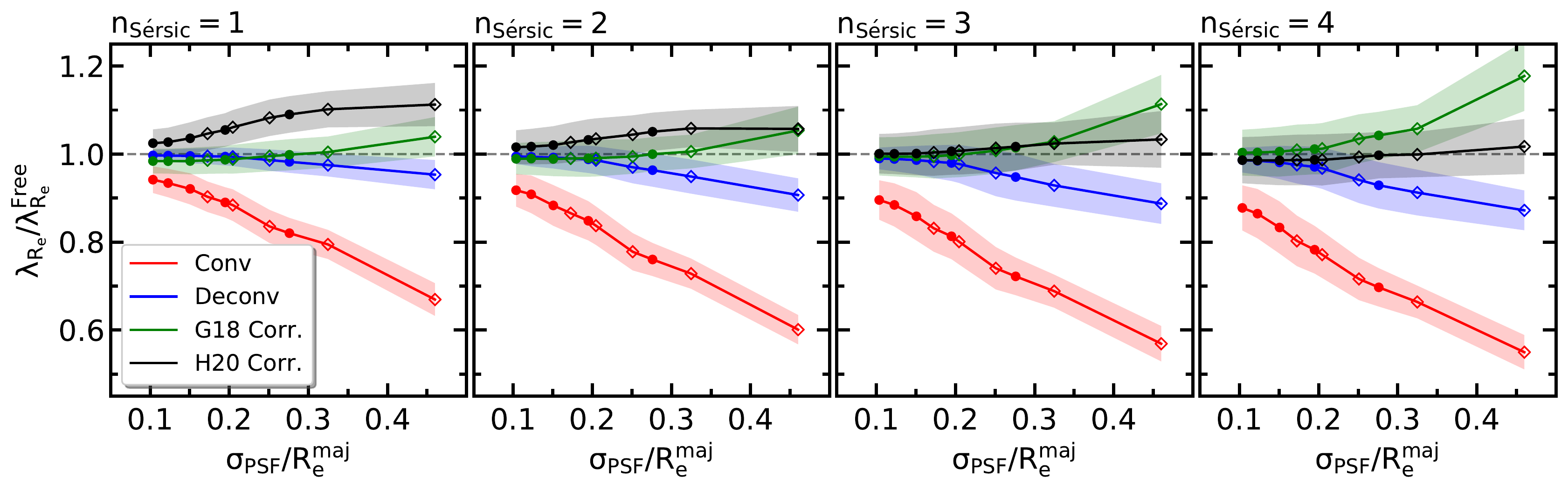}
\caption{\replaced{Ratio between the true and the measured/corrected spin parameter $\lre$ depending on $n_{S\acute{e}rsic}$, Radial coverage in $\Reff$, and IFU field of view. The left most panel shows the ratio measured from the Group 3 mock IFU data (\autoref{tab:mctb}) with  $n_{S\acute{e}rsic} = 1$ and Radial coverage = 1.5 $\Reff$, depending on mock IFU field of view (12 $\arcsec$ to 32 $\arcsec$). Similarly, other three panels show the result from mock IFU data with $n_{S\acute{e}rsic} = 1$ and Radial coverage = 2.5 $\Reff$, $n_{S\acute{e}rsic} = 4$ and Radial coverage = 1.5 $\Reff$, $n_{S\acute{e}rsic} = 4$ and Radial coverage = 2.5 $\Reff$, respectively. Shaded  region represents  1-$\rm\sigma$ range.}{Ratio between the true and the measured/corrected spin parameter $\lre$ depending on $n_{S\acute{e}rsic}$ and $\rm \sigma_{PSF}/R_{e}^{maj.}$. The left most panel shows the ratio measured from the Group 3 mock IFU data (\autoref{tab:mctb}) with  $n_{S\acute{e}rsic} = 1$. Other panels show the result from mock IFU data with $n_{S\acute{e}rsic}$= 2, 3, 4. Filled-circle points are estimated from mock IFU data with radial coverage = 1.5 {$\rm\Reff$}, and the open-diamond points are estimated from mock IFU data with IFU radial coverage = 2.5 {$\rm\Reff$}.}
}
\label{fig:spin_mock}
\end{figure*}
In this section, we investigate the reliability of the $\lambda_{R}$ parameter measured from the deconvolved IFU data.
$\lambda_{R}$ is a proxy of the spin parameter $\lambda$. It is calculated from the luminosity-weighted first and second velocity moments as in \citet{2007MNRAS.379..401E},
\begin{equation}
    \lambda_{R} \equiv \frac{\langle R |V| \rangle}{\langle R \sqrt{V^2 + \sigma^2} \rangle} = \frac{\Sigma ^{N}_{i=1} F_i R_i |V_i|}{\Sigma ^{N}_{i=1} F_i R_i \sqrt{V_i^2 + \sigma_i^2}},
    \label{eq:calspin}
\end{equation}
where $F_i, R_i, V_i, \sigma_i$ are flux, radius of the concentric ellipse, line-of-sight velocity and line-of-sight velocity dispersion at the $i$th spatial bin, respectively.
$\lambda_{R}$ is widely used in various applications, such as kinematic classification of galaxies \citep{2011MNRAS.414..888E, 2017ApJ...835..104V, 2016MNRAS.463..170C, 2018MNRAS.477.4711G}, measurement of the angular momentum of merger remnants \citep{2009MNRAS.397.1202J}, and the studies of the environmental dependence of galaxy spin \citep{2018ApJ...852...36G, 2018MNRAS.477.1567L}. 
\deleted{and the studies of the evolution of spin parameter using simulation data \citep{2017ApJ...837...68C, 2018ApJ...856..114C}. }
\added{
There are also studies investigating the origin of fast and slow rotators \citep{2014MNRAS.444.3357N, 2017MNRAS.468.3883P}, and the evolution of galaxy kinematics by using spin parameter ($\lambda_{R}$ or $V/\sigma$) and simulation data \citep{2017MNRAS.464.3850L, 2017ApJ...837...68C, 2018ApJ...856..114C, 2018MNRAS.480.2266M}. In particular, \citet{2019MNRAS.484..869V} provides comprehensive analysis on the dynamical parameter measured from both observation and simulation though using $V/\sigma$ instead of $\lambda_{R}$.
}

Typically $\lambda_{R}$ is calculated by using the information within galaxy half-light radius (equivalent to the $R_{e}^{major}$)\citep{2010MNRAS.401.1099H, 2013MNRAS.432.1709C}, and denoted as $\lre$. 
It is known that $\lre$ is mainly correlated with two parameters: inclination angle ($\epsilon$, as axis ratio) and FWHM of PSF that is convolved in data, because distribution of $F_i, R_i, V_i$, and $\sigma_i$ are much affected by those parameters \citep{2016ARA&A..54..597C,2018MNRAS.477.4711G}. 

There were several attempts to mitigate the effect of the PSF on $\lre$ measurement: 1) by correcting $\lre$ by 1/$\sqrt{\epsilon}$ \citep{2011MNRAS.414..888E, 2013MNRAS.429.1258D, 2016MNRAS.463..170C, 2018ApJ...852...36G} or 2) by applying an empirical correction function such as \defcitealias{2018MNRAS.477.4711G}{G18}\citet{2018MNRAS.477.4711G}(hereafter \citetalias{2018MNRAS.477.4711G}) and \defcitealias{2020MNRAS.497.2018H}{H20}\citet{2020MNRAS.497.2018H}(hereafter \citetalias{2020MNRAS.497.2018H}).
\added{
The empirical correction function provides convenient way to correct the PSF effect to a certain parameter. Generally these functions are derived based on galaxies from model or simulation. Therefore those functions have non-negligible dependence to the galaxy model or simulation used. In other words, the form or coefficient of those functions will be different if such function is generated using other model/simulated galaxies. Those correction functions should be used with caution because it can often under-correct or over-correct such a parameter when they are applied to the real galaxies. In addition, the equation should be used within a certain boundary conditions such as a range of S\'{e}rsic index. 
}

Here we show that our deconvolution method can also be used to accurately measure $\lre$. In \autoref{sub:spinmock}, we measure $\lre$ from 
the Group 3 Monte-Carlo mock IFU data ($\free$, $\ndc$, $\dc$) and compare the $\lre$ measured from each type of the mock IFU data. 
From the result, we find that the $\lre$ value measured from the deconvolved IFU data is close to the true $\lre$. \deleted{We also check the feasibility of the \citetalias{2018MNRAS.477.4711G} correction by using our mock data.} 
\added{
We also check the amount of change induced by \citetalias{2018MNRAS.477.4711G} and \citetalias{2020MNRAS.497.2018H} correction function using our mock data. 
Note that we use different definition of $R_i$ in \autoref{eq:lr} compared to \citetalias{2018MNRAS.477.4711G}. In \citetalias{2018MNRAS.477.4711G}, $R_i$ is defined as the circular radius, but in our work we define $R_i$ as the semi-major axis radius of an ellipse which pass through each spaxel. The definition in \citetalias{2018MNRAS.477.4711G} is the same as the original definition of $\lre$ in \citet{2007MNRAS.379..401E} as well as \citet{2018ApJ...852...36G}. The other definition (semi-major axis) was used in some other studies \citep{2016MNRAS.463..170C, 2017ApJ...835..104V, 2018MNRAS.477.1567L}. 
The difference between the two definitions is the different weighting of the kinematics around the minor axis. Generally, for regular rotators, this weighting is not a major contributor to the $\lre$ calculation because the line-of-sight velocity around the minor axis is small due to the projection effect. \citetalias{2020MNRAS.497.2018H} provides two forms of correction function for both definitions of $R_i$. In this study, we use the \citetalias{2020MNRAS.497.2018H} correction function with the $R_i$ defined as the semi-major axis radius. \citetalias{2020MNRAS.497.2018H} also showed that the difference between the $\lre$ value corrected by two forms of correction function is less than 0.02 dex. on average. 
Contrary to the seeing-correction of $\lre$ using empirical function, measurement of $\lre$ using the deconvolved IFU data is completely independent from any galaxy model.}
In \autoref{sub:mangaspin}, we measure $\lre$ from both MaNGA-$Ori$ and MaNGA-$Deconv$, and examine the differences. 
\added{Finally, in \autoref{sub:spin_verification_manga}, we test the validity of $\lre$ value measured from the deconvolved IFU data by using the actual MaNGA data as a proxy of PSF-Free data. We consider MaNGA DR15 data as a seeing-free ground truth, and generate PSF-convolved and PSF-deconvolved IFU data from the original MaNGA data. We present the difference between $\lre$ values measured from the original MaNGA IFU data and PSF-convolved/deconvolved data.
}

\subsection{Verification using Mock Data}\label{sub:spinmock}
\explain{Subsection title modified}
We calculate $\lre$ following \autoref{eq:calspin}, by using the reconstructed $r$-band flux, the velocity and the velocity dispersion distribution measured from the $\free$, $\ndc$, $\dc$ Monte-Carlo mock IFU data (Group 3, 40,000 IFU data each)(see \autoref{subsec:mockifs} and \autoref{tab:mctb}). \added{The area for the $\lre$ calculation is defined as the spaxels within an ellipse where the semi-major axis radius is $\Reff$. Semi-major axis radius of an ellipse that pass through} \deleted{Concentric elliptical radius at} each spaxel is calculated from the geometrical parameter of the $\free$ mock IFU data. 
We also calculate the corrected $\lre$ value by applying the correction function in \citetalias{2018MNRAS.477.4711G} to $\lre^{Conv}$ value to compare the result between the corrected value and the value measured from the deconvolved IFU data. To apply correction function of \citetalias{2018MNRAS.477.4711G} and \citetalias{2020MNRAS.497.2018H}, we use $n_{S\acute{e}rsic}^{Free}$, $R_e^{Free}$ and $\rm{FWHM}_{\rm{PSF}}$ at the $r$-band pivot wavelength (6231\AA). $R_e^{Free}$ should be used instead of  $R_e^{Conv}$. \deleted{because \citetalias{2018MNRAS.477.4711G} uses PSF-corrected $R_e$ value derived from Multi-Gaussian Expansion fitting \citep{1994A&A...285..723E, 2002MNRAS.333..400C} to the galaxy 2D flux distribution.} \added{This is because both correction functions assume that such a seeing-corrected $\Reff$ value is available from a high-resolution seeing-free image or obtained by a fitting method that calculates seeing-corrected $\Reff$ (e.g. Multi-Gaussian Expansion fitting \citep{1994A&A...285..723E, 2002MNRAS.333..400C}).}

First, we check the ratio between the calculated $\lre$ values ($\lre^{Conv}$, $\lre^{Deconv}$, $\lre^{G18\;Corr.}$, and $\lre^{H20\;Corr.}$) and the true $\lre$ value ($\lre^{Free}$), as a function of true $\lre^{Free}$ value. 
\deleted{It is known that those ratios have a strong dependence on $\nsrc$ and the ratio $\rm FWHM_{PSF}$/$\Reff^{maj}$ (or $\rm \sigma_{PSF}$/$\Reff^{maj}$, \citetalias{2018MNRAS.477.4711G}).}
In our mock IFU data, the angular size of $\Reff$ is determined by the combination of IFU FoV and radial coverage in $\Reff$ (\autoref{tab:mctb}, \autoref{app:mockgen}).
Thus, we divide the result depending on three parameters of mock IFU data, 1) $\nsrc$ (1, 2, 3, 4), 2) IFU FoV (12$\arcsec$ to 32 $\arcsec$), and 3) radial coverage in $\Reff$ (1.5 $\Reff$ and 2.5 $\Reff$). 
\deleted{In fact, $\rm FWHM_{PSF}$ is also different in each mock IFU data but within 2.6 $\arcsec \pm$ 0.3, so we decide not to divide the result depending on $\rm FWHM_{PSF}$ parameter.}
We plot the relation between the calculated ratios to the $\lre^{Free}$ of each divided result. \autoref{fig:spin_mock_matrix} is shown as an example of the result for the mock IFU data with $\nsrc$=1 and 4, and IFU FoV = 1.5 $\Reff$ and 2.5 $\Reff$. To illustrate the overall dependence of the ratio to the $\nsrc$ and the size of $\Reff$, we take the median of the ratios and the median of the standard deviation of the ratios at each bin (per $\rm \Delta\lre^{Free}$=0.1) in each panel of \autoref{fig:spin_mock_matrix}.
\added{We plot the median of the ratios and the median of the standard deviation of the ratios as a function of $\rm \sigma_{PSF}$/$\Reff^{maj}$ where $\rm \sigma_{PSF}=FWHM_{PSF}/2.355$. $\rm \sigma_{PSF}$/$\Reff^{maj}$ is calculated from each combination of IFU FoV and the radial coverage. } \replaced{For example, the left most red data point in the left most panel in \autoref{fig:spin_mock} (IFU FoV=12$\arcsec$ (=1.5 $\Reff$), $\lre^{Conv}/\lre^{Free}$=$\rm0.82\pm0.03$) is derived from the top left panel of \autoref{fig:spin_mock_matrix} by taking the median and the median of $\rm 1 \sigma$ of the binned relation.}{For example, the right most filled-circle red data point in the left most panel in \autoref{fig:spin_mock} ($\rm \sigma_{PSF}$/$\Reff^{maj}=0.276$, $\lre^{Conv}/\lre^{Free}$=$\rm0.82\pm0.03$)) is derived from the top left panel of \autoref{fig:spin_mock_matrix} ($\nsrc=1$, IFU FoV=12$\arcsec$ (=1.5 $\Reff$)) by taking the median and the median of $\rm 1 \sigma$ of the binned relation. Note that the actual $\rm \sigma_{PSF}$/$\Reff^{maj}$ of each combination of IFU FoV and the radial coverage is not a constant. This is because $\rm FWHM_{PSF}$ of the Group 3 IFU data is slightly different for each mock IFU data as 2.6 $\pm$ 0.3$\arcsec$. Since $\Reff$ is different for each combination of IFU FoV and the radial coverage, the difference caused by $\rm FWHM_{PSF}$ is also different for each combination of IFU FoV and the radial coverage (e.g. $\rm \Delta(\sigma_{PSF}/\Reff^{maj})=\pm 0.04$ for IFU FoV=12$\arcsec$ and radial coverage=2.5$\Reff$). Nevertheless, \autoref{fig:spin_mock} could still be used to show the dependence of $\lre^{Conv}/\lre^{Free}$ with respect to $\rm \sigma_{PSF}$/$\Reff^{maj}$.
}

\begin{figure*}
\includegraphics[width=\textwidth]{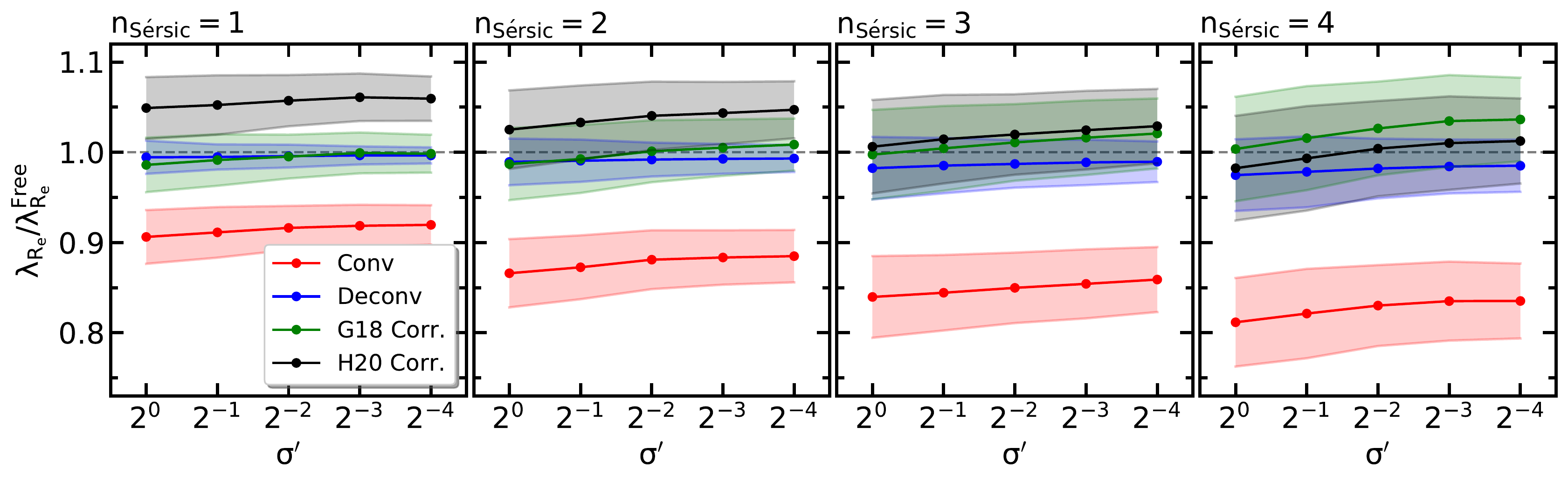}
\caption{Ratio between the true and the measured/corrected spin parameter $\lre$ depending on $n_{S\acute{e}rsic}$ and velocity dispersion profile coefficient $\sigma_1$. The left most panel shows the ratios measured from the Group 3-like mock IFU data set (\autoref{tab:mctb}, all with 32$\arcsec$ Field of view and radial coverage of 2.5 $\Reff$) with  $n_{S\acute{e}rsic} = 1$), depending on velocity dispersion profile coefficient $\sigma_1$ = 1, 0.5, 0.25, 0.125, 0.0625. Other panels show the result from mock IFU data with $n_{S\acute{e}rsic}$= 2, 3, 4. 
\label{fig:spin_sigma1}}
\end{figure*}

\autoref{fig:spin_mock} shows that $\lre^{Conv}$ deviates significantly from $\lre^{Free}$, and the amount of the deviation becomes larger as $\nsrc$ increases and as \replaced{relative size of $\rm FWHM_{PSF}$ to galaxy $\Reff$}{$\rm \sigma_{PSF}$/$\Reff^{maj}$} increases. \deleted{(Smaller IFU with wider coverage in $\Reff$ has higher $\rm FWHM_{PSF}$/$\Reff$ ratio. This is because $\Reff$ is proportional to IFU field of view in our mock IFU data, and mock IFU data with larger radial coverage in $\Reff$ has smaller $\Reff$ with fixed IFU angular field of view). }
On the other hand, $\lre^{Deconv}$ is strikingly well-restored to the correct value ($\lre^{Free}$), 
although there is some deviation when \replaced{the relative size of $\rm FWHM_{PSF}$ to $\Reff$}{$\rm \sigma_{PSF}$/$\Reff^{maj}$} greater than 0.2. When \replaced{IFU field of view is equal or greater than 22$\arcsec$,}{$\rm \sigma_{PSF}$/$\Reff^{maj}$ is less than 0.2,} the fractional difference between $\lre^{Deconv}$ and $\lre^{Free}$ is less than 3 percent with less than 4.6 percent point standard deviation. 
\replaced{We confirm that the corrected $\lre$ by \citetalias{2018MNRAS.477.4711G} correction function is also very close to the correct value. When IFU field of view is equal or greater than 22$\arcsec$, the fractional difference between $\lre^{G18 Corr.}$ and the correct value is also less then 2 percent with less than 3.3 percent point standard deviation.}{
We find that the corrected $\lre$ by \citetalias{2018MNRAS.477.4711G} or \citetalias{2020MNRAS.497.2018H} correction function is also close to the correct value compared to the uncorrected $\lre$. However, the fractional difference vs. $\rm \sigma_{PSF}$/$\Reff^{maj}$ shows different trends depend on $\nsrc$. \citetalias{2018MNRAS.477.4711G} correction works best when galaxy $\nsrc=1$. On contrary, \citetalias{2020MNRAS.497.2018H} correction works best when galaxy $\nsrc=4$. This discrepancy is most likely due to the difference of the galaxy model used between this study, \citetalias{2018MNRAS.477.4711G}, and \citetalias{2020MNRAS.497.2018H}. 
}

We conducted an additional test with different set of mock data that have slightly modified 2D velocity dispersion distribution profile. Group 3 mock data set is constructed with the velocity dispersion profile of only $\sigma^{\prime}=1$ where the velocity dispersion drops sharply between the center and $r=R_{1}$ (\autoref{eq:sig}, \autoref{tab:mctb}). However, the velocity dispersion profile of the actual galaxy does not always follow the same shape. For example, in \autoref{fig:sdssrc}, when we fit the \autoref{eq:sig} to the velocity dispersion profile, MaNGA data 8313-12705 (PLATE-IFU) is well described by $\sigma^{\prime}=1$. On the other hand, the velocity dispersion profile of MaNGA data 7495-12703 is not well-fitted by \autoref{eq:sig}, and for MaNGA data 8137-6103, the best-fit $\sigma^{\prime}$ value is around 0.44, which is fairly different compare to MaNGA data 8313-12705 case. 

To further explore the performance of deconvolution to the $\lre$ calculation on different velocity dispersion profile, we construct an additional mock IFU data set similar to Group 3 mock IFU data but with different $\sigma^{\prime}$ values and additional $\nsrc$ values. The additional data set is composed of $\sigma^{\prime}$=1, 0.5, 0.25, 0.125, 0.0625 ($\rm2^{0}$, $\rm2^{-1}$, $\rm2^{-2}$, $\rm2^{-3}$, $\rm2^{-4}$) and $\nsrc$= 1, 2, 3, 4. Smaller $\sigma^{\prime}$ value means less steep velocity dispersion profile. For example, if $\sigma^{\prime}$=0.1, then $\sigma_{r=R_{1}}/\sigma_{r=0}=0.91$, but if $\sigma^{\prime}$=1, then $\sigma_{r=R_{1}}/\sigma_{r=0}=0.5$. For each combination of $\sigma^{\prime}$ and $\nsrc$ (5 $\times$ 4 = 20), 4,000 mock IFU data are generated randomly in Monte-Carlo way just like Group 3 mock IFU data but with fixed IFU field of view as 32$\arcsec$ and IFU radial coverage as 2.5 $\Reff$, meaning that the ratio between $\rm FWHM_{PSF}$ and $\Reff$ is relatively fixed for this data set \added{($\rm \sigma_{PSF}$/$\Reff^{maj}\sim0.17$)}, compare to overall Group 3 mock IFU data. In total, 80,000 $\free$, 80,000 $\ndc$, 80,000 $\dc$ IFU data are produced as this additional test. This data set is analyzed as the same as Group 3 mock IFU data, and $\lre$ values ($\lre^{Free}$, $\lre^{Conv}$, $\lre^{Deconv}$, $\lre^{G18 Corr.}$, $\lre^{H20 Corr.}$) from this data set are calculated.  

First, we check the relation between the ratios ($\lre^{Conv}/\lre^{Free}$, $\lre^{Deconv}/\lre^{Free}$, $\lre^{G18 Corr.}/\lre^{Free}$, $\lre^{H20 Corr.}/\lre^{Free}$) to the true $\lre$ value ($\lre^{Free}$) as in \autoref{fig:spin_sigma1_matrix}. Then we plot \autoref{fig:spin_sigma1} in the same way as we did for \autoref{fig:spin_mock}. Again the result shows that $\lre^{Conv}$ deviates considerably from $\lre^{Free}$, and the amount of the deviation becomes larger as $\nsrc$ increases, with a mild dependence to the $\sigma^{\prime}$ value. Moreover, $\lre^{Deconv}$ is still well-restored to the correct value ($\lre^{Free}$) for all combinations of $\sigma^{\prime}$ and $\nsrc$, with the fractional difference between $\lre^{Deconv}$ and the correct value less then 1 percent and less than 1.9 percent point standard deviation. 

\replaced{On the other hand, $\lre^{G18 Corr.}$ shows some deviation from the correct value. In particular, the deviation shows noticeable dependence to the $\sigma^{\prime}$ value for all four $\nsrc$ cases, where the deviation increases as the $\sigma^{\prime}$ decreases. The reason for the deviation of $\lre^{G18 Corr.}$ would simply because \citetalias{2018MNRAS.477.4711G} used a fixed velocity dispersion profile with $\sigma^{\prime}$ = 1 to derive the empirical correction function.}
{On the other hand, $\lre^{G18 Corr.}$ and $\lre^{H20 Corr.}$ shows noticeable dependence to the $\sigma^{\prime}$ value for all four $\nsrc$ cases, where the deviation increases as the $\sigma^{\prime}$ decreases. The reason for the deviation of $\lre^{G18 Corr.}$ or $\lre^{H20 Corr.}$ could be because of the shape of the velocity dispersion profile covered by the model galaxy used by \citetalias{2018MNRAS.477.4711G} or \citetalias{2020MNRAS.497.2018H}, if the model galaxies did not include velocity dispersion distribution close to flat.} In \autoref{subsub:dcex}, we discuss the effect of PSF convolution to the velocity dispersion distribution. When there is a small gradient of the velocity dispersion profile (i.e smaller $\sigma^{\prime}$), the amount of smoothing effect to the velocity dispersion around the center will also be smaller, so the velocity dispersion around the center will be increased relatively less than the case of steep velocity dispersion gradient profile. Thus, the amount of change in $\lre$ value caused by PSF convolution will be smaller if the velocity dispersion profile has lower gradient. The dependence of $\lre^{Conv}$ to $\sigma^{\prime}$ shows the expected trend in \autoref{fig:spin_sigma1}, implies that the \citetalias{2018MNRAS.477.4711G} correction function over-corrects the $\lre^{Conv}$ value.

\begin{figure*}
\centering
\includegraphics[width=1\textwidth]{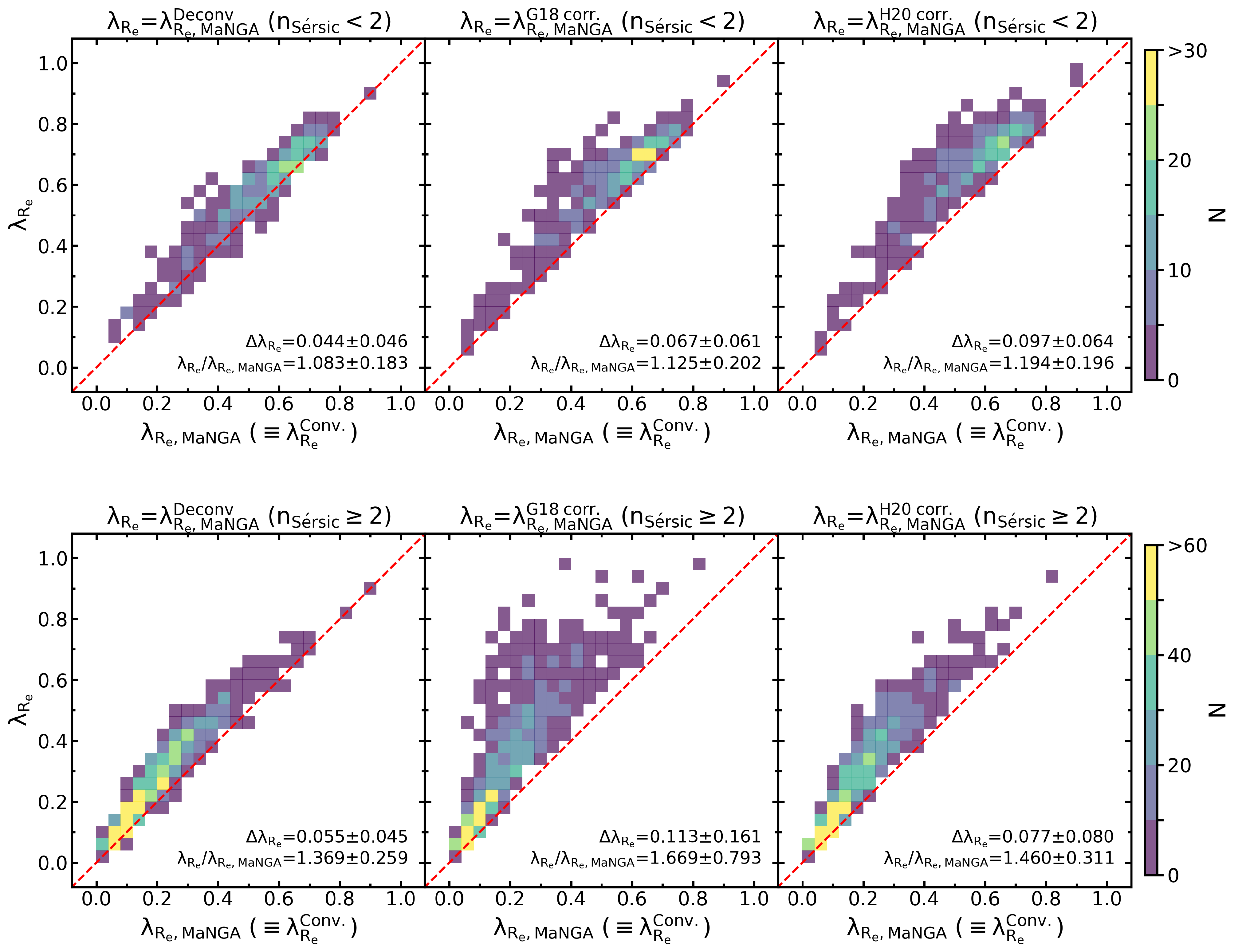}
\caption{(Top panels) Each panel shows comparison between the $\lambda_{R_e}$ measured from the original MaNGA IFU data and the $\lambda_{R_e}$ measured from the deconvolved MaNGA IFU data (left), the $\lambda_{R_e}$ corrected by following \citetalias{2018MNRAS.477.4711G} (center), and the $\lambda_{R_e}$ corrected by following \citetalias{2020MNRAS.497.2018H} (right), using the MaNGA galaxies with $\nsrc < 2$. Distribution of points are shown as 2D histogram. (Bottom panels) Same as the top panels but using the MaNGA galaxies with $\nsrc \ge 2$.
} \label{fig:spinmanga}
\end{figure*}

\subsection{Application to MaNGA Data}\label{sub:mangaspin}
We measure $\lre$ by using both the original and the PSF-deconvolved MaNGA data, and investigate the difference in the measured $\lre$ values that is induced by the deconvolution. We use 2D velocity and the velocity dispersion distribution (\autoref{sub:mangakin}) along with the reconstructed MaNGA $r$-band flux data of 4,426 MaNGA galaxies that we measured in \autoref{sec:appmanga}. 
We use 'NSA\_ELPETRO\_TH50\_R', 'NSA\_ELPETRO\_BA', 'NSA\_ELPETRO\_PHI', 'IFURA/IFUDEC' and 'OBJRA/OBJDEC' in the FITS header of each galaxy IFU data to evaluate the concentric ellipse of each. To ensure the quality of measured $\lre$, we did not include certain spaxels in the $\lre$ calculation when a spaxel has 1) median $\rm{S/N}_{pPXF} < 10$, or 2) velocity dispersion $<$ 40 km/s, following the prescription of \citet{2018MNRAS.477.1567L}. We also did not include spaxels with spurious kinematics to the $\lre$ calculation, where the absolute value of velocity is greater than 500 km/s or the velocity dispersion is less than 50 km/s. When the number fraction of the excluded spaxels within 1$\Reff$ becomes larger than 30\%, we do not use the $\lre$ from that galaxy for the further analysis. Although we use the same elliptical aperture for the measurement of both $\rm \lambda_{R_{e}, MaNGA}$ and $\rm \lambda_{R_{e}, MaNGA, Deconv.}$, the number of spaxels used for each measurement is not always identical because of $\rm{S/N}_{pPXF}$ and velocity dispersion criteria. 
We also exclude the galaxies flagged with 'CRITICAL' by the MaNGA Data Reduction Pipeline or Data Analysis Pipeline \citep{2016AJ....152...83L, 2019AJ....158..231W}. These criteria would be sufficient to observe the impact of deconvolution on $\lre$ for the real data. We note that more strict quality control criteria should be applied for the further analysis using $\lre$ \citep{2018MNRAS.477.1567L, 2018MNRAS.477.4711G}.
The number of galaxies having both quality assured $\rm \lambda_{R_{e}, MaNGA}$ and $\rm \lambda_{R_{e}, MaNGA, Deconv.}$ is 2,268.
We present the relation between the $\rm \lambda_{R_{e}, MaNGA}$ and the $\rm \lambda_{R_{e}, MaNGA, Deconv.}$ 
\replaced{in the upper panel of \autoref{fig:spinmanga}.}{in the left most panels of \autoref{fig:spinmanga} depending on {$\nsrc$} of the MaNGA galaxies (NSA\_SERSIC\_N).} Compare to the $\rm \lambda_{R_{e}, MaNGA}$, most of the $\rm\lambda_{R_{e}, MaNGA, Deconv.}$ values are moderately increased for both {$\nsrc$} ranges.
\replaced{The median and standard deviation of $\rm \Delta \lre$ is $\rm 0.065 \pm 0.045$, or median increase of 24 percent with 26 percent point standard deviation. We also check the correlation between the two ratios ($\rm \lambda_{R_{e}, MaNGA}/\lambda_{R_{e}, MaNGA, Deconv.}$ and $\rm FWHM_{PSF}/R_{e}$) and as expected from the result with mock IFU data, $\rm \lambda_{R_{e}, MaNGA, Deconv.}/\lambda_{R_{e}, MaNGA}$ increases as $\rm FWHM_{PSF}/R_{e}$ increases.}{The median and standard deviation of $\rm \Delta \lre$ is $\rm 0.044 \pm 0.046$, or median increase of 8 percent with 18 percent point standard deviation for galaxies with $\nsrc < 2$. The median and standard deviation of $\rm \Delta \lre$ is $\rm 0.055 \pm 0.045$, or median increase of 37 percent with 26 percent point standard deviation for galaxies with $\nsrc \ge 2$. Note that the fractional difference is larger for the galaxies with $\nsrc \ge 2$ because the majority of $\nsrc \ge 2$ galaxies have relatively small $\rm \lambda_{R_{e}}$ values compared to $\nsrc < 2$ galaxies. We also check the correlation between the two ratios ($\rm \lambda_{R_{e}, MaNGA}/\lambda_{R_{e}, MaNGA, Deconv.}$ and $\rm FWHM_{PSF}/R_{e}$) and as expected from the result with mock IFU data, $\rm \lambda_{R_{e}, MaNGA, Deconv.}/\lambda_{R_{e}, MaNGA}$ increases as $\rm FWHM_{PSF}/R_{e}$ increases.}

During the validity check of the calculated $\rm \lre$,
we find tens or more of galaxies that their $\lre$ do not seem measured correctly.
There are several such cases, for example, 
\begin{enumerate}
\item There are seven galaxies in \autoref{fig:spinmanga} that have both $\rm \lambda_{R_{e}, MaNGA}$ and $\rm \lambda_{R_{e}, MaNGA, Deconv.} >$ 0.8. However, all of those galaxies' systematic velocity are highly underestimated or overestimated, in other words, galaxy systematic velocity derived by the NSA redshift is not matching with the true systematic velocity. After correcting its systematic velocity, it turns out their $\lre$ value is significantly less than 0.8.
\item There are galaxies with foreground/background objects, either star or other galaxies, at or around the $\rm 1\Reff$ elliptical aperture. Some of them are already masked by MaNGA data reduction pipeline, but still there are tens of IFU data with unmasked interloper. Either masked or unmasked, the interloping object disrupt the kinematics measurement in particular at the border between the object of interest and the interloper.
\item Contrary to the sample definition of MaNGA galaxies, there are galaxies where their $\Reff$ size is comparable to the IFU field of view. This brings spaxels at the edge of the IFU field of view to the $\lre$ calculation. Since the kinematics measured at near the edge of the deconvolved IFU data could be different from the correct value, the calculated $\lre$ from such galaxy sample is not reliable.
\item IFU data with small field of view (i.e. 12$\arcsec$) often includes only a tens of spaxels to $\lre$ calculation. This means that even a small offset at its center position or systematic velocity can leads considerable change in the measured $\lre$ value. These suggest that more careful data quality assurance is required to assure the data with correctly measured $\lre$ value.
\end{enumerate}

\begin{figure*}
\includegraphics[width=\linewidth]{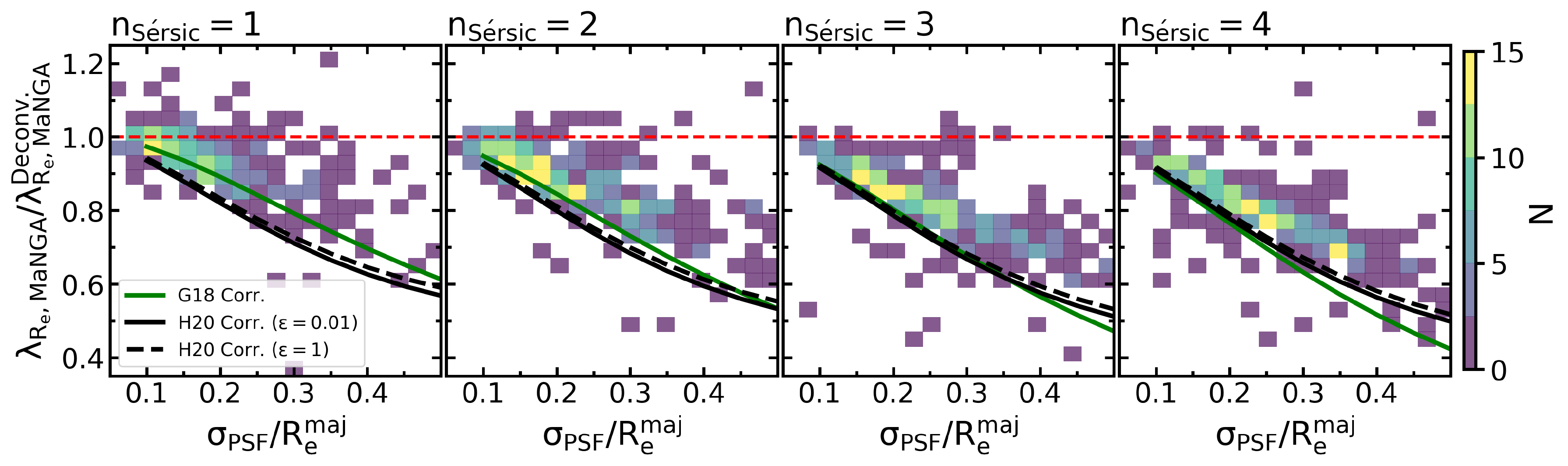}
\caption{The ratios between the $\lambda_{R_e}$ measured from the original MaNGA data and the $\lambda_{R_e}$ measured from PSF-deconvolved MaNGA data, depending on $\rm \sigma_{PSF}/R_{e}^{maj.}$. Each panel shows the 2D histogram of $\rm\lambda_{R_{e}, MaNGA}/\lambda_{R_{e}, MaNGA}^{Deconv.}$ versus $\rm \sigma_{PSF}/R_{e}^{maj.}$, when $\nsrc=$1,2,3, and 4. The overlapped lines show the inverse of \citetalias{2018MNRAS.477.4711G} and \citetalias{2020MNRAS.497.2018H} correction functions, which are equal to $\rm\lambda_{R_{e}, MaNGA}/\lambda_{R_{e}, MaNGA}^{G18\;Corr.}$ (green solid line) or  $\rm\lambda_{R_{e}, MaNGA}$/$\rm\lambda_{R_{e}, MaNGA}^{H20\;Corr.}$ (black lines). Two inverse \citetalias{2020MNRAS.497.2018H} functions are shown, one when $\rm\epsilon=$0.01 (black solid line) and the other when $\rm\epsilon=$1 (black dashed line). 
} \label{fig:spinmanga_ratio}
\end{figure*}

We also plot the relation between the $\rm \lambda_{R_{e}, MaNGA}$ and the corrected $\lre$ (following \citetalias{2018MNRAS.477.4711G} and \citetalias{2020MNRAS.497.2018H}) in \autoref{fig:spinmanga}. Compared to $\Delta \lre$ caused by the deconvolution method, the $\Delta \lre$ caused by \citetalias{2018MNRAS.477.4711G} correction is higher. Sometimes the corrected value becomes higher than $\lre = 1$ which is nonphysical. As noticed in \autoref{sub:spinmock}, the over-correction could be due to different shape of velocity dispersion profile of real galaxy compare to the profile used to derived the correction function, or some other unanticipated model-dependent bias.
\added{
For example, the data points in the Figure 4 of \citetalias{2018MNRAS.477.4711G} shows that the difference in $\lre$ caused by the PSF convolution is larger with model galaxies compared to the real (ATLAS$\rm ^{3D}$) galaxies. This shows that the \citetalias{2018MNRAS.477.4711G} correction function could over-correct the $\lre$ values when applied to the real galaxies. Compared to the $\Delta \lre$ caused by the deconvolution method, the $\Delta \lre$ caused by \citetalias{2020MNRAS.497.2018H} correction is also higher. The $\Delta \lre$ caused by \citetalias{2020MNRAS.497.2018H} correction is larger than the $\Delta \lre$ caused by \citetalias{2018MNRAS.477.4711G} when $\nsrc < 2$, but smaller than the $\Delta \lre$ caused by \citetalias{2018MNRAS.477.4711G} when $\nsrc \ge 2$. 

To better understand this $\nsrc$ dependence, we plot $\lambda_{R_{e}, MaNGA}$/$\lambda_{R_{e}, MaNGA}^{Deconv.}$ versus $\rm \sigma_{PSF}/R_{e}^{maj.}$ in \autoref{fig:spinmanga_ratio}, which is another representation of \autoref{fig:spinmanga}. In \autoref{fig:spinmanga_ratio}, the difference between the data points and the red dashed line ($\lambda_{R_{e}, MaNGA}$/$\lambda_{R_{e}, MaNGA}^{Deconv.}=1$) indicates the amount of $\lre$ difference caused by the deconvolution. When $\nsrc=1$, the $\lambda_{R_{e}, MaNGA}$/$\lambda_{R_{e}, MaNGA}^{Deconv.}$ distribution is similar to the green line which is \citetalias{2018MNRAS.477.4711G} correction. In other words, the amount of correction made by \citetalias{2018MNRAS.477.4711G} correction is similar to the amount of $\lre$ change made by the deconvolution method. Black lines in $\nsrc=1$ panel of \autoref{fig:spinmanga_ratio} are located below the green line, meaning that if \citetalias{2020MNRAS.497.2018H} correction is applied to $\lambda_{R_{e}, MaNGA}$, then the corrected value will be higher than the value corrected by \citetalias{2018MNRAS.477.4711G}. 
On contrary, when $\nsrc=4$, the black lines are closer to the $\lambda_{R_{e}, MaNGA}$/$\lambda_{R_{e}, MaNGA}^{Deconv.}$ distribution where as the green line is located below the black lines. This result aligns with the result in \autoref{fig:spin_mock}, where the $\lre$ corrected by \citetalias{2018MNRAS.477.4711G} correction is similar to the $\lre$ measured from deconvolved mock IFU data when $\nsrc=1$ and the $\lre$ corrected by \citetalias{2020MNRAS.497.2018H} correction is closer to the $\lre$ measured from deconvolved mock IFU data when $\nsrc=4$. In other words, the difference in $\lre$ caused by deconvolution in the mock IFU data and the MaNGA data shows similar trends, depending on $\nsrc$ and $\rm \sigma_{PSF}/R_{e}^{maj.}$. This result indicates that the \citetalias{2018MNRAS.477.4711G} or \citetalias{2020MNRAS.497.2018H} correction function should be used with caution, because \citetalias{2018MNRAS.477.4711G} correction could over-correct the $\lre$ when $\nsrc > 3$ and \citetalias{2020MNRAS.497.2018H} correction could over-correct the $\lre$ when $\nsrc < 3$. 
Note that the model galaxies used for \citetalias{2018MNRAS.477.4711G} were generated using Jeans Anisotropic Modelling method \citep{2008MNRAS.390...71C}, whereas the \citetalias{2020MNRAS.497.2018H} correction was made based on N-body galaxy models. Both \citetalias{2018MNRAS.477.4711G} and \citetalias{2020MNRAS.497.2018H} corrections are empirical correction function based on model/simulation. Therefore, any difference between the model galaxy and the real galaxy could cause a bias to the parameter values corrected by those functions when such functions are applied to a real data. \autoref{fig:spin_mock} and \autoref{fig:spinmanga_ratio} show that neither \citetalias{2018MNRAS.477.4711G} or \citetalias{2020MNRAS.497.2018H} correction could provide the correct $\lre$ value over the $\nsrc$ range from 1 to 4.
}
\deleted{Therefore, although the usage of analytic correction function is convenient, the kind of simple prescription should be used with caution.}

\begin{figure*}
\includegraphics[width=\linewidth]{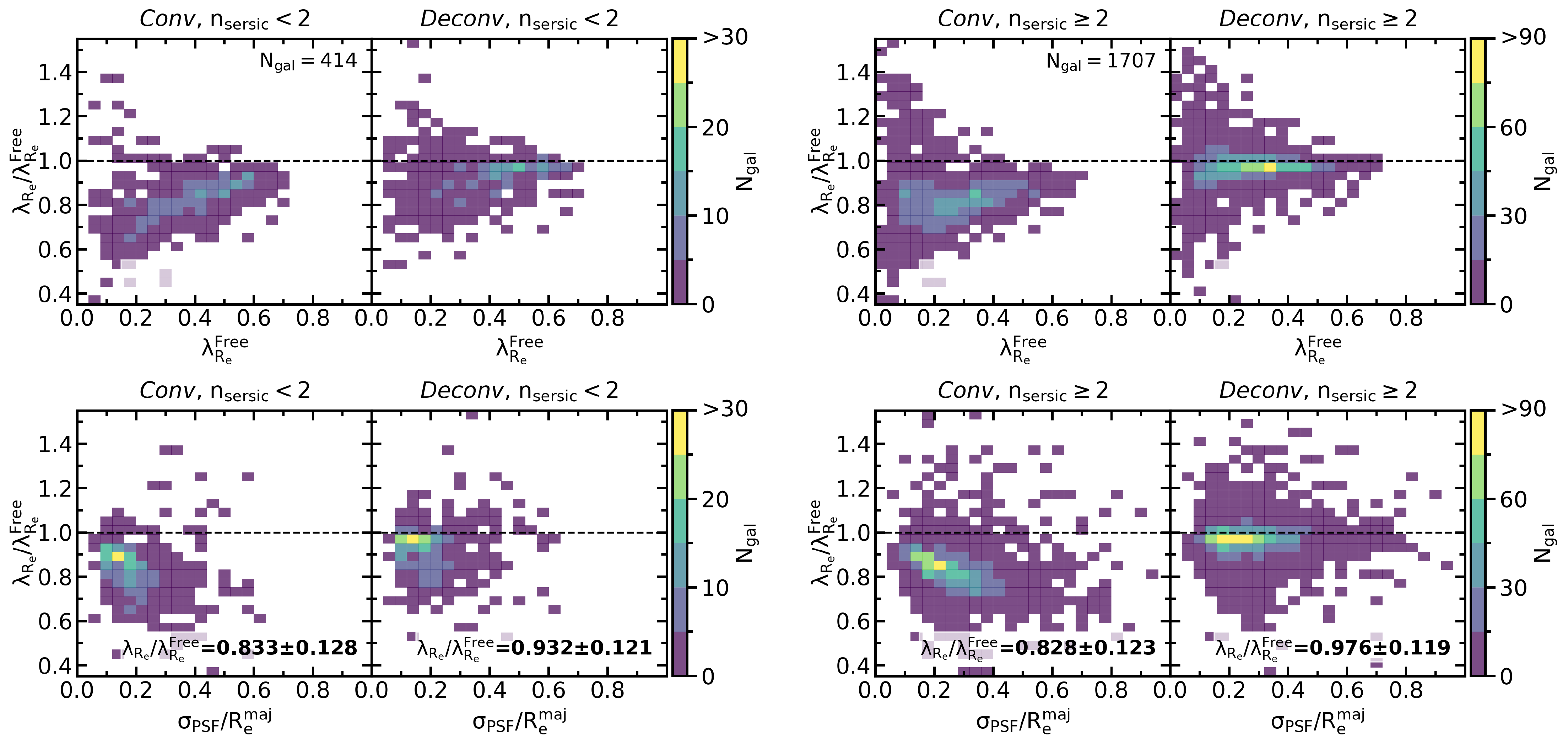}
\caption{Comparison between the ratios between $\lambda_{R_e}$ measured from PSF-Convolved or PSF-Deconvolved IFU data with respect to $\lambda_{R_e}^{Free}$ and $\rm \sigma_{PSF}/R_{e}^{maj.}$ using MaNGA DR15 galaxies as the PSF-Free data. Panels on left show $\lambda_{R_e}^{Conv}/\lambda_{R_e}^{Free}$ and $\lambda_{R_e}^{Deconv}/\lambda_{R_e}^{Free}$ as a function of $\lambda_{R_e}^{Free}$ or $\rm \sigma_{PSF}/R_{e}^{maj.}$ for the original MaNGA galaxies with $\nsrc < 2$. Panels on right show the same as the left but using the original MaNGA galaxies with $\nsrc \ge 2$. The median and standard deviation of $\lambda_{R_e}/\lambda_{R_e}^{Free}$ are noted in bottom panels.
} \label{fig:spin_corr_nsersic_sigre}
\end{figure*}

\added{
\subsection{Verification using MaNGA data}\label{sub:spin_verification_manga}
Verification of deconvolution method using mock data allows to test the method in more controlled samples. However, although the kinematics of the mock IFU data that we have generated is inspired by the kinematics of the real galaxies, the kinematics is described by empirical rotation curve model, not by a model based on physics. Nevertheless, it is also true that even the models based on physics are not sufficient to replace the kinematics of actual galaxy. The limitation of the empirical kinematics that we used could raise a question that whether the deconvolution method is still applicable to the real galaxy. To mitigate the gap between our empirical kinematics and the real galaxy kinematics, we examined the validity of our method by using the kinematics of real galaxies.

We use the MaNGA DR15 data as a ground truth (PSF-Free) and test our deconvolution method in the same way as we did with the mock data. Although the MaNGA data is already seeing-convolved, the flux and kinematic distribution is far more realistic than any empirical model or simulation. First, we generate PSF-convolved data (PSF-$\ndc$) by convolving fixed-size PSF (FWHM=2.6{\arcsec}) to MaNGA IFU data. Then, we again apply the deconvolution method to the PSF-convolved data and generate PSF-deconvolved data (PSF-$\dc$). Kinematic parameters (velocity and velocity distribution) are measured from all {$\free$}, {$\ndc$}, {$\dc$} in the same way as described in \autoref{sub:mangakin}, and $\lre$ values are also measured as described in \autoref{sub:mangaspin}. The velocity and velocity distribution of the seven selected nearby MaNGA galaxies (z $<$ 0.023 and IFU FoV $\ge$ 27{\arcsec}) are shown in \autoref{appfig:manga_deconv_test}. {$\lre$} values measured from {$\free$}, {$\ndc$}, {$\dc$} data of those seven galaxies are also presented.

In \autoref{fig:spin_corr_nsersic_sigre}, we plot the ratios between the {$\lre$} measured from {$\ndc$} and {$\free$} ($\lambda_{R_e}^{Conv}/\lambda_{R_e}^{Free}$) and {$\dc$} and {$\free$} ($\lambda_{R_e}^{Deconv}/\lambda_{R_e}^{Free}$) as a function of $\lre^{Free}$ and $\rm \sigma_{PSF}/R_{e}^{maj.}$. We only use the galaxies which passed quality control that we applied in \autoref{sub:mangaspin}. We further exclude the galaxies that show $\Delta \lambda_{R_e}^{Conv}>$0.05 and $\lambda_{R_e}^{Free}>$0.7, because visual inspection finds that that $\lre$ of those galaxies are not measured correctly, due to the reasons that we noted in  \autoref{sub:mangaspin}. The result is further divided depend on S\'{e}rsic index ($\nsrc$ $<$ 2 and $\nsrc \ge 2$). For galaxies with $\nsrc$ $<$ 2, 
$\lambda_{R_e}^{Conv}/\lambda_{R_e}^{Free}$=0.833 $\pm$ 0.128 (median and standard deviation) but $\lambda_{R_e}^{Deconv}/\lambda_{R_e}^{Free}$=0.932 $\pm$ 0.121.
For galaxies with $\nsrc$ $\ge$ 2, $\lambda_{R_e}^{Conv}/\lambda_{R_e}^{Free}$=0.828 $\pm$ 0.123 but $\lambda_{R_e}^{Deconv}/\lambda_{R_e}^{Free}$=0.976 $\pm$ 0.119.
In particular, $\lambda_{R_e}^{Deconv}$ is measured more accurately for the galaxies with $\nsrc \ge 2$. 
The recovered $\lambda_{R_e}^{Deconv}$ values do not show particular dependency with respect to the $\rm \sigma_{PSF}/R_{e}^{maj.}$. The standard deviation of $\lambda_{R_e}^{Deconv}/\lambda_{R_e}^{Free}$ is also decreased compared to the standard deviation of $\lambda_{R_e}^{Conv}/\lambda_{R_e}^{Free}$. 


The result presented in \autoref{fig:spin_corr_nsersic_sigre} shows that the $\lre$ values measured from the deconvolved data is well-restored to the correct value, when the deconvolution method is applied to data close to the real data. The result also show that the empirically derived correction functions are working well. However, the standard deviation of $\Delta \lambda$ values with the correction functions implies that the intrinsic model bias in the correction function may under or over-correct the $\lre$ values. In \autoref{fig:spin_fwhm_nsersic}, we plotted a figure which are the same as \autoref{fig:spin_corr_nsersic_sigre} but when $\fpsf$ value for the deconvolution or $\lre$ correction is different from the true value by $\rm \pm$0.3$\arcsec$. The results give an idea of the sensitivity of the $\lre$ values when incorrect $\fpsf$ is used for the deconvolution or correction function. Over the tested $\fpsf$ range ($\rm \pm$0.3$\arcsec$), the $\lambda_{R_e}/\lambda_{R_e}^{Free}$ values show the difference maximum 0.05 (5 percent point), whereas the $\lambda_{R_e}^{Conv}$ shows over 17 percent difference compared to $\lambda_{R_e}^{Free}$. Therefore, $\lre$ from the deconvolved data or corrected $\lre$ by the correction function still provides better values compared to the uncorrected $\lre$ value.



}

\section{Summary and Conclusion}\label{sec:summary}
\deleted{
We investigate the effects of the PSF deconvolution of the optical integral field spectroscopy data on the internal kinematics of galaxies. The Lucy-Richardson algorithm is used for the deconvolution. We develop the procedure to apply the algorithm to MaNGA IFU data, which deconvolves the given IFU data efficiently using only two parameters ($\niter$ and $\rm{FWHM}_{\rm{PSF}}$).
We generate a large number of mock data with varying S\'{e}rsic profile and the rotation curve model, and use them to check how well the deconvolution can restore the true kinematics when the input data is convolved with the PSF. The deconvolution is powerful in the sense that it can provide an unbiased (model independent) correction to any PSF-convolved IFU data. We apply the deconvolution to the real data, SDSS-IV MaNGA, and show that the deconvolution makes a noticeable difference in the 2D flux, the velocity, and the velocity dispersion distributions. Finally, we demonstrate that the $\lre$ spin parameter can be well-estimated compare to the true value by applying this technique.
}
\added{
We investigate the effects of the PSF deconvolution of the optical integral field spectroscopy data on galaxies' internal kinematics. We develop a procedure to apply the Lucy-Richardson algorithm to deconvolve IFU data. The procedure deconvolves a given IFU data efficiently using only two parameters ($\niter$ and $\rm{FWHM}_{\rm{PSF}}$). We generate a large number of mock data with varying S\'{e}rsic profile and the rotation curve model and use them to verify the feasibility of the deconvolution method. Using the mock data, we have shown that the deconvolution works well on the IFU data. Contrary to the kinematics measured from PSF-convolved mock IFU data, the kinematics measured from deconvolved mock IFU data is well restored to the true kinematics of mock galaxies. 

We deconvolve the mock IFU data with a wide range of a number of iteration (1 to 50). From the result, 
we have determined the number of iteration of the deconvolution procedure as 20. We also shows that the deconvolution is not highly sensitive to the PSF FWHM size used for the deconvolution, although the use of accurate PSF FWHM close to the one implied in the PSF-convolved IFU data is always desired. The result from various mock IFU data shows that the deconvolution method is working on a variety of galaxy photometric/kinematic distributions. The deconvolution is powerful in the sense that it can provide an unbiased (model-independent) correction to any PSF-convolved IFU data. We apply the deconvolution to the real data, SDSS-IV MaNGA. We show that the deconvolution makes a noticeable difference in the 2D flux, the velocity, and the velocity dispersion distributions. 

Finally, we demonstrate that the $\lre$ spin parameter can be well-estimated from the PSF-deconvolved IFU data. Using the mock IFU data, we show that the $\lre$ can be correctly measured from deconvolved IFU data within 3 percent, when the ratio between the convolved PSF size ($\rm \sigma_{PSF}$) and the galaxy effective radius ($\Reff$) is less than 0.2. We measure $\lre$ from the PSF-deconvolved MaNGA data. The $\lre$ measured from the original MaNGA data and the $\lre$ measured from the deconvolved MaNGA data show the moderate difference compared to the difference caused by empirically driven $\lre$ correction functions. We generate PSF-convolved/deconvolved mock IFU data by considering the original MaNGA IFU data as perfect (PSF-Free) data. The $\lre$ comparison result from the experiment using the original MaNGA IFU data as perfect data shows that $\lre$ values measured from the PSF deconvolved mock IFU data are close true $\lre$ values. Therefore, the deconvolution method is indeed working well with the realistic IFU data as well.
}

\section*{Acknowledgements}
We thank the anonymous referee for their detailed and constructive comments which led to the significant improvement of this manuscript.
We thank the Korea Institute for Advanced Study for providing computing resources (KIAS Center for Advanced Computation Linux Cluster System) for this work.  

Funding for the Sloan Digital Sky Survey IV has been provided by the Alfred P. Sloan Foundation, the U.S. Department of Energy Office of Science, and the Participating Institutions. SDSS-IV acknowledges
support and resources from the Center for High-Performance Computing at
the University of Utah. The SDSS web site is www.sdss.org.  

SDSS-IV is managed by the Astrophysical Research Consortium for the 
Participating Institutions of the SDSS Collaboration including the 
Brazilian Participation Group, the Carnegie Institution for Science, 
Carnegie Mellon University, the Chilean Participation Group, the French Participation Group, Harvard-Smithsonian Center for Astrophysics, 
Instituto de Astrof\'isica de Canarias, The Johns Hopkins University, Kavli Institute for the Physics and Mathematics of the Universe (IPMU) / 
University of Tokyo, the Korean Participation Group, Lawrence Berkeley National Laboratory, 
Leibniz Institut f\"ur Astrophysik Potsdam (AIP),  
Max-Planck-Institut f\"ur Astronomie (MPIA Heidelberg), 
Max-Planck-Institut f\"ur Astrophysik (MPA Garching), 
Max-Planck-Institut f\"ur Extraterrestrische Physik (MPE), 
National Astronomical Observatories of China, New Mexico State University, 
New York University, University of Notre Dame, 
Observat\'ario Nacional / MCTI, The Ohio State University, 
Pennsylvania State University, Shanghai Astronomical Observatory, 
United Kingdom Participation Group,
Universidad Nacional Aut\'onoma de M\'exico, University of Arizona, 
University of Colorado Boulder, University of Oxford, University of Portsmouth, 
University of Utah, University of Virginia, University of Washington, University of Wisconsin, Vanderbilt University, and Yale University.
\bibliography{ms}

\begin{thebibliography}{}
\expandafter\ifx\csname natexlab\endcsname\relax\def\natexlab#1{#1}\fi
\providecommand{\url}[1]{\href{#1}{#1}}
\providecommand{\dodoi}[1]{doi:~\href{http://doi.org/#1}{\nolinkurl{#1}}}
\providecommand{\doeprint}[1]{\href{http://ascl.net/#1}{\nolinkurl{http://ascl.net/#1}}}
\providecommand{\doarXiv}[1]{\href{https://arxiv.org/abs/#1}{\nolinkurl{https://arxiv.org/abs/#1}}}

\bibitem[{{Aguado} {et~al.}(2019){Aguado}, {Ahumada}, {Almeida}, {Anderson},
  {Andrews}, {Anguiano}, {Aquino Ort{\'\i}z}, {Arag{\'o}n-Salamanca},
  {Argudo-Fern{\'a}ndez}, {Aubert}, {Avila-Reese}, {Badenes}, {Barboza
  Rembold}, {Barger}, {Barrera-Ballesteros}, {Bates}, {Bautista}, {Beaton},
  {Beers}, {Belfiore}, {Bernardi}, {Bershady}, {Beutler}, {Bird}, {Bizyaev},
  {Blanc}, {Blanton}, {Blomqvist}, {Bolton}, {Boquien}, {Borissova}, {Bovy},
  {Brandt}, {Brinkmann}, {Brownstein}, {Bundy}, {Burgasser}, {Byler}, {Cano
  Diaz}, {Cappellari}, {Carrera}, {Cervantes Sodi}, {Chen}, {Cherinka}, {Choi},
  {Chung}, {Coffey}, {Comerford}, {Comparat}, {Covey}, {da Silva Ilha}, {da
  Costa}, {Dai}, {Damke}, {Darling}, {Davies}, {Dawson}, {de Sainte Agathe},
  {Deconto Machado}, {Del Moro}, {De Lee}, {Diamond-Stanic}, {Dom{\'\i}nguez
  S{\'a}nchez}, {Donor}, {Drory}, {du Mas des Bourboux}, {Duckworth}, {Dwelly},
  {Ebelke}, {Emsellem}, {Escoffier}, {Fern{\'a}ndez-Trincado}, {Feuillet},
  {Fischer}, {Fleming}, {Fraser-McKelvie}, {Freischlad}, {Frinchaboy}, {Fu},
  {Galbany}, {Garcia-Dias}, {Garc{\'\i}a-Hern{\'a}ndez}, {Garma Oehmichen},
  {Geimba Maia}, {Gil-Mar{\'\i}n}, {Grabowski}, {Gu}, {Guo}, {Ha},
  {Harrington}, {Hasselquist}, {Hayes}, {Hearty}, {Hernandez Toledo}, {Hicks},
  {Hogg}, {Holley-Bockelmann}, {Holtzman}, {Hsieh}, {Hunt}, {Hwang},
  {Ibarra-Medel}, {Jimenez Angel}, {Johnson}, {Jones}, {J{\"o}nsson},
  {Kinemuchi}, {Kollmeier}, {Krawczyk}, {Kreckel}, {Kruk}, {Lacerna}, {Lan},
  {Lane}, {Law}, {Lee}, {Li}, {Lian}, {Lin}, {Lin}, {Lintott}, {Long},
  {Longa-Pe{\~n}a}, {Mackereth}, {de la Macorra}, {Majewski}, {Malanushenko},
  {Manchado}, {Maraston}, {Mariappan}, {Marinelli}, {Marques-Chaves},
  {Masseron}, {Masters}, {McDermid}, {Medina Pe{\~n}a}, {Meneses-Goytia},
  {Merloni}, {Merrifield}, {Meszaros}, {Minniti}, {Minsley}, {Muna}, {Myers},
  {Nair}, {Correa do Nascimento}, {Newman}, {Nitschelm}, {Olmstead}, {Oravetz},
  {Oravetz}, {Ortega Minakata}, {Pace}, {Padilla}, {Palicio}, {Pan}, {Pan},
  {Parikh}, {Parker}, {Peirani}, {Penny}, {Percival}, {Perez-Fournon},
  {Peterken}, {Pinsonneault}, {Prakash}, {Raddick}, {Raichoor}, {Riffel},
  {Riffel}, {Rix}, {Robin}, {Roman-Lopes}, {Rose}, {Ross}, {Rossi}, {Rowlands},
  {Rubin}, {S{\'a}nchez}, {S{\'a}nchez-Gallego}, {Sayres}, {Schaefer},
  {Schiavon}, {Schimoia}, {Schlafly}, {Schlegel}, {Schneider}, {Schultheis},
  {Seo}, {Shamsi}, {Shao}, {Shen}, {Shetty}, {Simonian}, {Smethurst}, {Sobeck},
  {Souter}, {Spindler}, {Stark}, {Stassun}, {Steinmetz}, {Storchi-Bergmann},
  {Stringfellow}, {Su{\'a}rez}, {Sun}, {Taghizadeh-Popp}, {Talbot}, {Tayar},
  {Thakar}, {Thomas}, {Tissera}, {Tojeiro}, {Troup}, {Unda-Sanzana},
  {Valenzuela}, {Vargas-Maga{\~n}a}, {V{\'a}zquez-Mata}, {Wake}, {Weaver},
  {Weijmans}, {Westfall}, {Wild}, {Wilson}, {Woods}, {Yan}, {Yang}, {Zamora},
  {Zasowski}, {Zhang}, {Zheng}, {Zheng}, {Zhu}, {Zinn}, \&
  {Zou}}]{2019ApJS..240...23A}
{Aguado}, D.~S., {Ahumada}, R., {Almeida}, A., {et~al.} 2019, \apjs, 240, 23,
  \dodoi{10.3847/1538-4365/aaf651}

\bibitem[{{Allington-Smith} {et~al.}(2002){Allington-Smith}, {Murray},
  {Content}, {Dodsworth}, {Davies}, {Miller}, {Jorgensen}, {Hook}, {Crampton},
  \& {Murowinski}}]{2002PASP..114..892A}
{Allington-Smith}, J., {Murray}, G., {Content}, R., {et~al.} 2002, \pasp, 114,
  892, \dodoi{10.1086/341712}

\bibitem[{{Andersen} \& {Bershady}(2013)}]{2013ApJ...768...41A}
{Andersen}, D.~R., \& {Bershady}, M.~A. 2013, \apj, 768, 41,
  \dodoi{10.1088/0004-637X/768/1/41}

\bibitem[{{Bacon} {et~al.}(2001){Bacon}, {Copin}, {Monnet}, {Miller},
  {Allington-Smith}, {Bureau}, {Carollo}, {Davies}, {Emsellem}, {Kuntschner},
  {Peletier}, {Verolme}, \& {de Zeeuw}}]{2001MNRAS.326...23B}
{Bacon}, R., {Copin}, Y., {Monnet}, G., {et~al.} 2001, \mnras, 326, 23,
  \dodoi{10.1046/j.1365-8711.2001.04612.x}

\bibitem[{{Bacon} {et~al.}(2010){Bacon}, {Accardo}, {Adjali}, {Anwand},
  {Bauer}, {Biswas}, {Blaizot}, {Boudon}, {Brau-Nogue}, {Brinchmann},
  {Caillier}, {Capoani}, {Carollo}, {Contini}, {Couderc}, {Daguis{\'e}},
  {Deiries}, {Delabre}, {Dreizler}, {Dubois}, {Dupieux}, {Dupuy}, {Emsellem},
  {Fechner}, {Fleischmann}, {Fran{\c{c}}ois}, {Gallou}, {Gharsa}, {Glindemann},
  {Gojak}, {Guiderdoni}, {Hansali}, {Hahn}, {Jarno}, {Kelz}, {Koehler},
  {Kosmalski}, {Laurent}, {Le Floch}, {Lilly}, {Lizon}, {Loupias}, {Manescau},
  {Monstein}, {Nicklas}, {Olaya}, {Pares}, {Pasquini}, {P{\'e}contal-Rousset},
  {Pell{\'o}}, {Petit}, {Popow}, {Reiss}, {Remillieux}, {Renault}, {Roth},
  {Rupprecht}, {Serre}, {Schaye}, {Soucail}, {Steinmetz}, {Streicher}, {Stuik},
  {Valentin}, {Vernet}, {Weilbacher}, {Wisotzki}, \&
  {Yerle}}]{2010SPIE.7735E..08B}
{Bacon}, R., {Accardo}, M., {Adjali}, L., {et~al.} 2010, in Society of
  Photo-Optical Instrumentation Engineers (SPIE) Conference Series, Vol. 7735,
  Ground-based and Airborne Instrumentation for Astronomy III, ed. I.~S.
  {McLean}, S.~K. {Ramsay}, \& H.~{Takami}, 773508, \dodoi{10.1117/12.856027}

\bibitem[{{Bershady} {et~al.}(2010){Bershady}, {Verheijen}, {Swaters},
  {Andersen}, {Westfall}, \& {Martinsson}}]{2010ApJ...716..198B}
{Bershady}, M.~A., {Verheijen}, M. A.~W., {Swaters}, R.~A., {et~al.} 2010,
  \apj, 716, 198, \dodoi{10.1088/0004-637X/716/1/198}

\bibitem[{{Bongard} {et~al.}(2011){Bongard}, {Soulez}, {Thi{\'e}baut}, \&
  {Pecontal}}]{2011MNRAS.418..258B}
{Bongard}, S., {Soulez}, F., {Thi{\'e}baut}, {\'E}., \& {Pecontal}, {\'E}.
  2011, \mnras, 418, 258, \dodoi{10.1111/j.1365-2966.2011.19480.x}

\bibitem[{{Bouch{\'e}} {et~al.}(2015){Bouch{\'e}}, {Carfantan}, {Schroetter},
  {Michel-Dansac}, \& {Contini}}]{2015AJ....150...92B}
{Bouch{\'e}}, N., {Carfantan}, H., {Schroetter}, I., {Michel-Dansac}, L., \&
  {Contini}, T. 2015, \aj, 150, 92, \dodoi{10.1088/0004-6256/150/3/92}

\bibitem[{Bourguignon {et~al.}(2011)Bourguignon, Carfantan, Slezak, \&
  Mary}]{6080853}
Bourguignon, S., Carfantan, H., Slezak, E., \& Mary, D. 2011, in 2011 3rd
  Workshop on Hyperspectral Image and Signal Processing: Evolution in Remote
  Sensing (WHISPERS), 1--4, \dodoi{10.1109/WHISPERS.2011.6080853}

\bibitem[{{Bundy} {et~al.}(2015){Bundy}, {Bershady}, {Law}, {Yan}, {Drory},
  {MacDonald}, {Wake}, {Cherinka}, {S{\'a}nchez-Gallego}, {Weijmans}, {Thomas},
  {Tremonti}, {Masters}, {Coccato}, {Diamond-Stanic}, {Arag{\'o}n-Salamanca},
  {Avila-Reese}, {Badenes}, {Falc{\'o}n-Barroso}, {Belfiore}, {Bizyaev},
  {Blanc}, {Bland-Hawthorn}, {Blanton}, {Brownstein}, {Byler}, {Cappellari},
  {Conroy}, {Dutton}, {Emsellem}, {Etherington}, {Frinchaboy}, {Fu}, {Gunn},
  {Harding}, {Johnston}, {Kauffmann}, {Kinemuchi}, {Klaene}, {Knapen},
  {Leauthaud}, {Li}, {Lin}, {Maiolino}, {Malanushenko}, {Malanushenko}, {Mao},
  {Maraston}, {McDermid}, {Merrifield}, {Nichol}, {Oravetz}, {Pan}, {Parejko},
  {Sanchez}, {Schlegel}, {Simmons}, {Steele}, {Steinmetz}, {Thanjavur},
  {Thompson}, {Tinker}, {van den Bosch}, {Westfall}, {Wilkinson}, {Wright},
  {Xiao}, \& {Zhang}}]{2015ApJ...798....7B}
{Bundy}, K., {Bershady}, M.~A., {Law}, D.~R., {et~al.} 2015, \apj, 798, 7,
  \dodoi{10.1088/0004-637X/798/1/7}

\bibitem[{{Cappellari}(2002)}]{2002MNRAS.333..400C}
{Cappellari}, M. 2002, \mnras, 333, 400,
  \dodoi{10.1046/j.1365-8711.2002.05412.x}

\bibitem[{{Cappellari}(2008)}]{2008MNRAS.390...71C}
---. 2008, \mnras, 390, 71, \dodoi{10.1111/j.1365-2966.2008.13754.x}

\bibitem[{{Cappellari}(2016)}]{2016ARA&A..54..597C}
---. 2016, \araa, 54, 597, \dodoi{10.1146/annurev-astro-082214-122432}

\bibitem[{{Cappellari}(2017)}]{2017MNRAS.466..798C}
---. 2017, \mnras, 466, 798, \dodoi{10.1093/mnras/stw3020}

\bibitem[{{Cappellari} \& {Emsellem}(2004)}]{2004PASP..116..138C}
{Cappellari}, M., \& {Emsellem}, E. 2004, \pasp, 116, 138,
  \dodoi{10.1086/381875}

\bibitem[{{Cappellari} {et~al.}(2013){Cappellari}, {Scott}, {Alatalo}, {Blitz},
  {Bois}, {Bournaud}, {Bureau}, {Crocker}, {Davies}, {Davis}, {de Zeeuw},
  {Duc}, {Emsellem}, {Khochfar}, {Krajnovi{\'c}}, {Kuntschner}, {McDermid},
  {Morganti}, {Naab}, {Oosterloo}, {Sarzi}, {Serra}, {Weijmans}, \&
  {Young}}]{2013MNRAS.432.1709C}
{Cappellari}, M., {Scott}, N., {Alatalo}, K., {et~al.} 2013, \mnras, 432, 1709,
  \dodoi{10.1093/mnras/stt562}

\bibitem[{{Choi} \& {Yi}(2017)}]{2017ApJ...837...68C}
{Choi}, H., \& {Yi}, S.~K. 2017, \apj, 837, 68,
  \dodoi{10.3847/1538-4357/aa5e4b}

\bibitem[{{Choi} {et~al.}(2018){Choi}, {Yi}, {Dubois}, {Kimm}, {Devriendt}, \&
  {Pichon}}]{2018ApJ...856..114C}
{Choi}, H., {Yi}, S.~K., {Dubois}, Y., {et~al.} 2018, \apj, 856, 114,
  \dodoi{10.3847/1538-4357/aab08f}

\bibitem[{Chung(2021)}]{chung2021code}
Chung, H. 2021, {First Release of IFU data deconvolution code}, 1.0.0,  Zenodo,
  \dodoi{10.5281/zenodo.4783185}

\bibitem[{Cooley \& Tukey(1965)}]{cooley1965algorithm}
Cooley, J.~W., \& Tukey, J.~W. 1965, Mathematics of computation, 19, 297

\bibitem[{{Cortese} {et~al.}(2016){Cortese}, {Fogarty}, {Bekki}, {van de
  Sande}, {Couch}, {Catinella}, {Colless}, {Obreschkow}, {Taranu}, {Tescari},
  {Barat}, {Bland-Hawthorn}, {Bloom}, {Bryant}, {Cluver}, {Croom},
  {Drinkwater}, {d'Eugenio}, {Konstantopoulos}, {Lopez-Sanchez}, {Mahajan},
  {Scott}, {Tonini}, {Wong}, {Allen}, {Brough}, {Goodwin}, {Green}, {Ho},
  {Kelvin}, {Lawrence}, {Lorente}, {Medling}, {Owers}, {Richards}, {Sharp}, \&
  {Sweet}}]{2016MNRAS.463..170C}
{Cortese}, L., {Fogarty}, L.~M.~R., {Bekki}, K., {et~al.} 2016, \mnras, 463,
  170, \dodoi{10.1093/mnras/stw1891}

\bibitem[{{Courbin} {et~al.}(2000){Courbin}, {Magain}, {Kirkove}, \&
  {Sohy}}]{2000ApJ...529.1136C}
{Courbin}, F., {Magain}, P., {Kirkove}, M., \& {Sohy}, S. 2000, \apj, 529,
  1136, \dodoi{10.1086/308291}

\bibitem[{{D'Eugenio} {et~al.}(2013){D'Eugenio}, {Houghton}, {Davies}, \&
  {Dalla Bont{\`a}}}]{2013MNRAS.429.1258D}
{D'Eugenio}, F., {Houghton}, R.~C.~W., {Davies}, R.~L., \& {Dalla Bont{\`a}},
  E. 2013, \mnras, 429, 1258, \dodoi{10.1093/mnras/sts406}

\bibitem[{{Dressler} {et~al.}(2011){Dressler}, {Bigelow}, {Hare}, {Sutin},
  {Thompson}, {Burley}, {Epps}, {Oemler}, {Bagish}, {Birk}, {Clardy},
  {Gunnels}, {Kelson}, {Shectman}, \& {Osip}}]{2011PASP..123..288D}
{Dressler}, A., {Bigelow}, B., {Hare}, T., {et~al.} 2011, \pasp, 123, 288,
  \dodoi{10.1086/658908}

\bibitem[{{Emsellem} {et~al.}(1994){Emsellem}, {Monnet}, \&
  {Bacon}}]{1994A&A...285..723E}
{Emsellem}, E., {Monnet}, G., \& {Bacon}, R. 1994, \aap, 285, 723

\bibitem[{{Emsellem} {et~al.}(2007){Emsellem}, {Cappellari}, {Krajnovi{\'c}},
  {van de Ven}, {Bacon}, {Bureau}, {Davies}, {de Zeeuw}, {Falc{\'o}n-Barroso},
  {Kuntschner}, {McDermid}, {Peletier}, \& {Sarzi}}]{2007MNRAS.379..401E}
{Emsellem}, E., {Cappellari}, M., {Krajnovi{\'c}}, D., {et~al.} 2007, \mnras,
  379, 401, \dodoi{10.1111/j.1365-2966.2007.11752.x}

\bibitem[{{Emsellem} {et~al.}(2011){Emsellem}, {Cappellari}, {Krajnovi{\'c}},
  {Alatalo}, {Blitz}, {Bois}, {Bournaud}, {Bureau}, {Davies}, {Davis}, {de
  Zeeuw}, {Khochfar}, {Kuntschner}, {Lablanche}, {McDermid}, {Morganti},
  {Naab}, {Oosterloo}, {Sarzi}, {Scott}, {Serra}, {van de Ven}, {Weijmans}, \&
  {Young}}]{2011MNRAS.414..888E}
---. 2011, \mnras, 414, 888, \dodoi{10.1111/j.1365-2966.2011.18496.x}

\bibitem[{{Falc{\'o}n-Barroso} {et~al.}(2011){Falc{\'o}n-Barroso},
  {S{\'a}nchez-Bl{\'a}zquez}, {Vazdekis}, {Ricciardelli}, {Cardiel}, {Cenarro},
  {Gorgas}, \& {Peletier}}]{2011A&A...532A..95F}
{Falc{\'o}n-Barroso}, J., {S{\'a}nchez-Bl{\'a}zquez}, P., {Vazdekis}, A.,
  {et~al.} 2011, \aap, 532, A95, \dodoi{10.1051/0004-6361/201116842}

\bibitem[{{Feng} \& {Gallo}(2011)}]{2011RAA....11.1429F}
{Feng}, J.~Q., \& {Gallo}, C.~F. 2011, Research in Astronomy and Astrophysics,
  11, 1429, \dodoi{10.1088/1674-4527/11/12/005}

\bibitem[{{Girardi} {et~al.}(2000){Girardi}, {Bressan}, {Bertelli}, \&
  {Chiosi}}]{2000A&AS..141..371G}
{Girardi}, L., {Bressan}, A., {Bertelli}, G., \& {Chiosi}, C. 2000, \aaps, 141,
  371, \dodoi{10.1051/aas:2000126}

\bibitem[{{Graham} {et~al.}(2018){Graham}, {Cappellari}, {Li}, {Mao},
  {Bershady}, {Bizyaev}, {Brinkmann}, {Brownstein}, {Bundy}, {Drory}, {Law},
  {Pan}, {Thomas}, {Wake}, {Weijmans}, {Westfall}, \&
  {Yan}}]{2018MNRAS.477.4711G}
{Graham}, M.~T., {Cappellari}, M., {Li}, H., {et~al.} 2018, \mnras, 477, 4711,
  \dodoi{10.1093/mnras/sty504}

\bibitem[{{Greene} {et~al.}(2018){Greene}, {Leauthaud}, {Emsellem}, {Ge},
  {Arag{\'o}n-Salamanca}, {Greco}, {Lin}, {Mao}, {Masters}, {Merrifield},
  {More}, {Okabe}, {Schneider}, {Thomas}, {Wake}, {Pan}, {Bizyaev}, {Oravetz},
  {Simmons}, {Yan}, \& {van den Bosch}}]{2018ApJ...852...36G}
{Greene}, J.~E., {Leauthaud}, A., {Emsellem}, E., {et~al.} 2018, \apj, 852, 36,
  \dodoi{10.3847/1538-4357/aa9bde}

\bibitem[{{Harborne} {et~al.}(2020){Harborne}, {van de Sande}, {Cortese},
  {Power}, {Robotham}, {Lagos}, \& {Croom}}]{2020MNRAS.497.2018H}
{Harborne}, K.~E., {van de Sande}, J., {Cortese}, L., {et~al.} 2020, \mnras,
  497, 2018, \dodoi{10.1093/mnras/staa1847}

\bibitem[{{Henault} {et~al.}(2003){Henault}, {Bacon}, {Bonneville}, {Boudon},
  {Davies}, {Ferruit}, {Gilmore}, {Le F{\`e}vre}, {Lemonnier}, {Lilly},
  {Morris}, {Prieto}, {Steinmetz}, \& {de Zeeuw}}]{2003SPIE.4841.1096H}
{Henault}, F., {Bacon}, R., {Bonneville}, C., {et~al.} 2003, in Society of
  Photo-Optical Instrumentation Engineers (SPIE) Conference Series, Vol. 4841,
  Instrument Design and Performance for Optical/Infrared Ground-based
  Telescopes, ed. M.~{Iye} \& A.~F.~M. {Moorwood}, 1096--1107,
  \dodoi{10.1117/12.462334}

\bibitem[{{Hopkins} {et~al.}(2010){Hopkins}, {Bundy}, {Hernquist}, {Wuyts}, \&
  {Cox}}]{2010MNRAS.401.1099H}
{Hopkins}, P.~F., {Bundy}, K., {Hernquist}, L., {Wuyts}, S., \& {Cox}, T.~J.
  2010, \mnras, 401, 1099, \dodoi{10.1111/j.1365-2966.2009.15699.x}

\bibitem[{{Jesseit} {et~al.}(2009){Jesseit}, {Cappellari}, {Naab}, {Emsellem},
  \& {Burkert}}]{2009MNRAS.397.1202J}
{Jesseit}, R., {Cappellari}, M., {Naab}, T., {Emsellem}, E., \& {Burkert}, A.
  2009, \mnras, 397, 1202, \dodoi{10.1111/j.1365-2966.2009.14984.x}

\bibitem[{{Kelz} {et~al.}(2006){Kelz}, {Verheijen}, {Roth}, {Bauer}, {Becker},
  {Paschke}, {Popow}, {S{\'a}nchez}, \& {Laux}}]{2006PASP..118..129K}
{Kelz}, A., {Verheijen}, M. A.~W., {Roth}, M.~M., {et~al.} 2006, \pasp, 118,
  129, \dodoi{10.1086/497455}

\bibitem[{{Lagos} {et~al.}(2017){Lagos}, {Theuns}, {Stevens}, {Cortese},
  {Padilla}, {Davis}, {Contreras}, \& {Croton}}]{2017MNRAS.464.3850L}
{Lagos}, C. d.~P., {Theuns}, T., {Stevens}, A. R.~H., {et~al.} 2017, \mnras,
  464, 3850, \dodoi{10.1093/mnras/stw2610}

\bibitem[{{Law} {et~al.}(2015){Law}, {Yan}, {Bershady}, {Bundy}, {Cherinka},
  {Drory}, {MacDonald}, {S{\'a}nchez-Gallego}, {Wake}, {Weijmans}, {Blanton},
  {Klaene}, {Moran}, {Sanchez}, \& {Zhang}}]{2015AJ....150...19L}
{Law}, D.~R., {Yan}, R., {Bershady}, M.~A., {et~al.} 2015, \aj, 150, 19,
  \dodoi{10.1088/0004-6256/150/1/19}

\bibitem[{{Law} {et~al.}(2016){Law}, {Cherinka}, {Yan}, {Andrews}, {Bershady},
  {Bizyaev}, {Blanc}, {Blanton}, {Bolton}, {Brownstein}, {Bundy}, {Chen},
  {Drory}, {D'Souza}, {Fu}, {Jones}, {Kauffmann}, {MacDonald}, {Masters},
  {Newman}, {Parejko}, {S{\'a}nchez-Gallego}, {S{\'a}nchez}, {Schlegel},
  {Thomas}, {Wake}, {Weijmans}, {Westfall}, \& {Zhang}}]{2016AJ....152...83L}
{Law}, D.~R., {Cherinka}, B., {Yan}, R., {et~al.} 2016, \aj, 152, 83,
  \dodoi{10.3847/0004-6256/152/4/83}

\bibitem[{{Le F{\`e}vre} {et~al.}(2005){Le F{\`e}vre}, {Vettolani}, {Garilli},
  {Tresse}, {Bottini}, {Le Brun}, {Maccagni}, {Picat}, {Scaramella},
  {Scodeggio}, {Zanichelli}, {Adami}, {Arnaboldi}, {Arnouts}, {Bardelli},
  {Bolzonella}, {Cappi}, {Charlot}, {Ciliegi}, {Contini}, {Foucaud},
  {Franzetti}, {Gavignaud}, {Guzzo}, {Ilbert}, {Iovino}, {McCracken}, {Marano},
  {Marinoni}, {Mathez}, {Mazure}, {Meneux}, {Merighi}, {Paltani}, {Pell{\`o}},
  {Pollo}, {Pozzetti}, {Radovich}, {Zamorani}, {Zucca}, {Bondi}, {Bongiorno},
  {Busarello}, {Lamareille}, {Mellier}, {Merluzzi}, {Ripepi}, \&
  {Rizzo}}]{2005A&A...439..845L}
{Le F{\`e}vre}, O., {Vettolani}, G., {Garilli}, B., {et~al.} 2005, \aap, 439,
  845, \dodoi{10.1051/0004-6361:20041960}

\bibitem[{{Lee} {et~al.}(2018){Lee}, {Hwang}, \& {Chung}}]{2018MNRAS.477.1567L}
{Lee}, J.~C., {Hwang}, H.~S., \& {Chung}, H. 2018, \mnras, 477, 1567,
  \dodoi{10.1093/mnras/sty729}

\bibitem[{{Lucy}(1974)}]{1974AJ.....79..745L}
{Lucy}, L.~B. 1974, \aj, 79, 745, \dodoi{10.1086/111605}

\bibitem[{{Lucy} \& {Walsh}(2003)}]{2003AJ....125.2266L}
{Lucy}, L.~B., \& {Walsh}, J.~R. 2003, \aj, 125, 2266, \dodoi{10.1086/368144}

\bibitem[{{Magain} {et~al.}(1998){Magain}, {Courbin}, \&
  {Sohy}}]{1998ApJ...494..472M}
{Magain}, P., {Courbin}, F., \& {Sohy}, S. 1998, \apj, 494, 472,
  \dodoi{10.1086/305187}

\bibitem[{{Martin} {et~al.}(2018){Martin}, {Kaviraj}, {Devriendt}, {Dubois}, \&
  {Pichon}}]{2018MNRAS.480.2266M}
{Martin}, G., {Kaviraj}, S., {Devriendt}, J.~E.~G., {Dubois}, Y., \& {Pichon},
  C. 2018, \mnras, 480, 2266, \dodoi{10.1093/mnras/sty1936}

\bibitem[{{Naab} {et~al.}(2014){Naab}, {Oser}, {Emsellem}, {Cappellari},
  {Krajnovi{\'c}}, {McDermid}, {Alatalo}, {Bayet}, {Blitz}, {Bois}, {Bournaud},
  {Bureau}, {Crocker}, {Davies}, {Davis}, {de Zeeuw}, {Duc}, {Hirschmann},
  {Johansson}, {Khochfar}, {Kuntschner}, {Morganti}, {Oosterloo}, {Sarzi},
  {Scott}, {Serra}, {van de Ven}, {Weijmans}, \& {Young}}]{2014MNRAS.444.3357N}
{Naab}, T., {Oser}, L., {Emsellem}, E., {et~al.} 2014, \mnras, 444, 3357,
  \dodoi{10.1093/mnras/stt1919}

\bibitem[{{Penoyre} {et~al.}(2017){Penoyre}, {Moster}, {Sijacki}, \&
  {Genel}}]{2017MNRAS.468.3883P}
{Penoyre}, Z., {Moster}, B.~P., {Sijacki}, D., \& {Genel}, S. 2017, \mnras,
  468, 3883, \dodoi{10.1093/mnras/stx762}

\bibitem[{Press {et~al.}(2007)Press, Teukolsky, Vetterling, \&
  Flannery}]{press2007numerical}
Press, W.~H., Teukolsky, S.~A., Vetterling, W.~T., \& Flannery, B.~P. 2007,
  Numerical recipes 3rd edition: The art of scientific computing (Cambridge
  university press)

\bibitem[{{Puech} {et~al.}(2008){Puech}, {Flores}, {Hammer}, {Yang}, {Neichel},
  {Lehnert}, {Chemin}, {Nesvadba}, {Epinat}, {Amram}, {Balkowski}, {Cesarsky},
  {Dannerbauer}, {di Serego Alighieri}, {Fuentes-Carrera}, {Guiderdoni},
  {Kembhavi}, {Liang}, {{\"O}stlin}, {Pozzetti}, {Ravikumar}, {Rawat},
  {Vergani}, {Vernet}, \& {Wozniak}}]{2008A&A...484..173P}
{Puech}, M., {Flores}, H., {Hammer}, F., {et~al.} 2008, \aap, 484, 173,
  \dodoi{10.1051/0004-6361:20079313}

\bibitem[{{Richardson}(1972)}]{1972JOSA...62...55R}
{Richardson}, W.~H. 1972, Journal of the Optical Society of America
  (1917-1983), 62, 55

\bibitem[{{Rodet} {et~al.}(2008){Rodet}, {Orieux}, {Giovannelli}, \&
  {Abergel}}]{2008ISTSP...2..802R}
{Rodet}, T., {Orieux}, F., {Giovannelli}, J.-F., \& {Abergel}, A. 2008, IEEE
  Journal of Selected Topics in Signal Processing, 2, 802,
  \dodoi{10.1109/JSTSP.2008.2006392}

\bibitem[{{S{\'a}nchez} {et~al.}(2016){S{\'a}nchez}, {Garc{\'\i}a-Benito},
  {Zibetti}, {Walcher}, {Husemann}, {Mendoza}, {Galbany}, {Falc{\'o}n-Barroso},
  {Mast}, {Aceituno}, {Aguerri}, {Alves}, {Amorim}, {Ascasibar},
  {Barrado-Navascues}, {Barrera-Ballesteros}, {Bekerait{\`e}},
  {Bland-Hawthorn}, {Cano D{\'\i}az}, {Cid Fernandes}, {Cavichia}, {Cortijo},
  {Dannerbauer}, {Demleitner}, {D{\'\i}az}, {Dettmar}, {de
  Lorenzo-C{\'a}ceres}, {del Olmo}, {Galazzi}, {Garc{\'\i}a-Lorenzo}, {Gil de
  Paz}, {Gonz{\'a}lez Delgado}, {Holmes}, {Igl{\'e}sias-P{\'a}ramo}, {Kehrig},
  {Kelz}, {Kennicutt}, {Kleemann}, {Lacerda}, {L{\'o}pez Fern{\'a}ndez},
  {L{\'o}pez S{\'a}nchez}, {Lyubenova}, {Marino}, {M{\'a}rquez},
  {Mendez-Abreu}, {Moll{\'a}}, {Monreal-Ibero}, {Ortega Minakata},
  {Torres-Papaqui}, {P{\'e}rez}, {Rosales-Ortega}, {Roth},
  {S{\'a}nchez-Bl{\'a}zquez}, {Schilling}, {Spekkens}, {Vale Asari}, {van den
  Bosch}, {van de Ven}, {Vilchez}, {Wild}, {Wisotzki}, {Y{\i}ld{\i}r{\i}m}, \&
  {Ziegler}}]{2016A&A...594A..36S}
{S{\'a}nchez}, S.~F., {Garc{\'\i}a-Benito}, R., {Zibetti}, S., {et~al.} 2016,
  \aap, 594, A36, \dodoi{10.1051/0004-6361/201628661}

\bibitem[{{S{\'a}nchez-Bl{\'a}zquez} {et~al.}(2006){S{\'a}nchez-Bl{\'a}zquez},
  {Peletier}, {Jim{\'e}nez-Vicente}, {Cardiel}, {Cenarro},
  {Falc{\'o}n-Barroso}, {Gorgas}, {Selam}, \& {Vazdekis}}]{2006MNRAS.371..703S}
{S{\'a}nchez-Bl{\'a}zquez}, P., {Peletier}, R.~F., {Jim{\'e}nez-Vicente}, J.,
  {et~al.} 2006, \mnras, 371, 703, \dodoi{10.1111/j.1365-2966.2006.10699.x}

\bibitem[{{Scott} {et~al.}(2018){Scott}, {van de Sande}, {Croom}, {Groves},
  {Owers}, {Poetrodjojo}, {D'Eugenio}, {Medling}, {Barat}, {Barone},
  {Bland-Hawthorn}, {Brough}, {Bryant}, {Cortese}, {Foster}, {Green}, {Oh},
  {Colless}, {Drinkwater}, {Driver}, {Goodwin}, {Gunawardhana}, {Federrath},
  {Harischandra}, {Jin}, {Lawrence}, {Lorente}, {Mannering}, {O'Toole},
  {Richards}, {Sanchez}, {Schaefer}, {Sealey}, {Sharp}, {Sweet}, {Taranu}, \&
  {Varidel}}]{2018MNRAS.481.2299S}
{Scott}, N., {van de Sande}, J., {Croom}, S.~M., {et~al.} 2018, \mnras, 481,
  2299, \dodoi{10.1093/mnras/sty2355}

\bibitem[{{Sharples} {et~al.}(2013){Sharples}, {Bender}, {Agudo Berbel},
  {Bezawada}, {Castillo}, {Cirasuolo}, {Davidson}, {Davies}, {Dubbeldam},
  {Fairley}, {Finger}, {F{\"o}rster Schreiber}, {Gonte}, {Hess}, {Jung},
  {Lewis}, {Lizon}, {Muschielok}, {Pasquini}, {Pirard}, {Popovic}, {Ramsay},
  {Rees}, {Richter}, {Riquelme}, {Rodrigues}, {Saviane}, {Schlichter},
  {Schmidtobreick}, {Segovia}, {Smette}, {Szeifert}, {van Kesteren}, {Wegner},
  \& {Wiezorrek}}]{2013Msngr.151...21S}
{Sharples}, R., {Bender}, R., {Agudo Berbel}, A., {et~al.} 2013, The Messenger,
  151, 21

\bibitem[{Shepp \& Vardi(1982)}]{4307558}
Shepp, L.~A., \& Vardi, Y. 1982, IEEE Transactions on Medical Imaging, 1, 113,
  \dodoi{10.1109/TMI.1982.4307558}

\bibitem[{Soulez {et~al.}(2011)Soulez, Bongard, Thiebaut, \& Bacon}]{6080865}
Soulez, F., Bongard, S., Thiebaut, E., \& Bacon, R. 2011, in 2011 3rd Workshop
  on Hyperspectral Image and Signal Processing: Evolution in Remote Sensing
  (WHISPERS), 1--4, \dodoi{10.1109/WHISPERS.2011.6080865}

\bibitem[{{van de Sande} {et~al.}(2017){van de Sande}, {Bland-Hawthorn},
  {Fogarty}, {Cortese}, {d'Eugenio}, {Croom}, {Scott}, {Allen}, {Brough},
  {Bryant}, {Cecil}, {Colless}, {Couch}, {Davies}, {Elahi}, {Foster},
  {Goldstein}, {Goodwin}, {Groves}, {Ho}, {Jeong}, {Jones}, {Konstantopoulos},
  {Lawrence}, {Leslie}, {L{\'o}pez-S{\'a}nchez}, {McDermid}, {McElroy},
  {Medling}, {Oh}, {Owers}, {Richards}, {Schaefer}, {Sharp}, {Sweet}, {Taranu},
  {Tonini}, {Walcher}, \& {Yi}}]{2017ApJ...835..104V}
{van de Sande}, J., {Bland-Hawthorn}, J., {Fogarty}, L. M.~R., {et~al.} 2017,
  \apj, 835, 104, \dodoi{10.3847/1538-4357/835/1/104}

\bibitem[{{van de Sande} {et~al.}(2019){van de Sande}, {Lagos}, {Welker},
  {Bland-Hawthorn}, {Schulze}, {Remus}, {Bah{\'e}}, {Brough}, {Bryant},
  {Cortese}, {Croom}, {Devriendt}, {Dubois}, {Goodwin}, {Konstantopoulos},
  {Lawrence}, {Medling}, {Pichon}, {Richards}, {Sanchez}, {Scott}, \&
  {Sweet}}]{2019MNRAS.484..869V}
{van de Sande}, J., {Lagos}, C. D.~P., {Welker}, C., {et~al.} 2019, \mnras,
  484, 869, \dodoi{10.1093/mnras/sty3506}

\bibitem[{{Vazdekis} {et~al.}(1996){Vazdekis}, {Casuso}, {Peletier}, \&
  {Beckman}}]{1996ApJS..106..307V}
{Vazdekis}, A., {Casuso}, E., {Peletier}, R.~F., \& {Beckman}, J.~E. 1996,
  \apjs, 106, 307, \dodoi{10.1086/192340}

\bibitem[{{Vazdekis} {et~al.}(2010){Vazdekis}, {S{\'a}nchez-Bl{\'a}zquez},
  {Falc{\'o}n-Barroso}, {Cenarro}, {Beasley}, {Cardiel}, {Gorgas}, \&
  {Peletier}}]{2010MNRAS.404.1639V}
{Vazdekis}, A., {S{\'a}nchez-Bl{\'a}zquez}, P., {Falc{\'o}n-Barroso}, J.,
  {et~al.} 2010, \mnras, 404, 1639, \dodoi{10.1111/j.1365-2966.2010.16407.x}

\bibitem[{{Villeneuve} \& {Carfantan}(2014)}]{2014ITIP...23.4322V}
{Villeneuve}, E., \& {Carfantan}, H. 2014, IEEE Transactions on Image
  Processing, 23, 4322, \dodoi{10.1109/TIP.2014.2343461}

\bibitem[{{Wake} {et~al.}(2017){Wake}, {Bundy}, {Diamond-Stanic}, {Yan},
  {Blanton}, {Bershady}, {S{\'a}nchez-Gallego}, {Drory}, {Jones}, {Kauffmann},
  {Law}, {Li}, {MacDonald}, {Masters}, {Thomas}, {Tinker}, {Weijmans}, \&
  {Brownstein}}]{2017AJ....154...86W}
{Wake}, D.~A., {Bundy}, K., {Diamond-Stanic}, A.~M., {et~al.} 2017, \aj, 154,
  86, \dodoi{10.3847/1538-3881/aa7ecc}

\bibitem[{{Westfall} {et~al.}(2019){Westfall}, {Cappellari}, {Bershady},
  {Bundy}, {Belfiore}, {Ji}, {Law}, {Schaefer}, {Shetty}, {Tremonti}, {Yan},
  {Andrews}, {Brownstein}, {Cherinka}, {Coccato}, {Drory}, {Maraston},
  {Parikh}, {S{\'a}nchez-Gallego}, {Thomas}, {Weijmans}, {Barrera-Ballesteros},
  {Du}, {Goddard}, {Li}, {Masters}, {Ibarra Medel}, {S{\'a}nchez}, {Yang},
  {Zheng}, \& {Zhou}}]{2019AJ....158..231W}
{Westfall}, K.~B., {Cappellari}, M., {Bershady}, M.~A., {et~al.} 2019, \aj,
  158, 231, \dodoi{10.3847/1538-3881/ab44a2}

\bibitem[{{Yan} {et~al.}(2016){Yan}, {Bundy}, {Law}, {Bershady}, {Andrews},
  {Cherinka}, {Diamond-Stanic}, {Drory}, {MacDonald}, {S{\'a}nchez-Gallego},
  {Thomas}, {Wake}, {Weijmans}, {Westfall}, {Zhang}, {Arag{\'o}n-Salamanca},
  {Belfiore}, {Bizyaev}, {Blanc}, {Blanton}, {Brownstein}, {Cappellari},
  {D'Souza}, {Emsellem}, {Fu}, {Gaulme}, {Graham}, {Goddard}, {Gunn},
  {Harding}, {Jones}, {Kinemuchi}, {Li}, {Li}, {Maiolino}, {Mao}, {Maraston},
  {Masters}, {Merrifield}, {Oravetz}, {Pan}, {Parejko}, {Sanchez}, {Schlegel},
  {Simmons}, {Thanjavur}, {Tinker}, {Tremonti}, {van den Bosch}, \&
  {Zheng}}]{2016AJ....152..197Y}
{Yan}, R., {Bundy}, K., {Law}, D.~R., {et~al.} 2016, \aj, 152, 197,
  \dodoi{10.3847/0004-6256/152/6/197}

\end{thebibliography}

\appendix
\renewcommand\thefigure{\thesection.\arabic{figure}}    
\setcounter{figure}{0}    

\section{Mock IFU Data Generation}\label{app:mockgen}
In order to quantitatively examine the change made by the deconvolution, the mock IFU data should be generated correctly as per the model galaxy parameters. Here we describe the generation process of each type of mock IFU data ($\free$, $\ndc$, $\dc$) in detail. Initially, an ideal IFU data ($\free$, without any seeing effect) is produced for each set of galaxy model parameters. 
Then the PSF-convolved IFU data ($\ndc$) is made by the convolution of a wavelength dependent PSF on the 2D image at each wavelength slice with addition of Gaussian random noise. $\dc$ IFU data is produced from $\ndc$ IFU data by applying the deconvolution method.

An arbitrary synthetic spectrum, composed by single-stellar populations with three different age (1 Gyr (15\%), 5 Gyr (60\%), 10 Gyr (25\%)) from 
 MILES stellar library \citep{2006MNRAS.371..703S, 2011A&A...532A..95F, 2010MNRAS.404.1639V} is chosen as a rest-frame model spectrum (using unimodal initial mass function \citep{1996ApJS..106..307V} and Padova+00 isochrones \citep{2000A&AS..141..371G}; $\Delta \lambda = 2.51$ \AA, $\lambda$ range from 3,540 to 7,410\AA)

For each set of galaxy model parameters (\autoref{subsec:mockifs}), we generate the $\free$ and $\ndc$ mock IFU data, according to the below steps.  

\begin{enumerate}
  \item  The spatial and spectral sampling size of the mock IFU data is determined. Following the sampling size and the data structure of MaNGA IFU data, we choose spatial sampling size as 0.5$\arcsec$, and in spectral direction we use a logarithmic wavelength sampling from $\rm log_{10}\lambda = $ 3.5589 to 4.0151, with total number of 4,563 wavelength bins.

  \item\label{step2dgen} 2D maps of flux (S\'{e}rsic profile), line-of-sight velocity with respect to the galaxy center (\autoref{eq:vr}), velocity dispersion (\autoref{eq:sig}), and S/N distribution (set by S/N at 1 effective radius) are identified as per a set of galaxy model parameters. Angular size of $\Reff$ is determined by two parameters, IFU field of view and IFU radial coverage in $\Reff$, by dividing half of the IFU field of view by the IFU radial coverage in $\Reff$. We assume that all three maps follow the identical geometry as defined by the inclination angle, position angle, $x_{cent}$ and $y_{cent}$. 
In case of S/N map, a relative S/N map is generated as per the Sersic profile then scaled to have a S/N at 1 $\Reff$ as per the galaxy model parameter.
  
  \item At each 2D pixel (Spaxel), a rest-frame spectrum is shifted and broadened in the spectral direction as per the respective line-of-sight velocity and the velocity dispersion value in the 2D map. First, the spectrum is convolved by a Gaussian function as per the velocity dispersion value. Second, the spectrum is redshifted by $z$ of a model galaxy. Third, a spectrum at each spaxel is blue- or red-shifted according to the corresponding line-of-sight velocity value with respect to the galaxy center.
  
  \item ($\ndc$ IFU data only) A 2D Gaussian PSF is convolved to the 2D image slice at each wavelength bin. The size of the Gaussian PSF FWHM is determined according to the model FWHM coefficient parameters. For example, in case of $c_0 = 2.6 \arcsec$ and $c_1 = -1.2 \times 10^{-5} \arcsec/$\AA, $\fndc$ at $g$-band effective wavelength (4770 \AA) is $2.52 \arcsec$ (which is median $g$-band PSF FWHM size of the MaNGA galaxies (\autoref{fig:fwhm})).
  
  \item A constant spectral resolution (2.9 \AA) is applied at each spaxel as a proxy of instrument resolution of the real IFU data. It is done by the convolution of Gaussian function (FWHM=1.45 \AA) to each spectrum. The FWHM size of applied Gaussian function is determined by quadratic difference between the instrument spectral resolution and the intrinsic resolution of the model synthetic spectrum (2.51 \AA).
  
  \item\label{gennoisespec} Noise spectrum at each spaxel is calculated. First, a relative S/N spectrum is calculated from the flux spectrum (assuming Poisson noise), and the relative S/N spectrum is scaled so that the S/N value of the median flux value would be matched to the S/N value in the 2D S/N map (generated in step \ref{step2dgen}). The noise spectrum is calculated by dividing the flux spectrum by the scaled S/N spectrum. The noise spectrum is not added to the flux spectrum at this stage.
  
  \item Hexagonal shape mask is applied to the IFU data to resemble the MaNGA-like IFU data.
  
  \item ($\ndc$ IFU data only) Gaussian random noise is applied to the IFU data using noise spectrum from the step \ref{gennoisespec}. At each spaxel, the noise spectrum is multiplied by the Gaussian random value (-3 to 3) and then added to the flux spectrum.
  
  \item Generated mock spectra are saved in 3D cube FITS format. Each noise spectrum is converted to an inverse variance spectrum (=1/noise$^2$) before saved. Flux, inverse variance, mask, wavelength, and spectral resolution data are saved in FITS extension similar to the actual MaNGA IFU data.
\end{enumerate}

\section{Validity of Rotation Curve Model Fitting}\label{sec:rcmodelvalid}
\setcounter{figure}{0} 

\begin{figure}
\centering
\includegraphics[width=\linewidth]{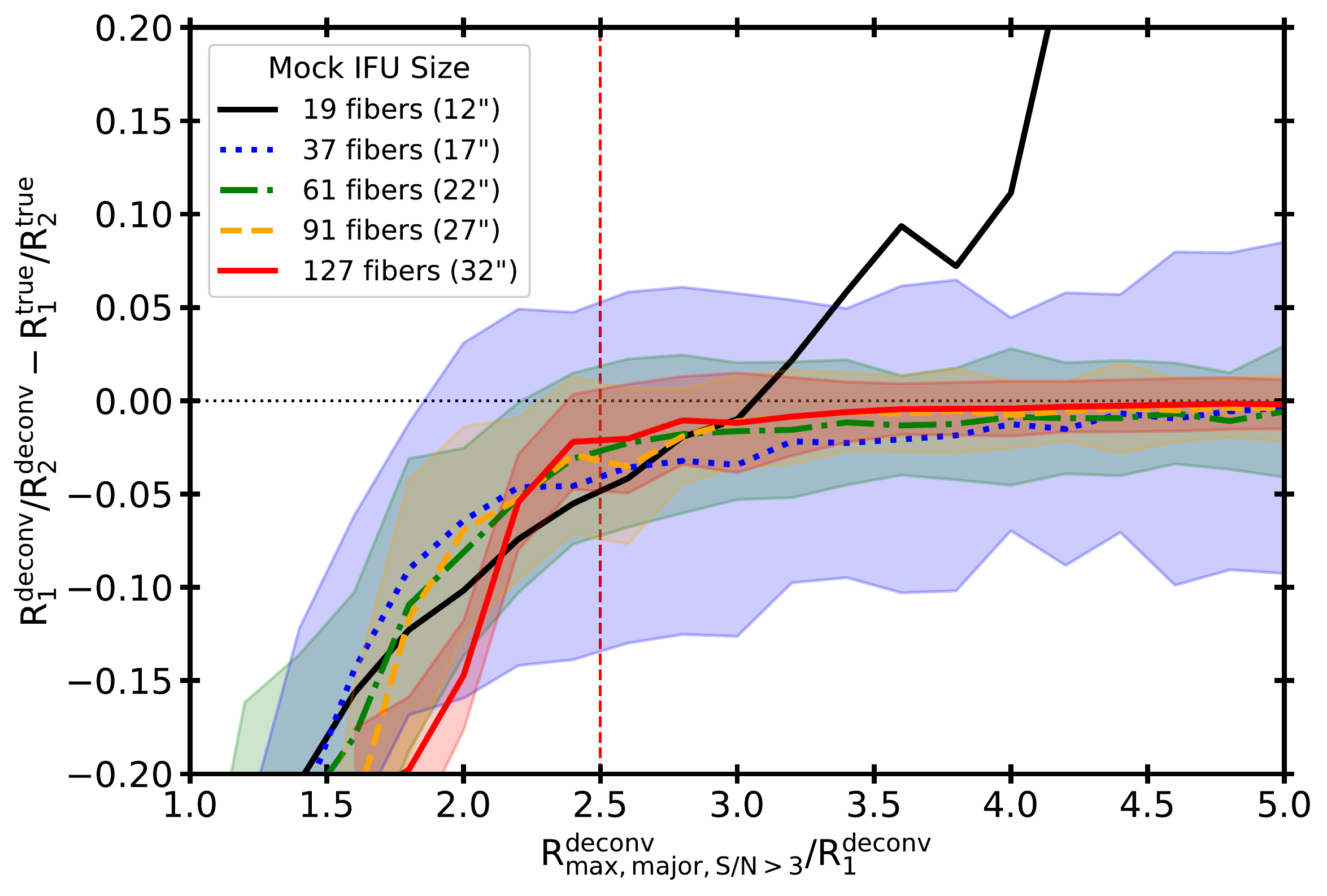}
\caption{The difference between the true and fitted RC outer radius slope ($\rm R_{1}/R_{2}$) from the deconvolved mock IFU cubes with respect to the R-ratio ($\rm R_{max, S/N > 3, major}/R_{1}$). Color represents the size of mock IFU. Each line represents the median of the normalized RC outer radius slope differences at each bin. Shaded region shows 1-$\sigma$ range of the differences. 1-$\sigma$ range of the 19-fiber size IFU is not shown because it is significantly higher than others. The vertical red-dashed line indicates our choice of R-ratio criteria, which is R-ratio $\rm> 2.5$.}
\label{fig:rratio_cut}
\end{figure}

In \autoref{sub:mockkm}, we notice that there are cases where the RC model fitting could give an unreliable result under certain circumstances. One case is where the 2D velocity data does not have sufficient radial coverage to constrain the RC slope at the outer radius. For example, in \autoref{fig:vrfunc}, the model would not be able to constrain the RC outer radius slope if the data covers only up to r = 4$\arcsec$. The other case is where the geometry of the fitted galaxy is close to both edge or face-on. In this case, both fitted RC velocity amplitude ($\vrot$) and inclination angle ($i$) becomes unreliable. We investigate this two cases in details using use the Group 3 mock IFU data (\autoref{subsec:mockifs}) which represents various combinations of galaxy parameters that mimic the actual IFU data. From the result, we estimate the criteria that the result of RC model fitting can be considered as valid.

First, we calculate R-ratio, a ratio between the maximum radial distance along the major axis to the $R_1$ parameter value (=$\rm R_{major, S/N > 3}/R_1$).
\replaced{We define the maximum radial distance as the farthest radial distance among the radial distances of the spaxels satisfying $\rm S/N_{pPXF} > 3$, the spaxels which are located within the $\rm \pm 5\arcdeg$ from the major axis, and the median $\rm S/N$ of the spectrum that are used for line-of-sight velocity distribution fitting (\ppxf$\;$routine).}{We define the maximum radial distance as the farthest radial distance among the radial distances of the spaxels which are located within the $\rm \pm 5\arcdeg$ from the major axis and satisfying $\rm S/N_{pPXF} > 3$. $\rm S/N_{pPXF}$ is the median $\rm S/N$ of the spectrum that are used for line-of-sight velocity distribution fitting (\ppxf$\;$routine).}
Then we plot the relation between the R-ratio and the fitting accuracy of the normalized RC outer slope value as in \autoref{fig:rratio_cut}. Although there are multiple factor which affecting R-ratio including the size of IFU, S/N cut, and S\'{e}rsic index, 
we divide the result depend on its IFU size only because the size causes the most significant systematic difference to the 1-$\sigma$ variation of the normalized RC outer slope accuracy. 
In \autoref{fig:rratio_cut}, the accuracy of the normalized RC outer radius slope ($\sratio$) shows strong dependence to the r-ratio. Regardless of the IFU size, the median difference between the true and the fitted $\sratio$ is large at low R-ratio, and the difference becomes smaller at higher R-ratio, except for the 19-fiber size IFU which is the smallest in its size. The 1-$\sigma$ variation of the $\sratio$ difference becomes smaller as the mock IFU size increases, mainly because the larger IFU have more spaxels so naturally it can better constraint the parameter values. We didn't plot the 1-$\sigma$ range of the 19-fiber size IFU because the range is larger than the height of the plot. From the result, we set a criteria of R-ratio $>$ 2.5, to determine whether the measured $\sratio$ can be considered as valid. Because the difference between the true and the fitted $\sratio$ becomes stable and small at R-ratio $>$ 2.5 compare to the difference at  R-ratio $<$ 2.5 
We also find that the $\sratio$ value measured from 19-fiber size IFU (Field of view equal to 12$\arcsec$) should not be used. This is because the measured $\sratio$ value remains inaccurate even at R-ratio $>$ 2.5.

\begin{figure}
\centering
\includegraphics[width=\linewidth]{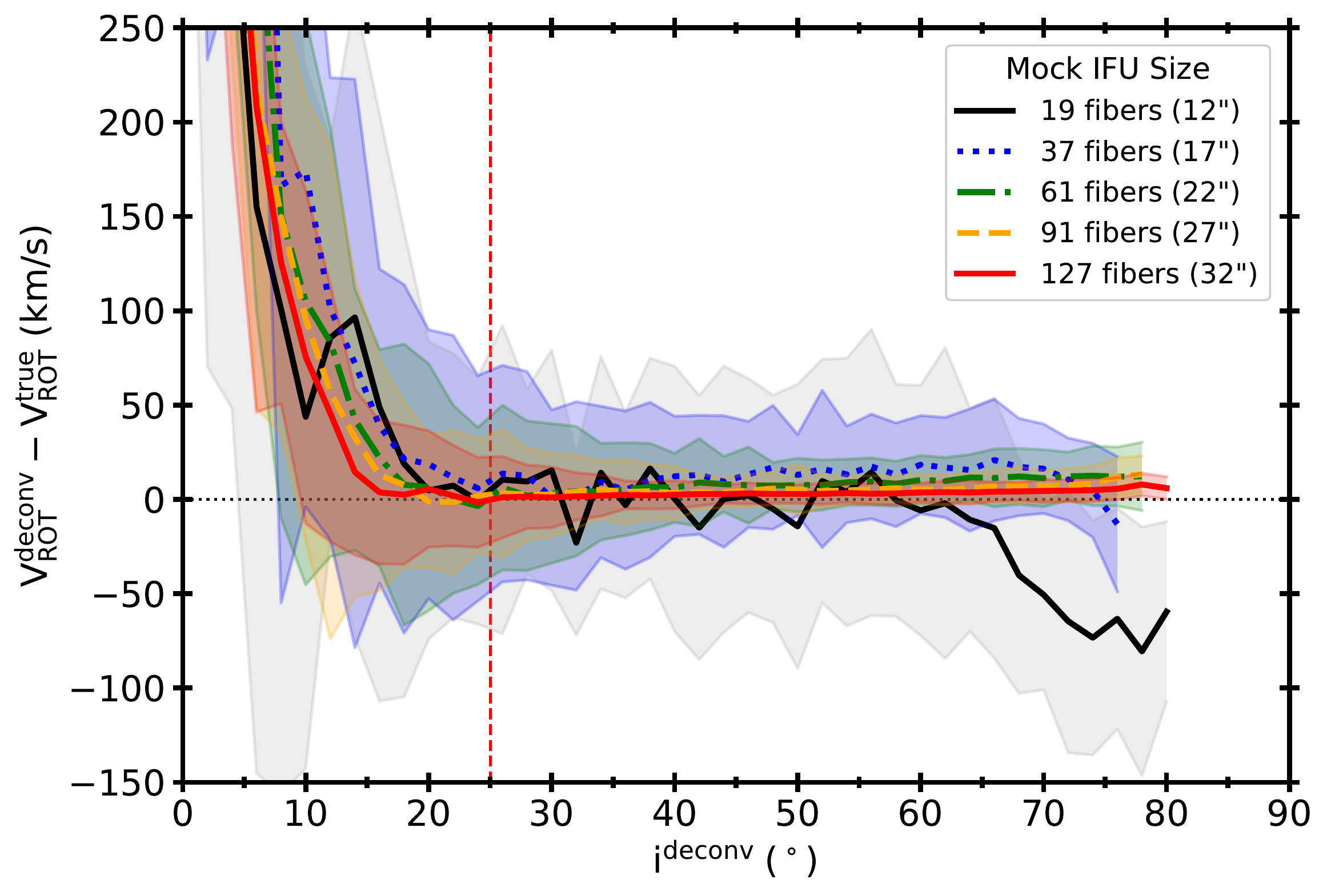}
\caption{The difference between the true and fitted RC velocity amplitude ($\rm V_{ROT}$) from the deconvolved mock IFU cubes with respect to the fitted inclination angle. Color represents the size of mock IFU. Solid line represents the median of the RC velocity amplitude differences at each bin. Shaded region shows 1-$\sigma$ range of the differences. The vertical red-dashed line indicates the lower end of our inclination angle criteria choice ($\rm 25 \arcdeg < i^{deconv} < 75 \arcdeg$).  }
\label{fig:incl_cut}
\end{figure}

We also analyze the relation between the R-ratio and the fitting accuracy of the galaxy kinematic inclination angle in \autoref{fig:incl_cut}. This result is plotted with the IFU data which are satisfying a criteria of R-ratio $>$ 2.5 only. The result shows that the fitted RC velocity amplitude ($\vrot$) is highly uncertain when the fitted inclination angle is low. In addition, 1-$\sigma$ of the median $\vrot$ also decreases when the fitted inclination value gets higher. Again we notice that the result of 19-fiber size IFU is not reliable due to its small number of spaxels. There is a slight hint that the fitted result may not be reliable at the higher inclination side, because the 1-$\sigma$ range is getting increased when the inclination angle is high. It can be explained by the low number of total spaxel elements when the inclination angle is high. From the shape of the curves and the 1-$\sigma$ range, we set a conservative criteria of $\rm 25 \arcdeg < i^{deconv} < 75 \arcdeg$ and consider the fitted $\vrot$ value as valid when the result meets those criteria.

\section{Deconvolution Effect Examples}\label{app:dcex}
\setcounter{figure}{0} 
In \autoref{fig:dcex}, we presented the example of the effects of PSF convolution and deconvolution to the IFU data. Here we show more examples from our mock IFU data to illustrate the effects of the deconvolution in various mock galaxy parameter space. Examples are taken from Group 1 mock IFU data.
\autoref{appfig:n1r1i55s10} and \autoref{appfig:n4r1i55s10} show the result of the deconvolution at low S/N ($\rm{S/N}$@1$\Reff$ = 10) when $\nsrc$ = 1 and 4, $R_1$ = 3$\arcsec$, $i$ = 55$\arcdeg$. 
\autoref{appfig:n1r1i55s20}, \autoref{appfig:n4r1i55s20}, \autoref{appfig:n1r1i70s20}, and \autoref{appfig:n4r1i70s20} show the result of the deconvolution at different combinations of $i$ (55, 70$\arcdeg$) and $\nsrc$ (1,4), when $\rm{S/N}$ at 1 $\Reff$ = 20 and $R_1$ = 3$\arcsec$. The forth columns of the all figures represent the significant difference between the maps from $\free$ and $\ndc$. The effect of PSF convolution is crucial in the distribution of Flux, velocity and the velocity dispersion. The fifth columns of the all figures show that the changes made by the PSF convolution are significantly restored by the deconvolution method. However, the restoration is not very effective at the outer radius where S/N becomes low, and also the flux distribution shows non-negligible artifacts around the center of galaxies, in particular when $\nsrc$ = 4. Nevertheless, the velocity and the velocity dispersion are generally well recovered even when the $\nsrc$ = 4.

\begin{figure*}[b]
\centering
\includegraphics[height=0.445\textheight]{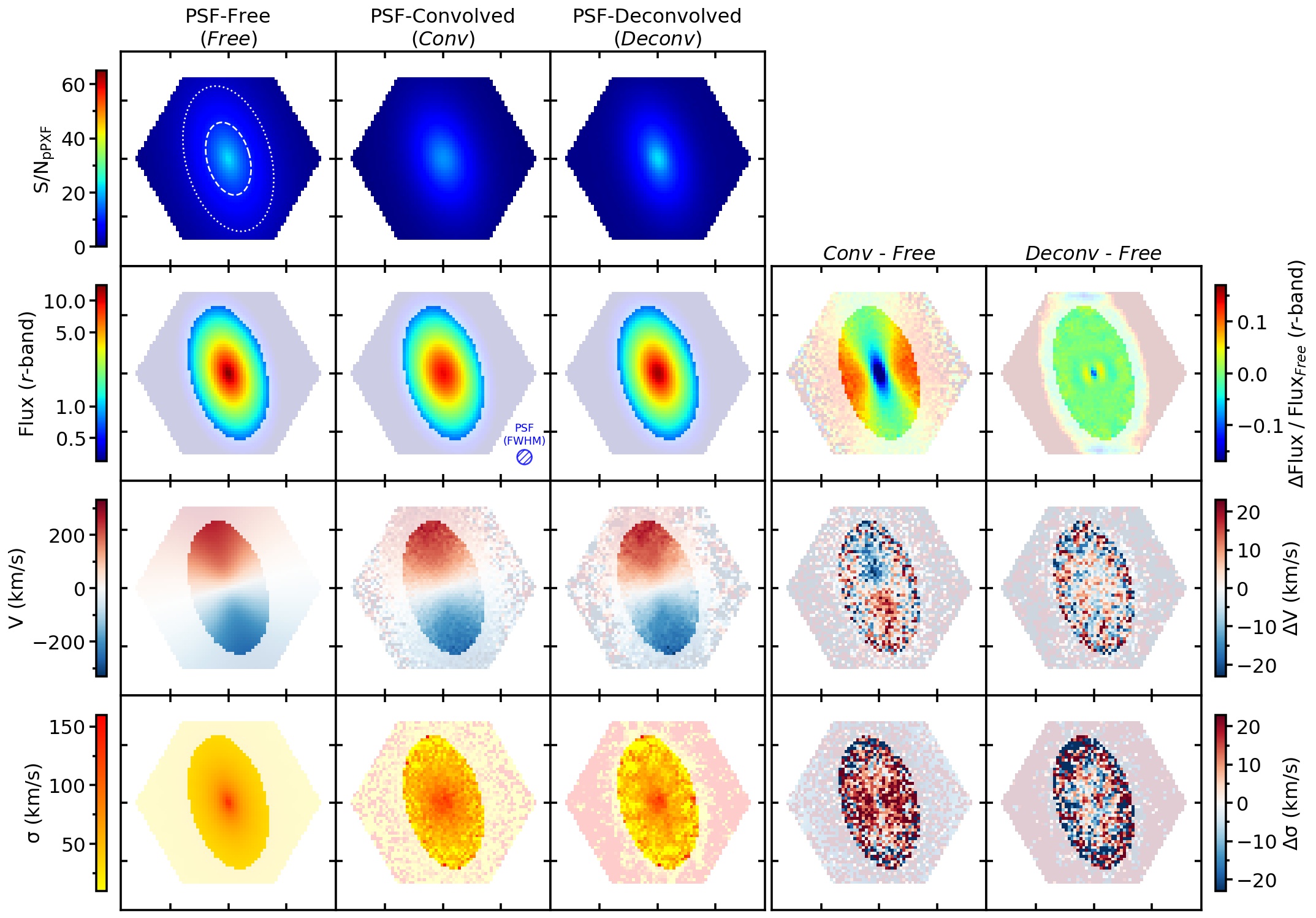}
\caption{Plots similar to \autoref{fig:dcex} but for 
$\nsrc$=1, $R_1=3\arcsec$, $1/R_2=0.05/\arcsec$, $i=55\arcdeg$, and $\rm{S/N}$@1$\Reff$ = 10} \label{appfig:n1r1i55s10}
\end{figure*}

\begin{figure*}
\centering
\includegraphics[height=0.445\textheight]{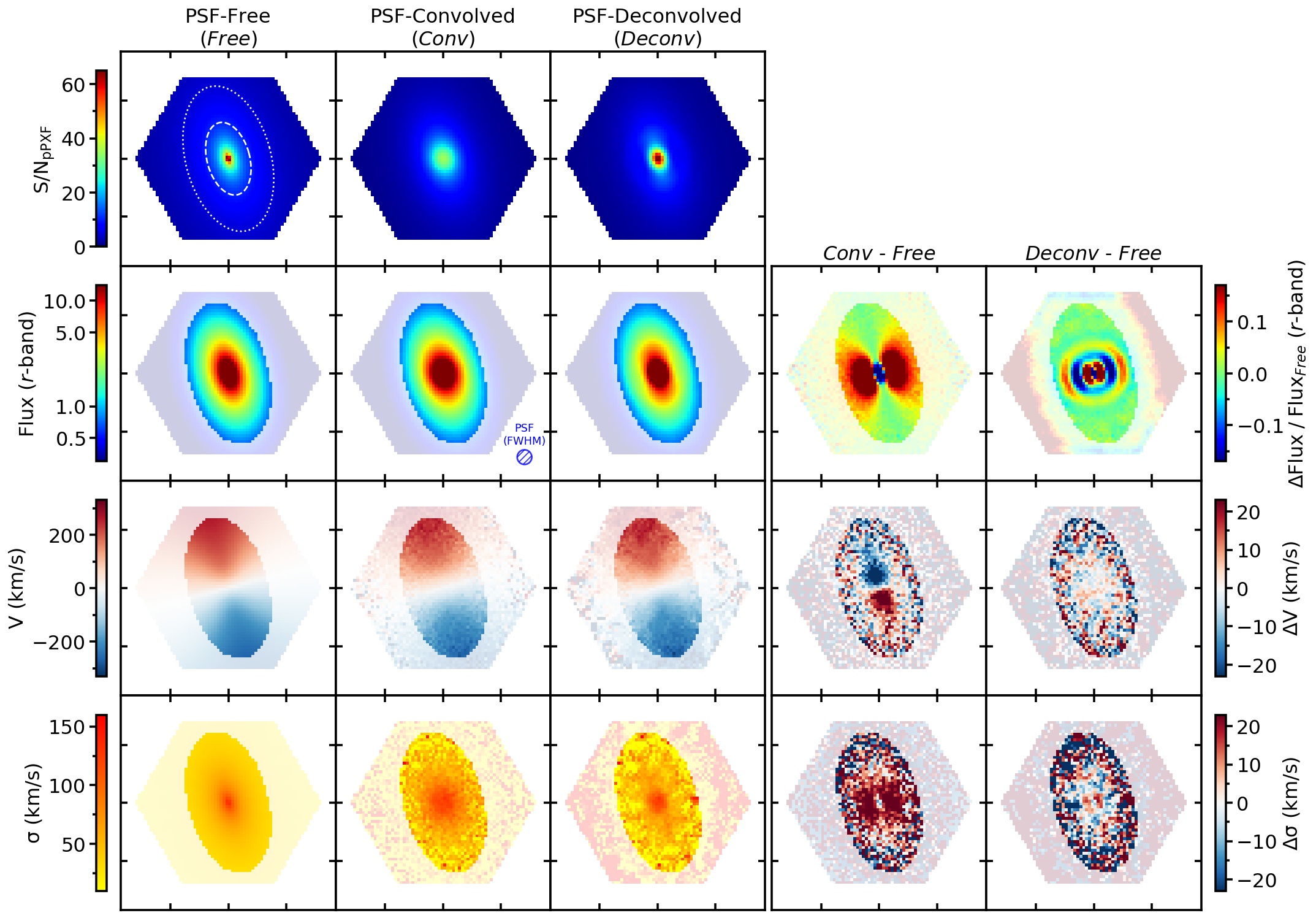}
\caption{Plots similar to \autoref{fig:dcex} but for 
$\rm \nsrc$=4, $R_1=3\arcsec$, $1/R_2=0.05/\arcsec$, $i=55\arcdeg$, and $\rm{S/N}$@1$\Reff$ = 10} \label{appfig:n4r1i55s10}
\end{figure*}

\begin{figure*}
\centering
\includegraphics[height=0.445\textheight]{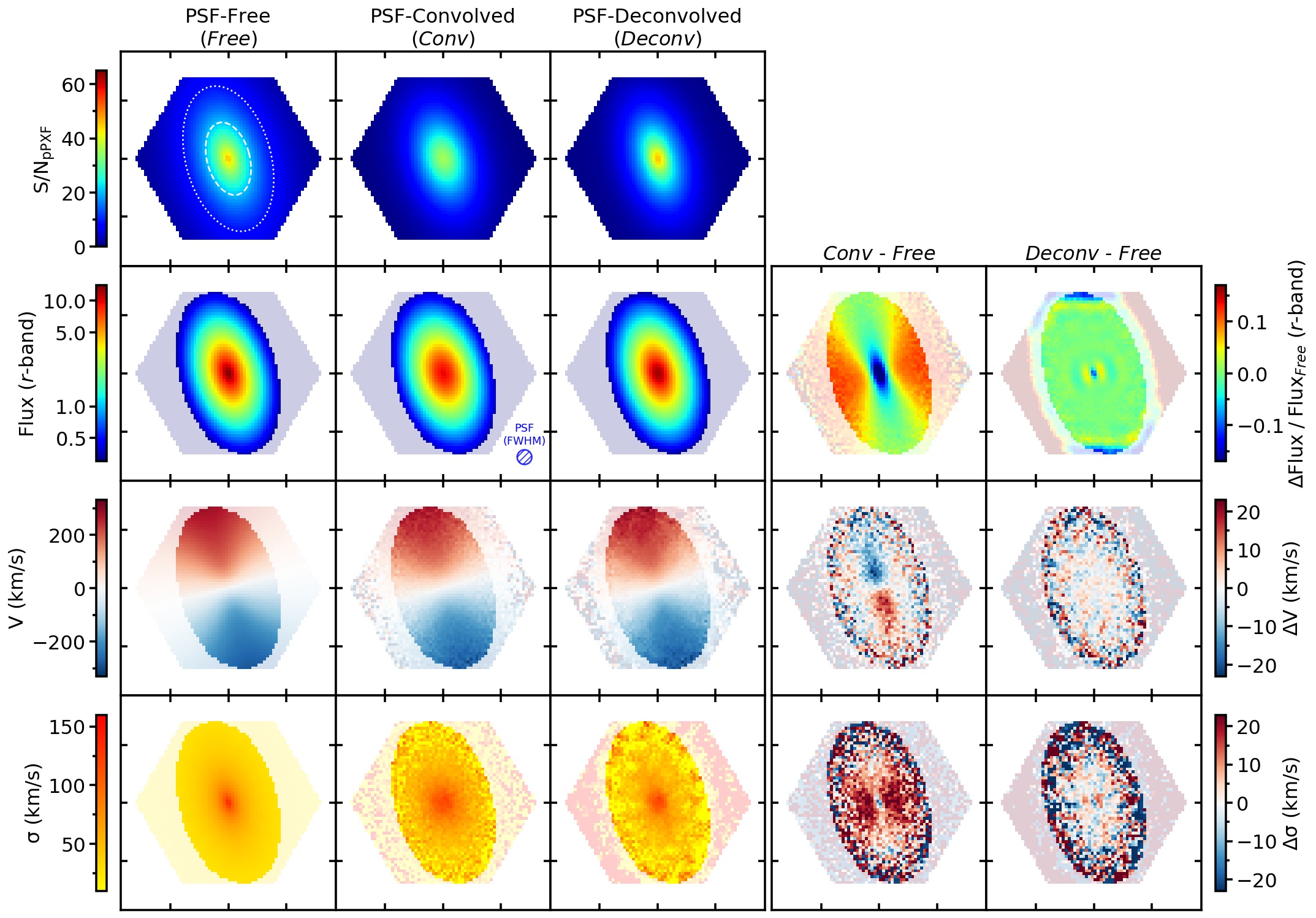}
\caption{Plots similar to \autoref{fig:dcex} but for 
$\rm \nsrc$=1, $R_1=3\arcsec$, $1/R_2=0.05/\arcsec$, $i=55\arcdeg$, and $\rm{S/N}$@1$\Reff$ = 20} \label{appfig:n1r1i55s20}
\end{figure*}

\begin{figure*}
\centering
\includegraphics[height=0.445\textheight]{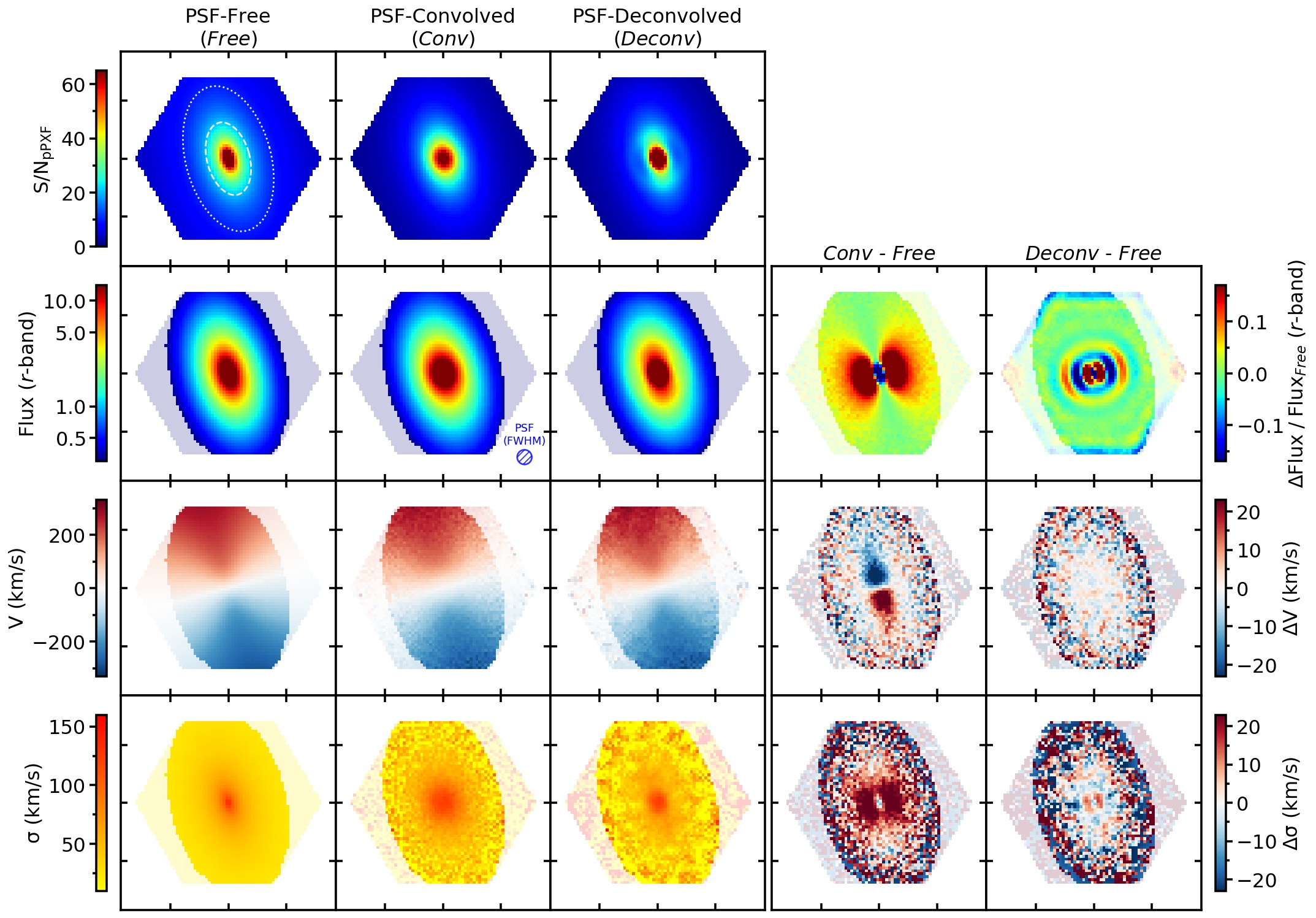}
\caption{Plots similar to \autoref{fig:dcex} but for 
$\rm \nsrc$=4, $R_1=3\arcsec$, $1/R_2=0.05/\arcsec$, $i=55\arcdeg$, and $\rm{S/N}$@1$\Reff$ = 20 }\label{appfig:n4r1i55s20}
\end{figure*}

\begin{figure*}
\centering
\includegraphics[height=0.445\textheight]{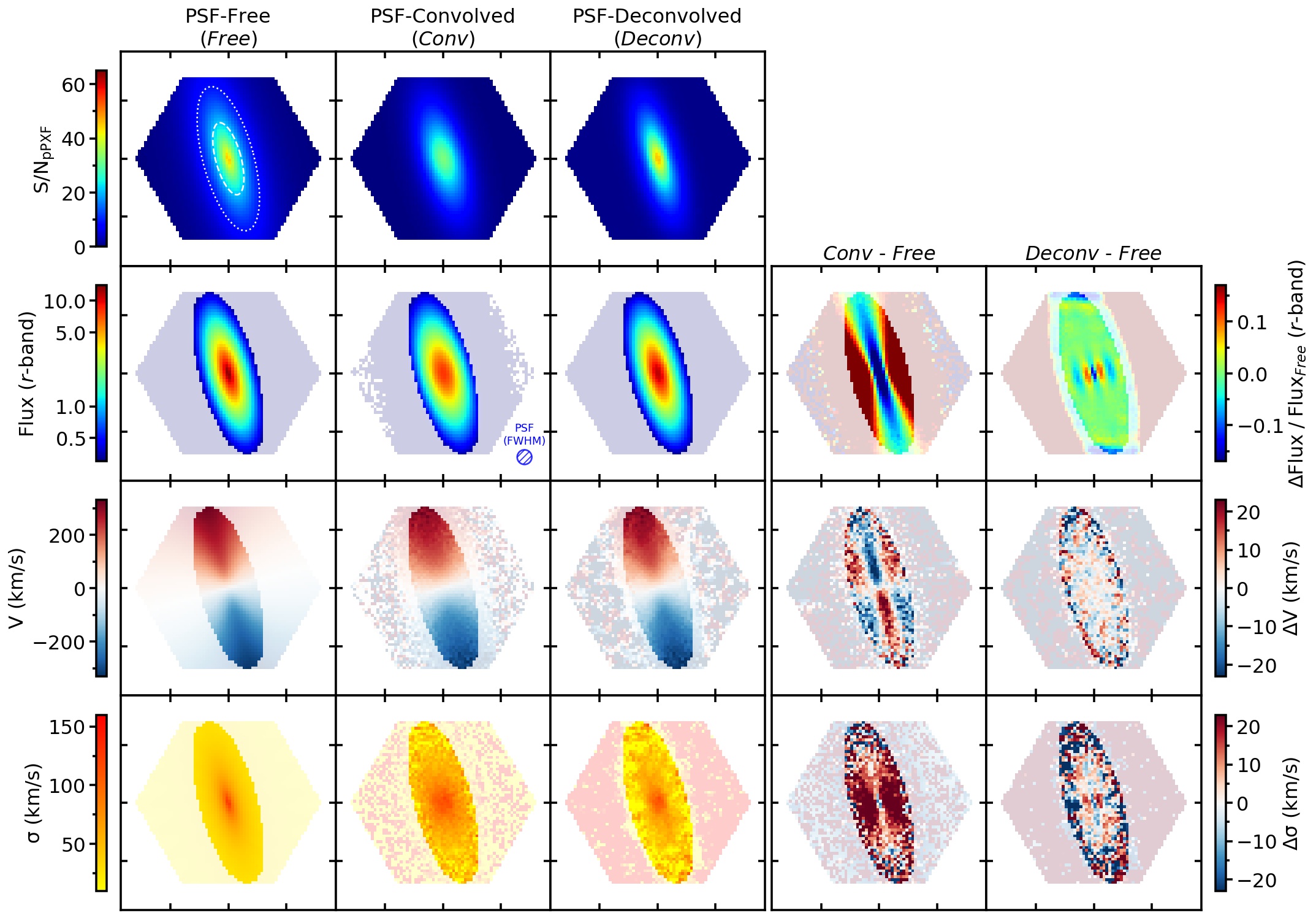}
\caption{Plots similar to \autoref{fig:dcex} but for 
$\rm \nsrc$=1, $R_1=3\arcsec$, $1/R_2=0.05/\arcsec$, $i=70\arcdeg$, and $\rm{S/N}$@1$\Reff$ = 20 }\label{appfig:n1r1i70s20}
\end{figure*}

\begin{figure*}
\centering
\includegraphics[height=0.445\textheight]{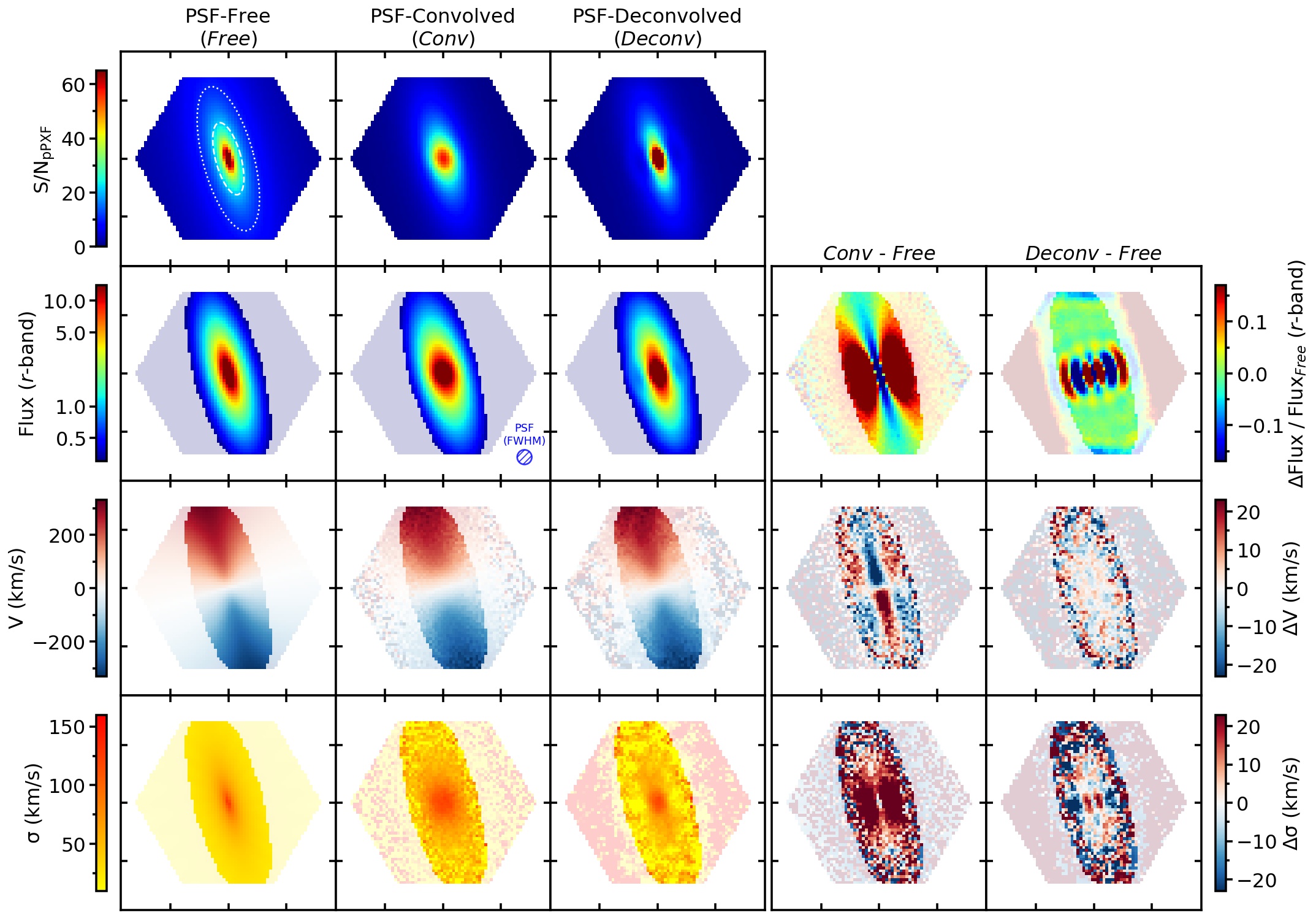}
\caption{Plots similar to \autoref{fig:dcex} but for 
$\rm \nsrc$=4, $R_1=3\arcsec$, $1/R_2=0.05/\arcsec$, $i=70\arcdeg$, and $\rm{S/N}$@1$\Reff$ = 20}\label{appfig:n4r1i70s20}
\end{figure*}

\section{Dependence on Deconvolution Parameters}
\setcounter{figure}{0} 
\subsection{Number of Deconvolution Iterations}\label{app:niter}
In section \ref{subsub:dcparams}, we described the relationship between $\niter$ and the restored model kinematic parameters (\autoref{fig:niter_diffincl}). Here we give similar plots with model galaxies of different parameters to provide more insight into the determination of $\niter$ to the readers. \autoref{niter:r1ih0}, \autoref{niter:r1ih1}, \autoref{niter:r0ih2}, \autoref{niter:r2ih2} are complementary figures to the \autoref{fig:niter_diffincl}. The figures show the relations between the fitted RC model parameter and the $\niter$ for different $R_1$ (2, 3, 4 ($\arcsec$)) and $1/R_2$ (-0.05, 0, 0.05 ($1/\arcsec$)). The result is consistent with \autoref{fig:niter_diffincl} thus the $\niter$ = 20 is an adequate choice for the deconvolution.

\subsection{Size of PSF FWHM}\label{app:fwhm}
Here we show the relation between $\rm{FWHM_{Deconv}}$ and the restored model kinematic parameters. We present plots similar to the \autoref{fig:fwhm_diffincl} but with different model galaxies as well as different $\rm{FWHM_{Conv}}$.

\autoref{fwhm:f1r1ih0}, \autoref{fwhm:f1r1ih1}, \autoref{fwhm:f1r0ih2},
and \autoref{fwhm:f1r2ih2} are complementary figures to the \autoref{fig:fwhm_diffincl}.
The figures show the relation between the fitted RC model parameter and the $\fdc$ with different $R_1$ (2, 3, 4 ($\arcsec$)) and $1/R_2$ (-0.05, 0, 0.05 ($1/\arcsec$)). The result is consistent with \autoref{fig:fwhm_diffincl} thus the result of the deconvolution is consistent when $|\fdc-\fndc|$ is smaller then the $\fpsf$ measurement error (0.2$\arcsec$). 

\autoref{fwhm:f0r1ih2} and \autoref{fwhm:f2r1ih2} show the result of the deconvolution with different $\fndc$ (2.3, 2.9 ($\arcsec$)). Again, the result of the deconvolution is consistent when $|\fdc-\fndc|$ is small.

\begin{figure*}
\centering
\includegraphics[width=\textwidth]{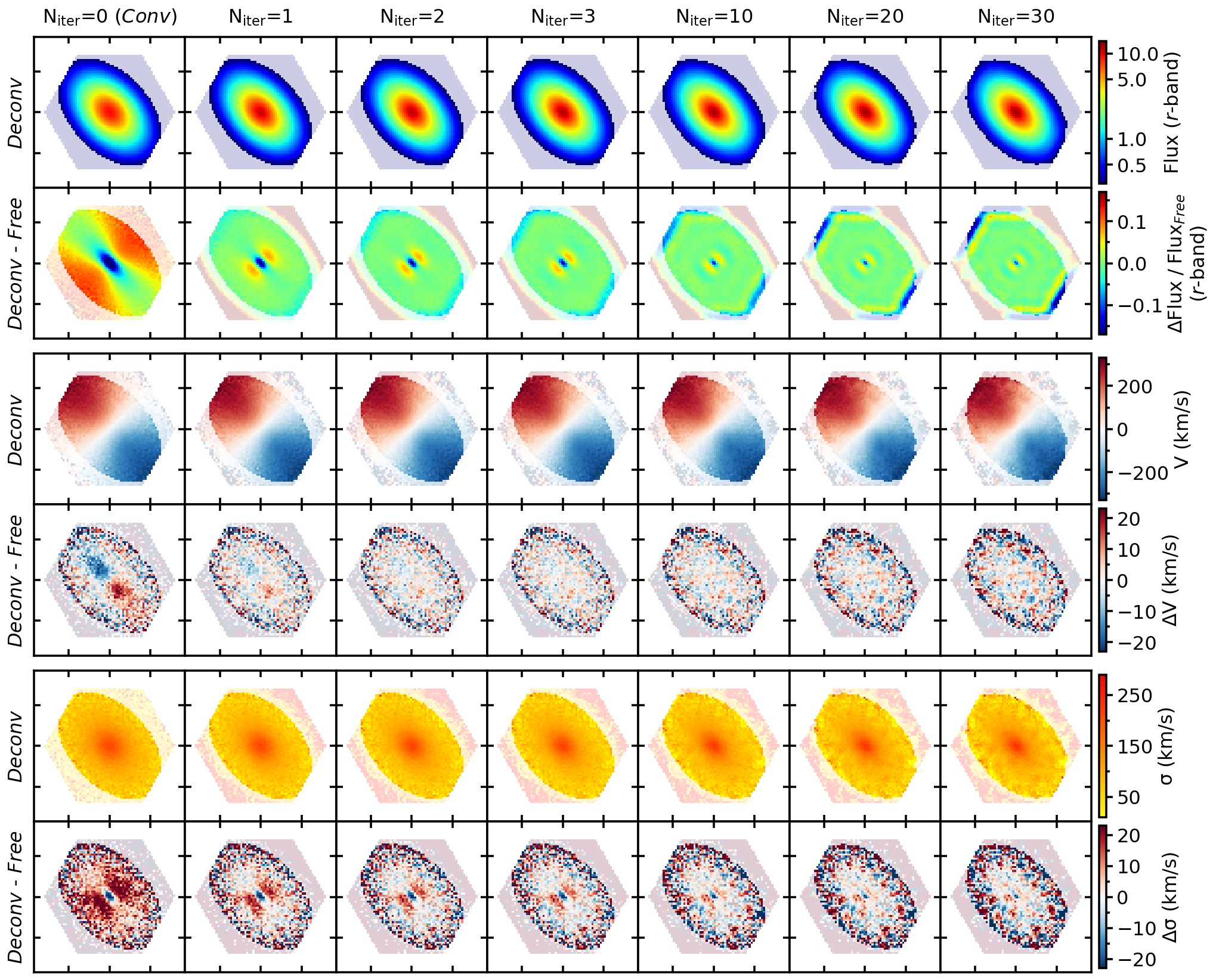}
\caption{
Effect of the number of deconvolution iterations ($\niter$) to the 2D maps of $r$-band flux, line of sight velocity (V) and velocity dispersion ($\sigma$). The first and second rows show 2D flux map from the deconvolved mock IFU data and the difference between the 2D flux map from the deconvolved mock IFU data and the PSF-free mock IFU data at selected $\niter$ = 0, 1, 2, 3, 10, 20, 30, respectively. The third and fourth rows show the same for the 2D line of sight velocity map, and the fifth and sixth rows show the same for the 2D velocity dispersion map. The 2D maps for $\niter$=0 and $\niter$=20 are the same as in the \autoref{fig:dcex}. 
}\label{niter:img}
\end{figure*}

\begin{figure*}
\centering
\includegraphics[height=0.445\textheight]{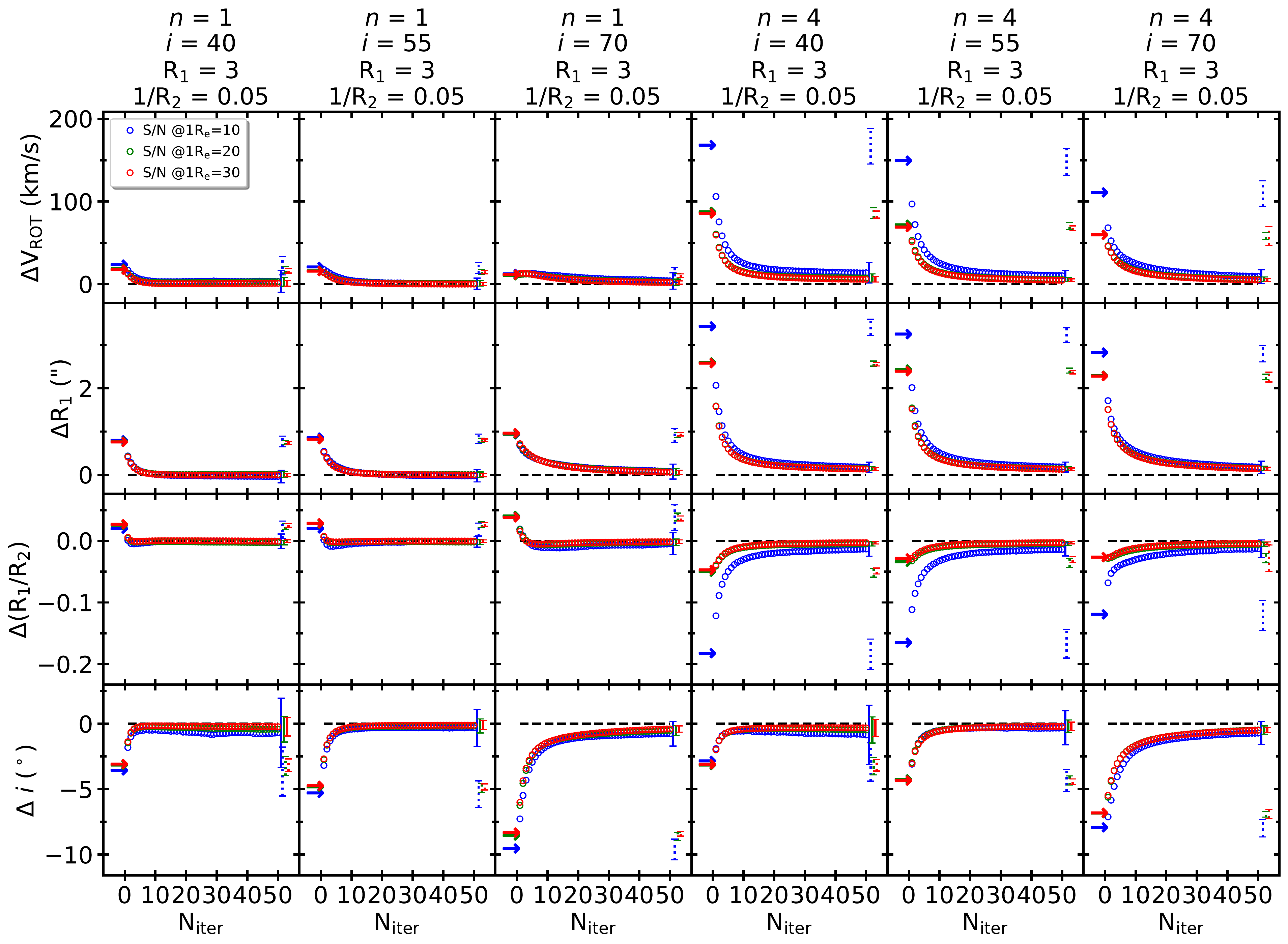}
\caption{Relations between the RC model parameters and $\niter$ for
$\rm R_1=3, 1/R_2=-0.05$}\label{niter:r1ih0}
\end{figure*}

\begin{figure*}
\centering
\includegraphics[height=0.445\textheight]{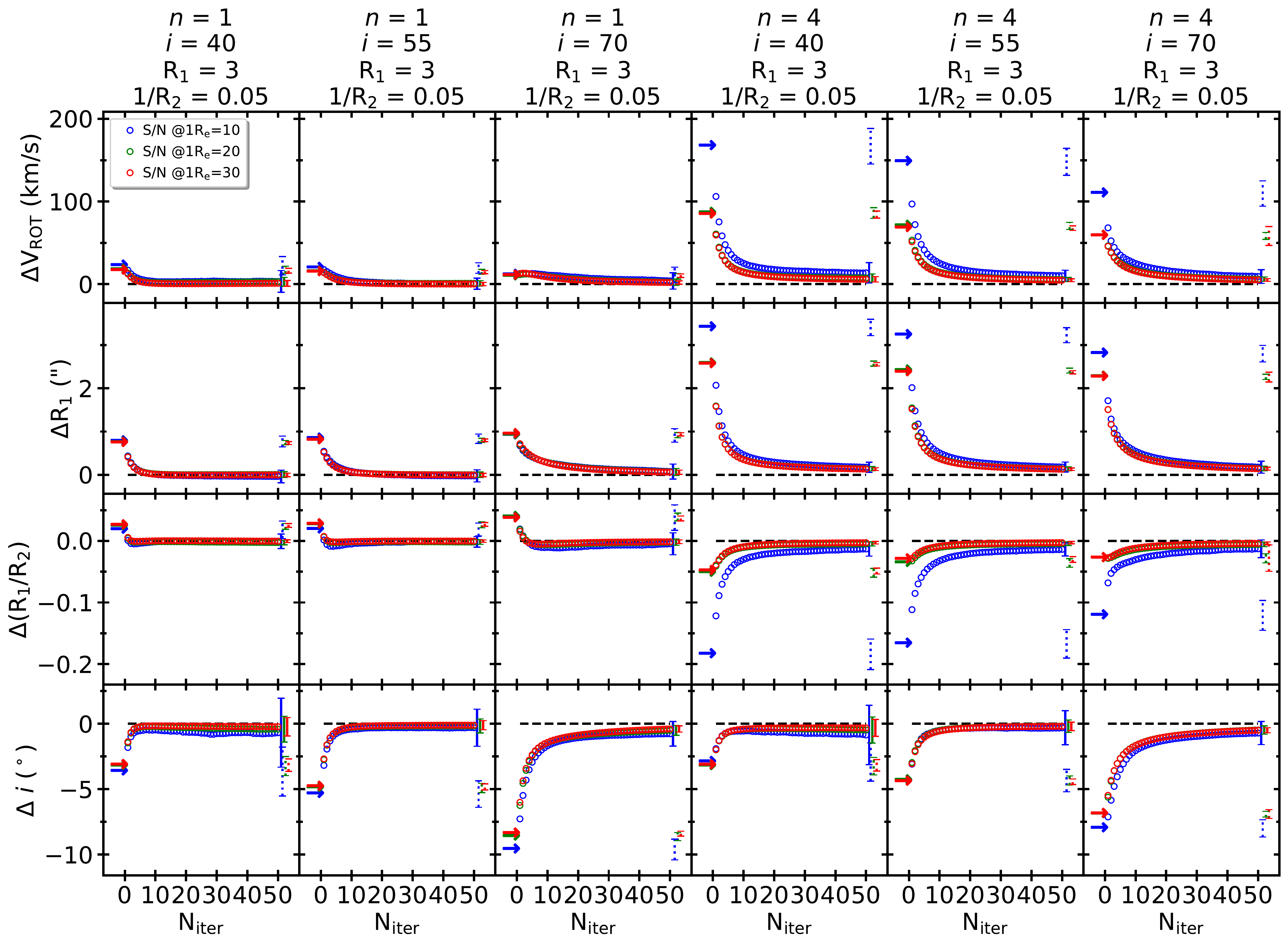}
\caption{Plots similar to \autoref{niter:r1ih0} but for $\rm R_1=3, 1/R_2=0$}\label{niter:r1ih1}
\end{figure*}

\begin{figure*}
\centering
\includegraphics[height=0.445\textheight]{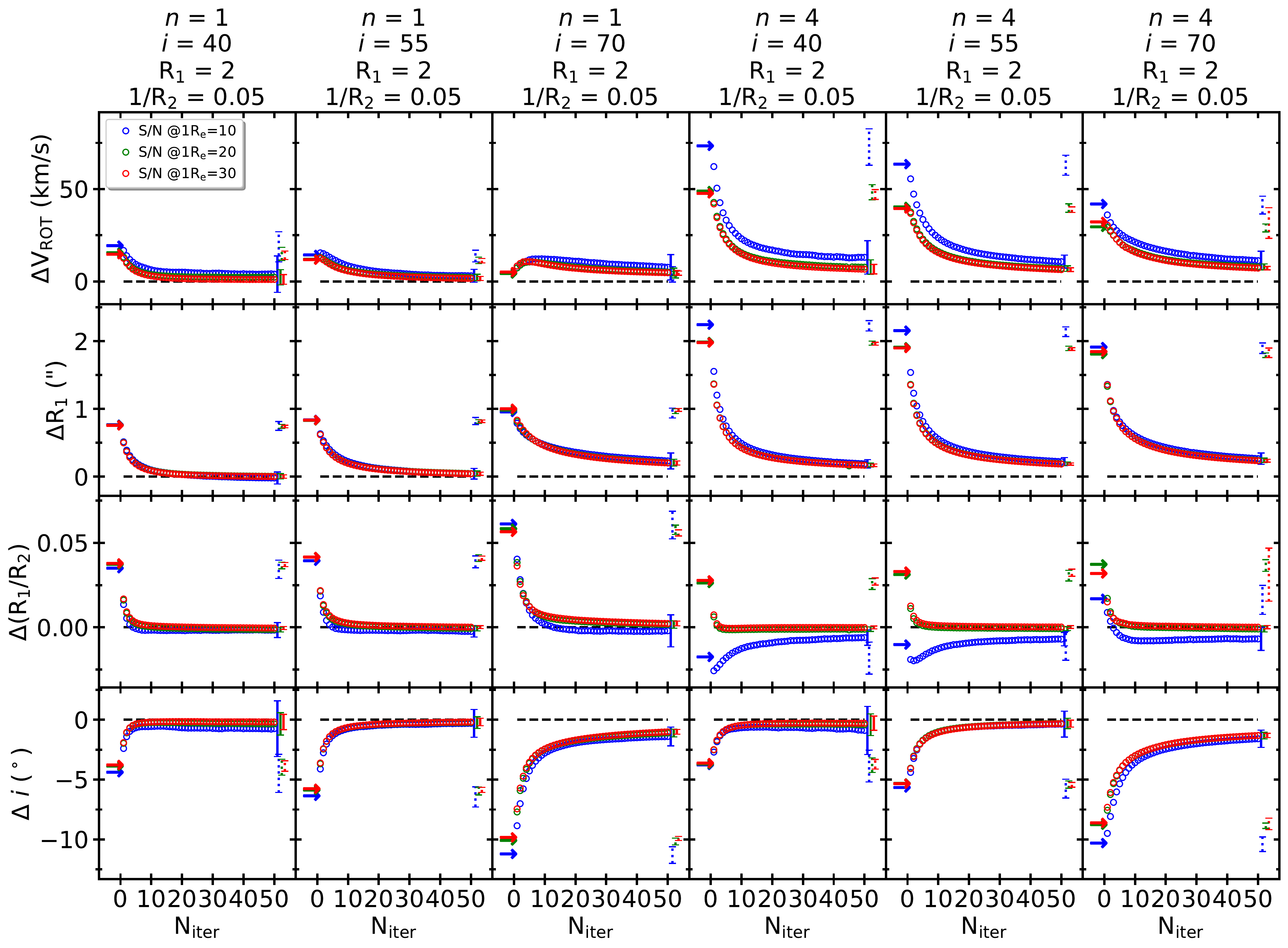}
\caption{Plots similar to \autoref{niter:r1ih0} but for $\rm R_1=2, 1/R_2=0.05$}\label{niter:r0ih2}
\end{figure*}

\begin{figure*}
\centering
\includegraphics[height=0.445\textheight]{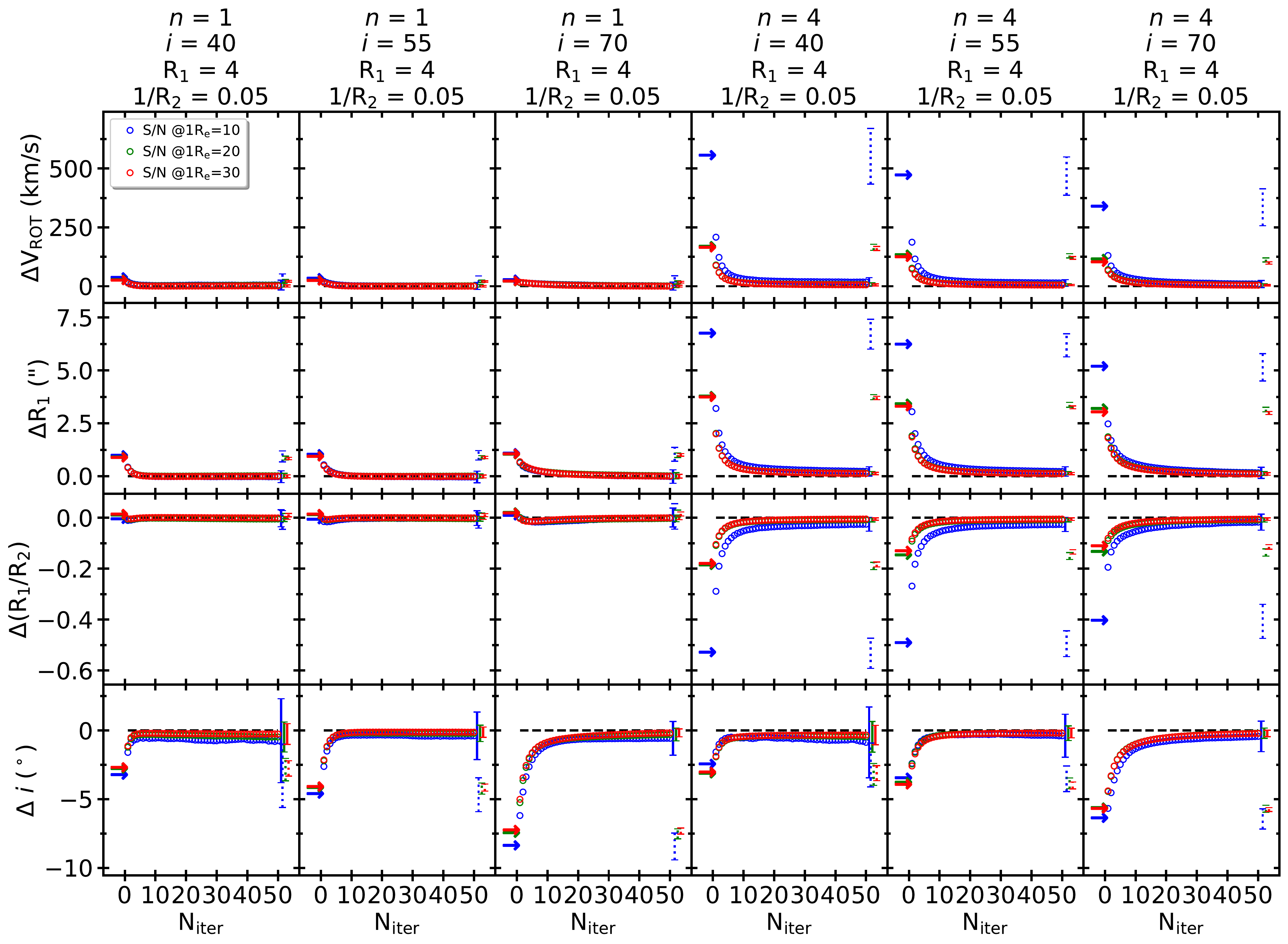}
\caption{Plots similar to \autoref{niter:r1ih0} but for $\rm R_1=4, 1/R_2=0.05$}\label{niter:r2ih2}
\end{figure*}

\begin{figure*}
\centering
\includegraphics[height=0.445\textheight]{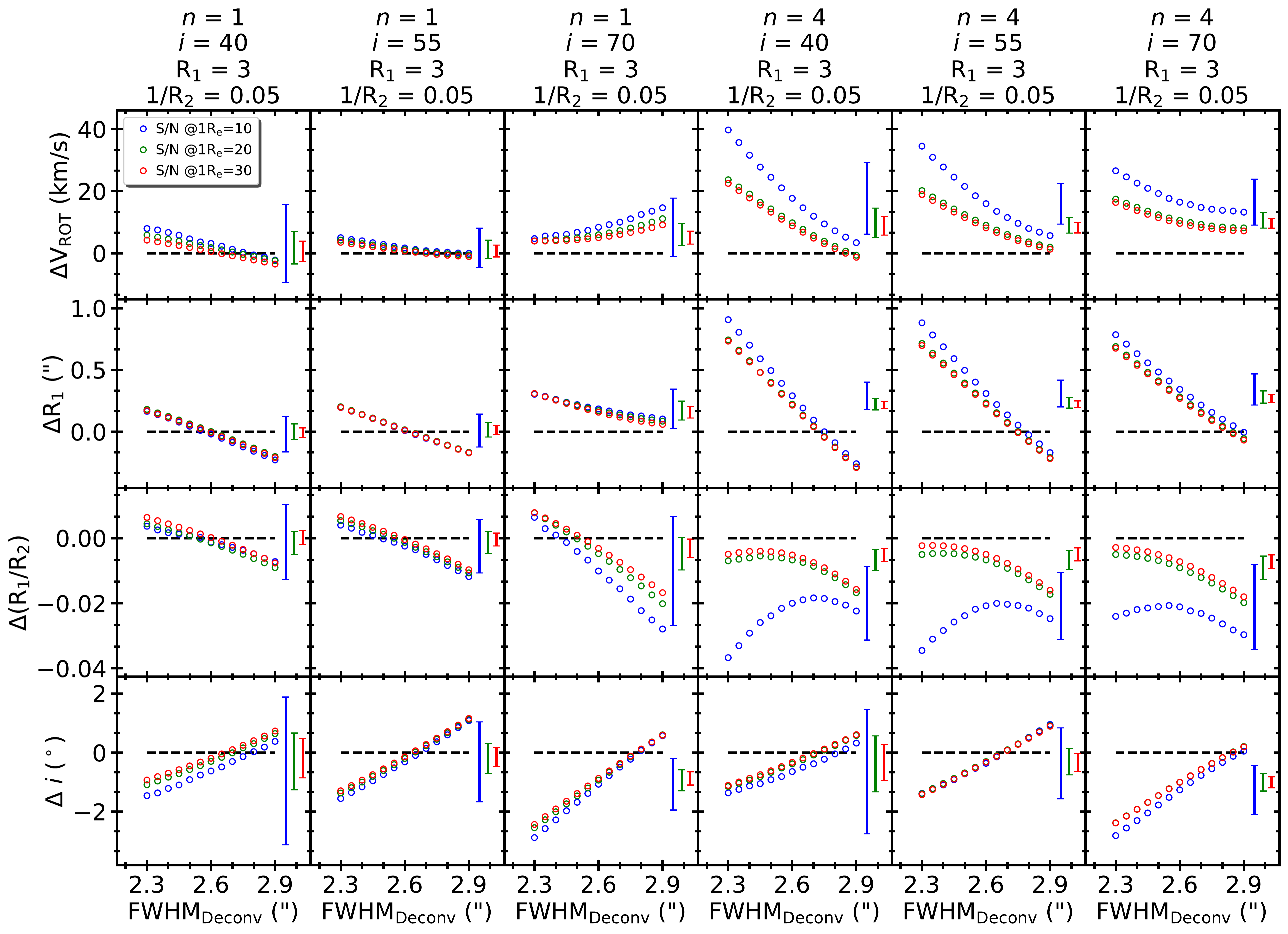}
\caption{Relations between the RC model parameters and $\fdc$ for
$\rm R_1=3, 1/R_2=-0.05, FWHM c_0=2.6$}\label{fwhm:f1r1ih0}
\end{figure*}

\begin{figure*}
\centering
\includegraphics[height=0.445\textheight]{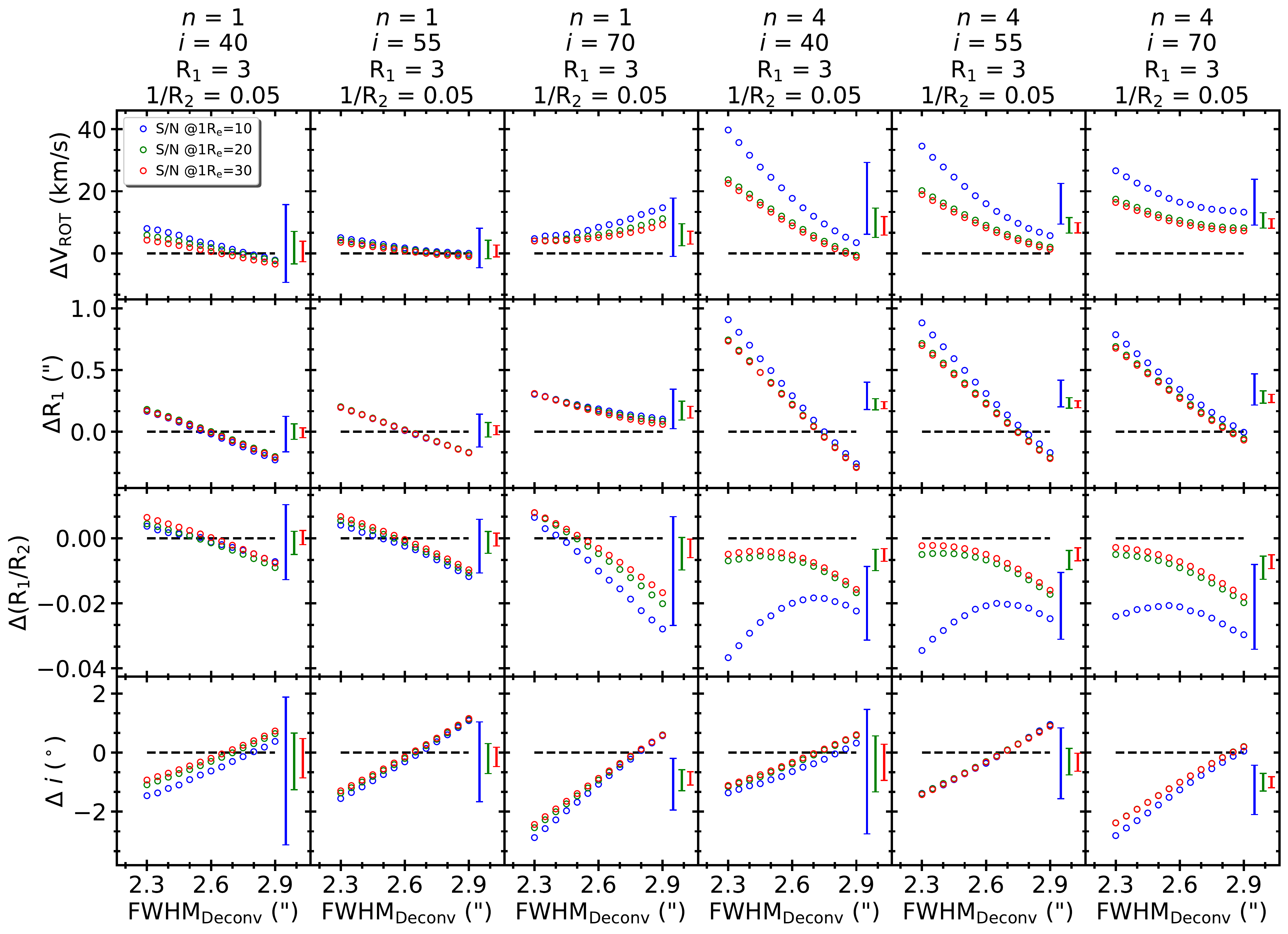}
\caption{Plots similar to \autoref{fwhm:f1r1ih0} but for
$\rm R_1=3, 1/R_2=0.05, FWHM c_0=2.6$}\label{fwhm:f1r1ih1}
\end{figure*}

\begin{figure*}
\centering
\includegraphics[height=0.445\textheight]{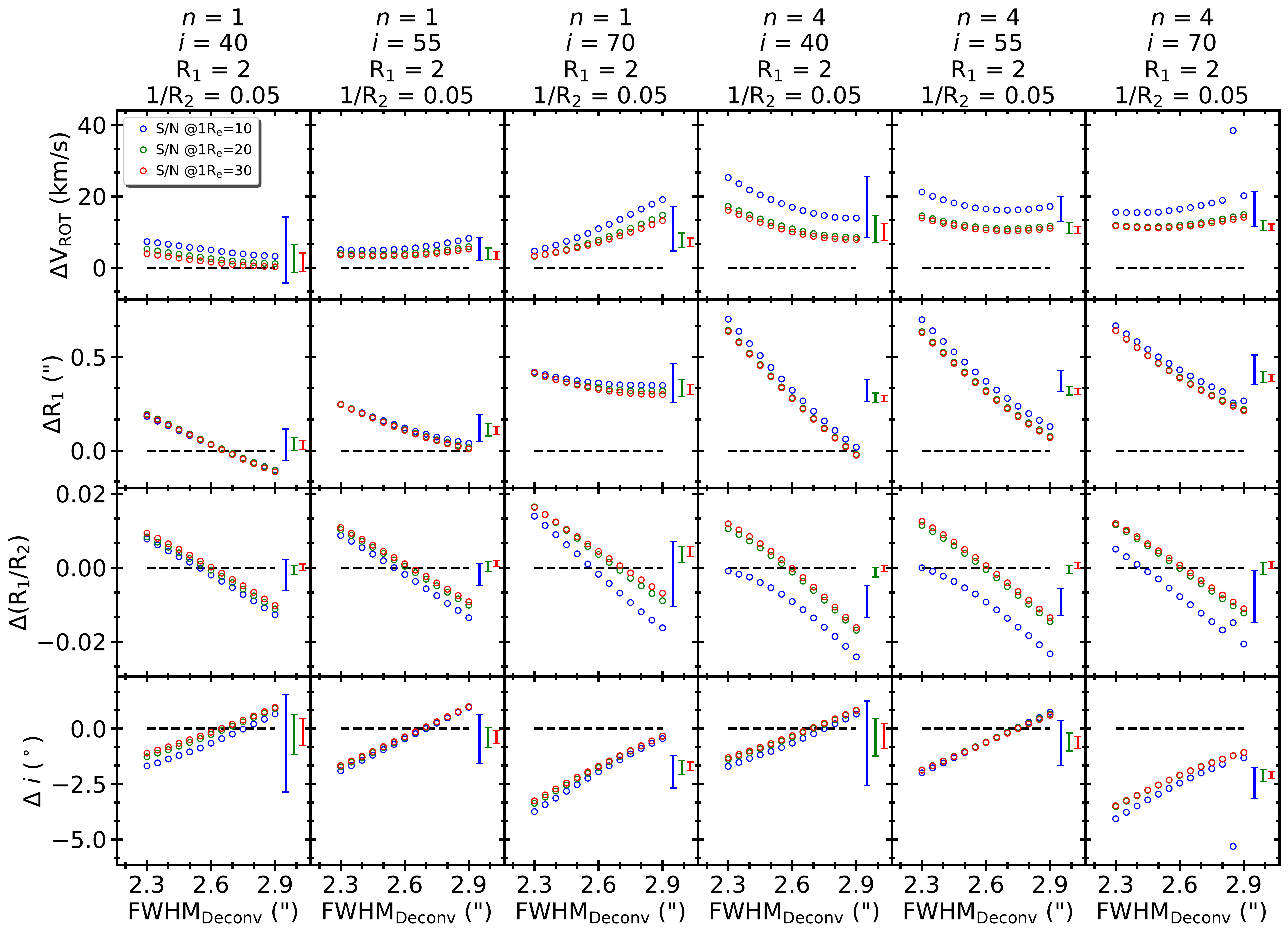}
\caption{Plots similar to \autoref{fwhm:f1r1ih0} but for
$\rm R_1=2, 1/R_2=0.05, FWHM c_0=2.6$}\label{fwhm:f1r0ih2}
\end{figure*}

\begin{figure*}
\centering
\includegraphics[height=0.445\textheight]{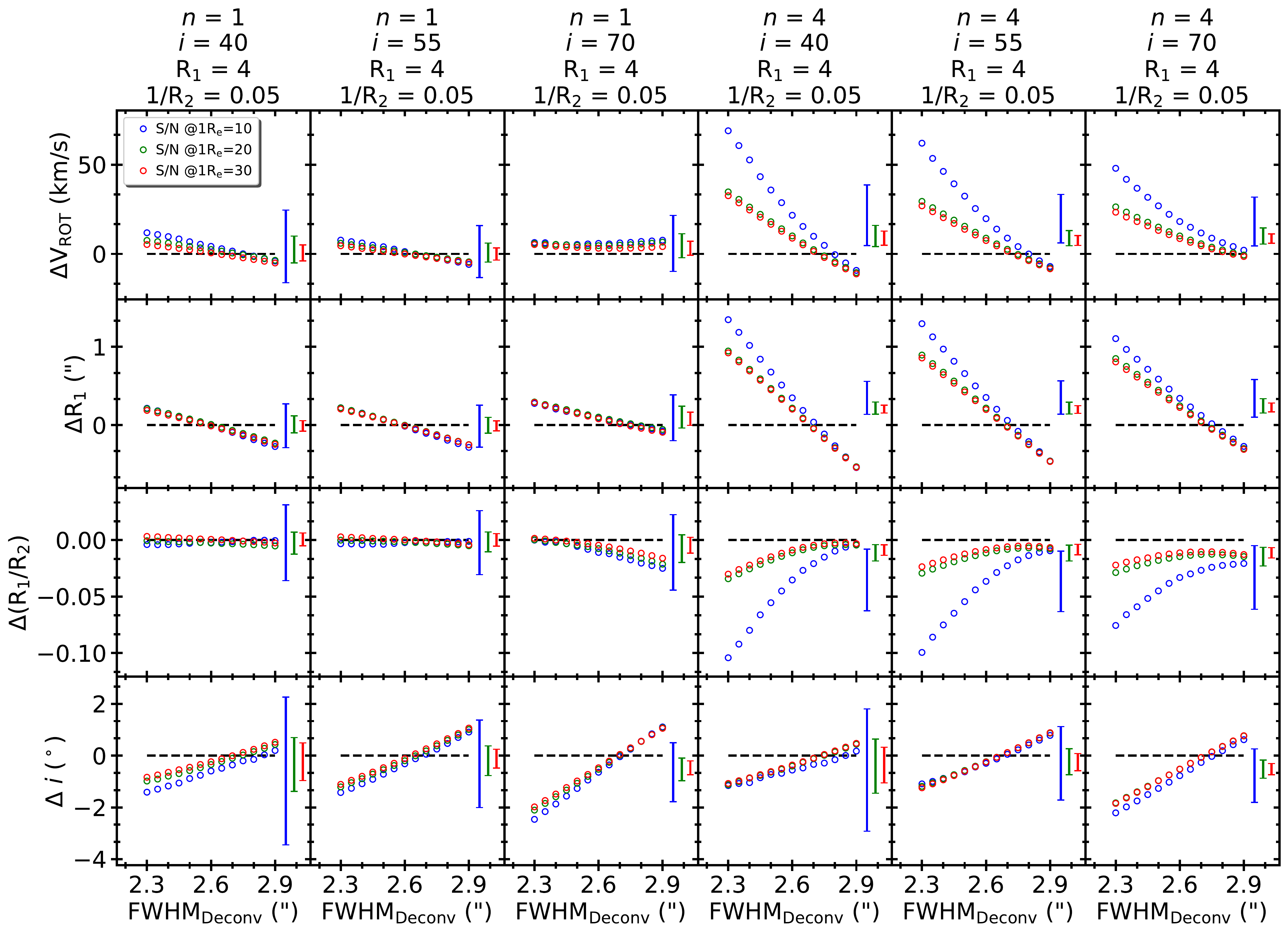}
\caption{Plots similar to \autoref{fwhm:f1r1ih0} but for
$\rm R_1=4, 1/R_2=0.05, FWHM c_0=2.6$}\label{fwhm:f1r2ih2}
\end{figure*}

\begin{figure*}
\centering
\includegraphics[height=0.445\textheight]{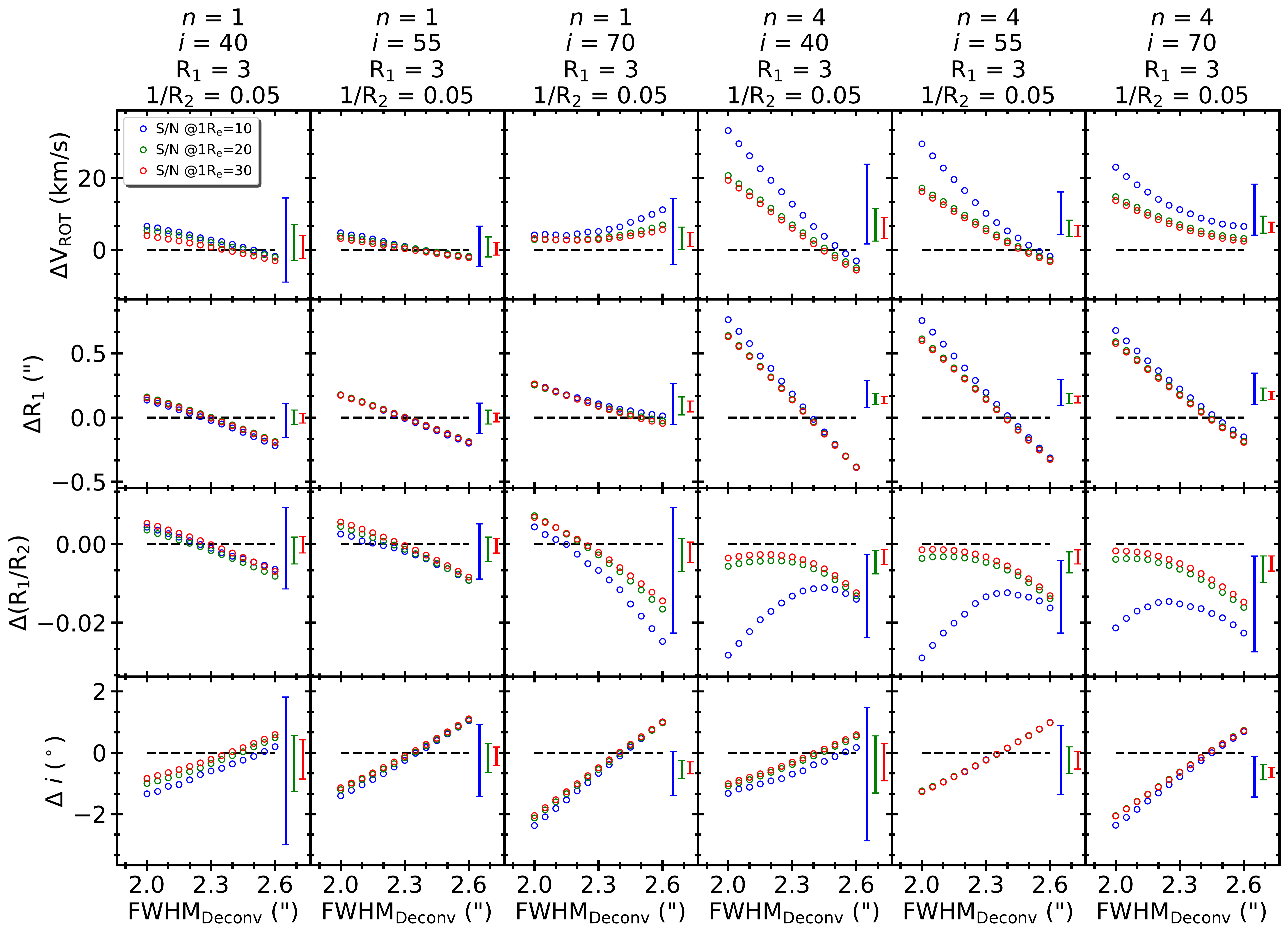}
\caption{Plots similar to \autoref{fwhm:f1r1ih0} but for
$\rm R_1=3, 1/R_2=0.05, FWHM c_0=2.3$}\label{fwhm:f0r1ih2}
\end{figure*}

\begin{figure*}
\centering
\includegraphics[height=0.445\textheight]{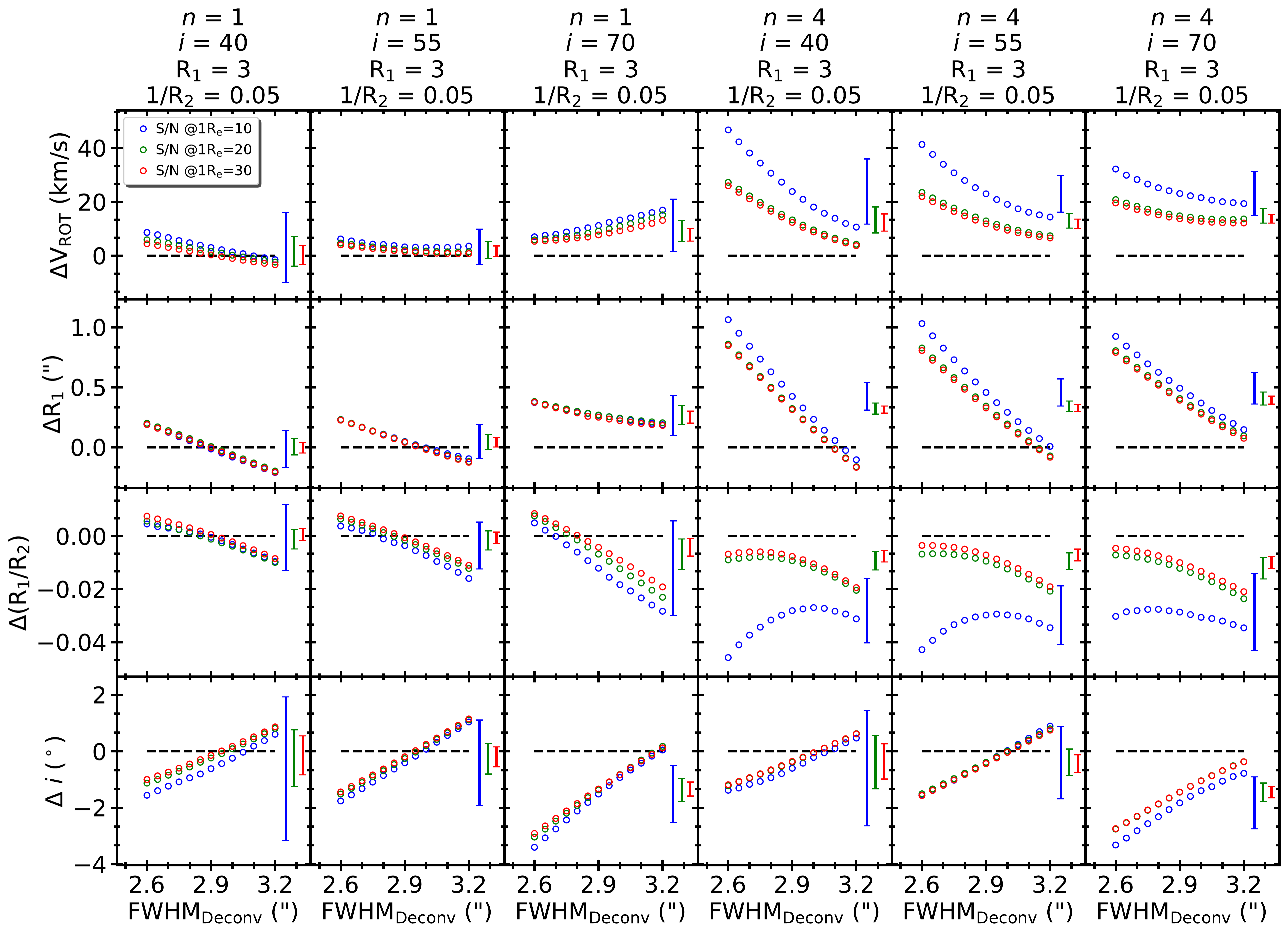}
\caption{Plots similar to \autoref{fwhm:f1r1ih0} but for
$\rm R_1=3, 1/R_2=0.05, FWHM c_0=2.9$}\label{fwhm:f2r1ih2}
\end{figure*}

\section{Effect of PSF Convolution to the Spin Parameter Measurement}\label{app:spin}
In \autoref{fig:spin_mock_matrix}, we plot the relations between the 
$\lre$ ratios ($\lre^{Conv}/\lre^{Free}$, $\lre^{Deconv}/\lre^{Free}$, $\lre^{G18 Corr.}/\lre^{Free}$) and the true $\lre$ value ($\lre^{Free}$), depends on three mock IFU parameters, IFU field of view, $\nsrc$, and IFU radial coverage in $\Reff$, using Group 3 mock IFU data (\autoref{subsec:rcmodel}).
Most of the panel of \autoref{fig:spin_mock_matrix} shows that $\lre$ ratios have little or negligible dependence on $\lre^{Free}$, except when $\lre^{Free}$ $<$ 0.1. The ratio and its standard deviation at $\lre^{Free}$ $<$ 0.1 looks different compare to the ratios at $\lre^{Free}$ $>$ 0.1, but this is simply an effect of small denominator when $\lre^{Free}$ $<$ 0.1. Since the denominator ($\lre^{Free}$) is already small, the actual deviation of $\lre$ values to the true value ($\lre-\lre^{Free}$) is also small.
The median and the median of standard deviation of each binned relation ($\rm \Delta\lre^{Free}$=0.1) is used to show the overall dependence of the ratios to the mock IFU parameters as in \autoref{fig:spin_mock}. Unlike average value of entire points, use of median of the binned relations could avoid the contribution from large difference and the standard deviation from the points at $\lre^{Free} <$ 0.1. 
In \autoref{fig:spin_sigma1_matrix} We plot the relation between the 
$\lre$ ratios and the true $\lre$ value, depend on $\nsrc$ and $\sigma^{\prime}$ parameters using the additional set of mock IFU data (\autoref{sub:spinmock}). Again the median and the median of standard deviation of each binned relation ($\rm \Delta\lre^{Free}$=0.1) is used to show the overall dependence of the ratios to the mock IFU parameters as in \autoref{fig:spin_sigma1}.

\setcounter{figure}{0} 
\begin{figure*}
\centering
\includegraphics[width=\textwidth]{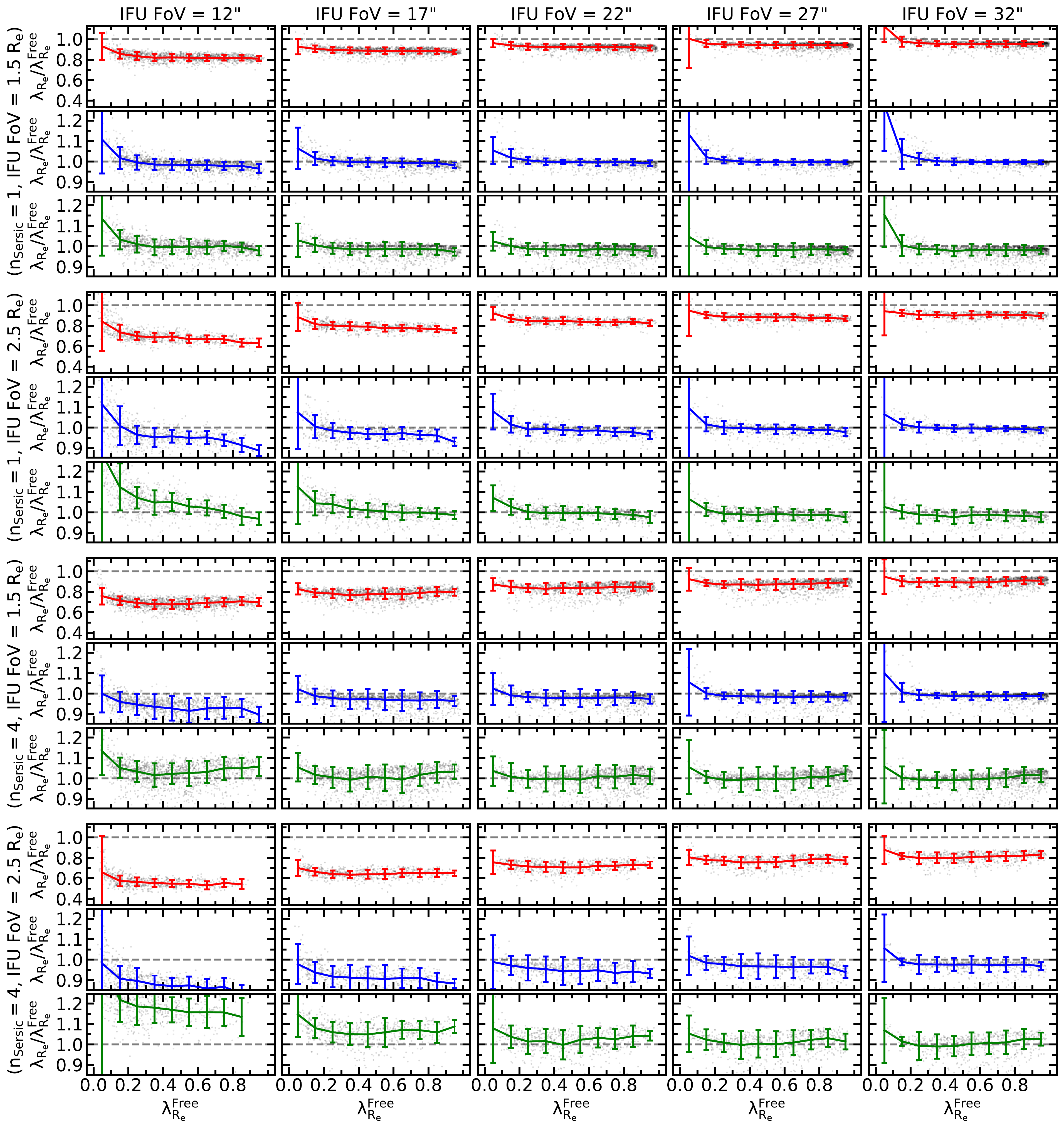}
\caption{Relations between the ratio of $\lre$ values ($\lre^{Conv}$, $\lre^{Deconv}$, $\lre^{G18 Corr.}$) and the true $\lre$ value ($\lre^{Free}$) as a function of $\lre^{Free}$. 
 Each panel with red, blue, and green lines represents data points with $\lre^{Conv}/\lre^{Free}$, $\lre^{Deconv}/\lre^{Free}$, and $\lre^{G18 Corr.}/\lre^{Free}$, respectively. 
Each column represents the field of view of mock IFU data used for each panel (IFU field of view = 12$\arcsec$, 17$\arcsec$, 22$\arcsec$, 27$\arcsec$, 32$\arcsec$). Each of continuous three rows represent the different combination of $\nsrc$ and the IFU radial coverage in $\Reff$ ($\nsrc$=1 \& radial coverage of 1.5 $\Reff$, $\nsrc$=1 \& radial coverage of 2.5 $\Reff$, $\nsrc$=4 \& radial coverage of 1.5 $\Reff$, $\nsrc$=4 \& radial coverage of 2.5 $\Reff$). Data points are plotted as grey dots in the background. Color lines and the corresponding error bars are the median and the standard deviation of the data points from each bin with the bin size of $\rm \Delta\lre^{Free}$=0.1.}
\label{fig:spin_mock_matrix}
\end{figure*}

\begin{figure*}
\centering
\includegraphics[width=\textwidth]{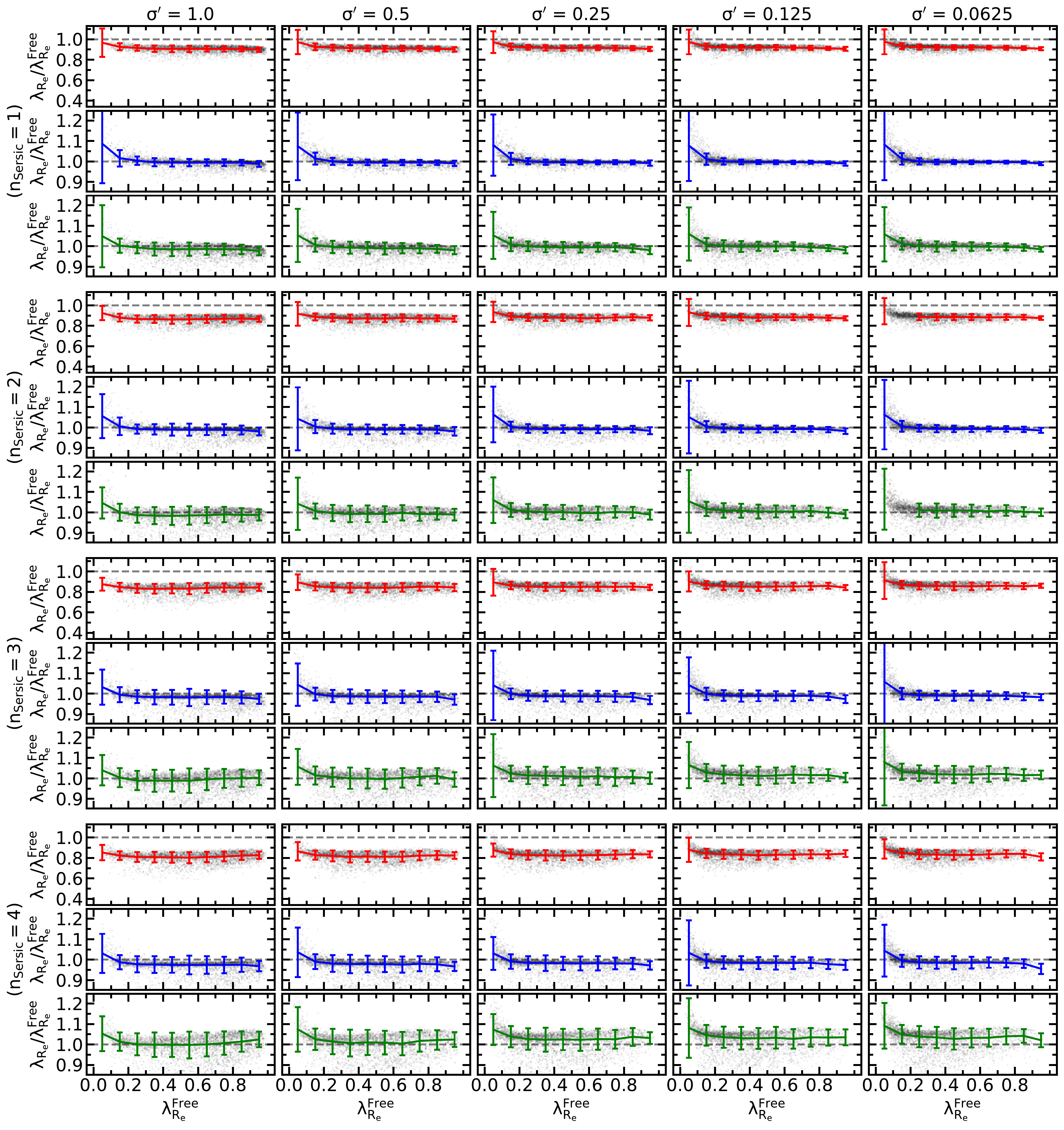}
\caption{
Relations between the ratio of $\lre$ values ($\lre^{Conv}$, $\lre^{Deconv}$, $\lre^{G18 Corr.}$) and the true $\lre$ value ($\lre^{Free}$) as a function of $\lre^{Free}$, similar to \autoref{fig:spin_mock_matrix} but using different set of mock IFU data (\autoref{sub:spinmock}). Each column represents different velocity dispersion profile steepness of the mock IFU data used for each panel ($\sigma^{\prime}$ = 1, 0.5, 0.25, 0.125, 0.0625). Each of continuous three rows represents different $\nsrc$ of the mock IFU data ($\nsrc$=1,2,3,4). Data points, color lines and the corresponding error bars are plotted in the same way as \autoref{fig:spin_mock_matrix}.
}
\label{fig:spin_sigma1_matrix}
\end{figure*}

\begin{figure*}
\centering
\includegraphics[width=\textwidth]{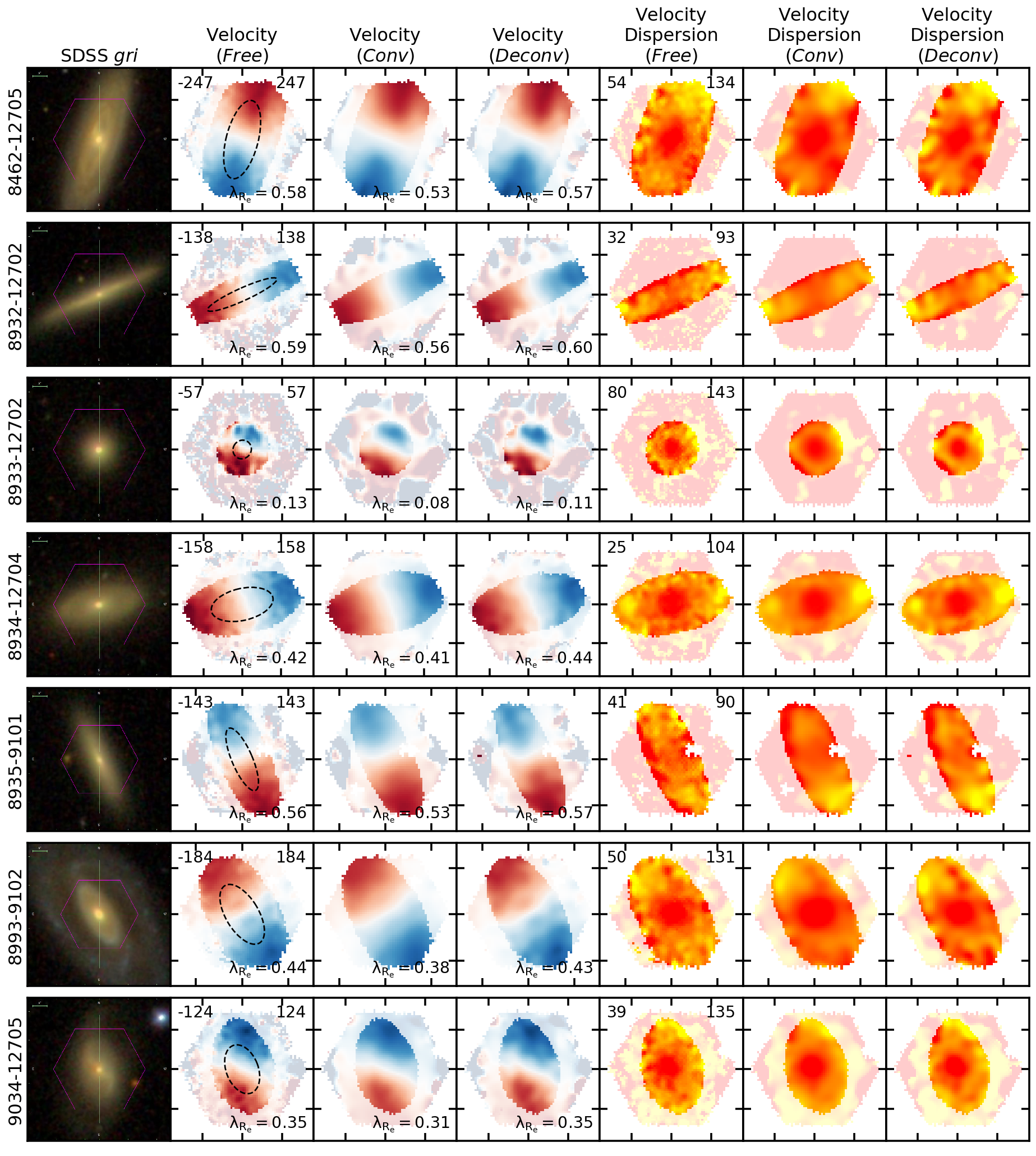}
\caption{The velocity and velocity distribution of selected  nearby MaNGA galaxies (z$<$0.023 and IFU FoV $\ge$27{\arcsec}). Distributions in {\free} column are velocity and velocity dispersion distribution measured from the original MaNGA IFU data. Distributions in the other columns ({\ndc} and {\dc}) are measured from PSF convolved and deconvolved (PSF FWHM=2.6{\arcsec}) to the original MaNGA IFU data. The ellipses (black-dashed) in the left most column show 1 $\rm R_e$ aperture where the {$\lre$} values are measured. The values in the top left and top right at each panel in the second and fifth columns indicate the minimum and the maximum range of color, where the color bars are the same as in the \autoref{fig:dcex} (Blue to red for the velocity distribution and yellow to red for the velocity dispersion distribution). Each major tick interval corresponds to 10$\arcsec$.
}\label{appfig:manga_deconv_test}
\end{figure*}


\begin{figure*}
\includegraphics[width=\linewidth]{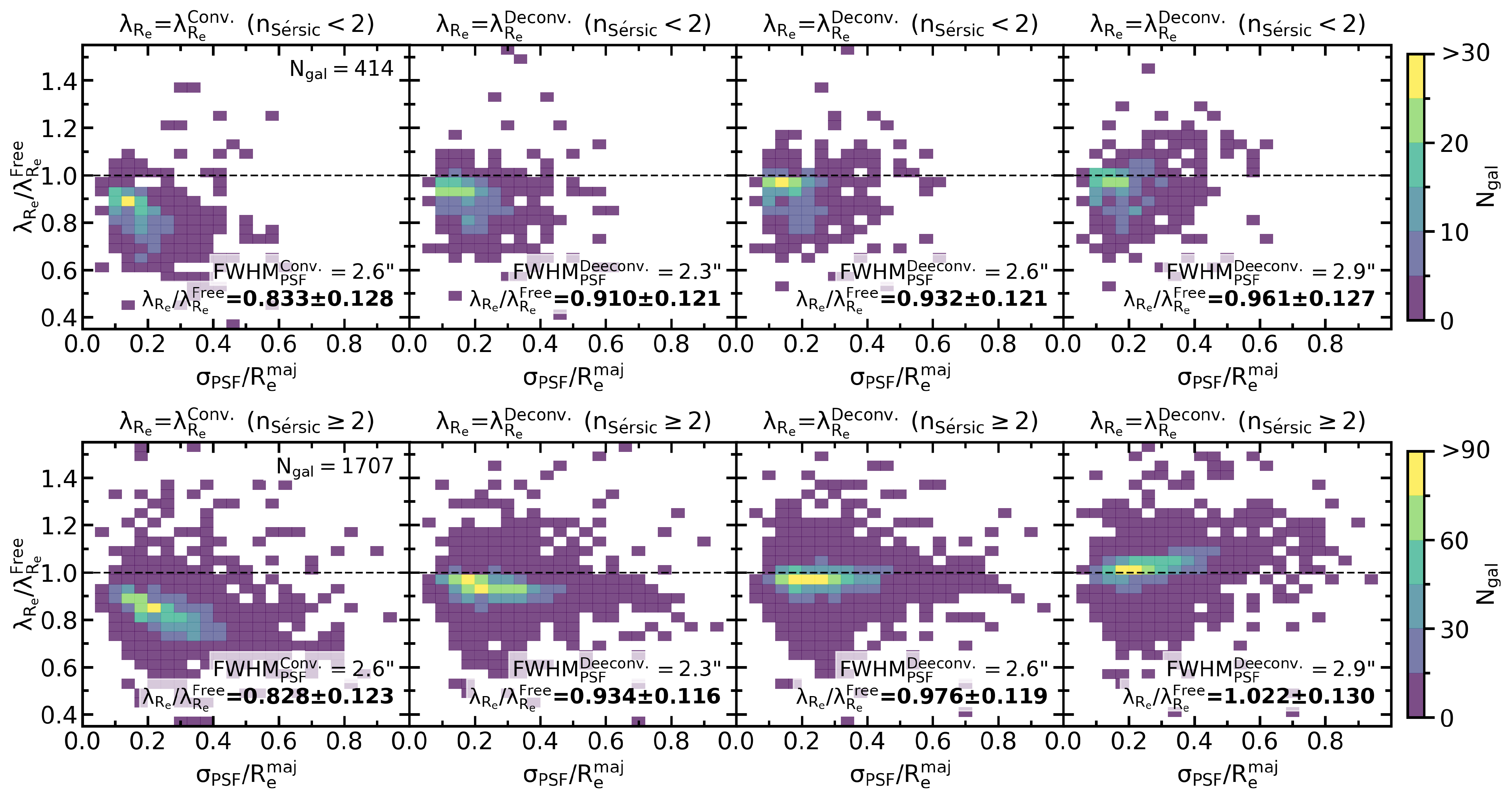}
\caption{
Comparison between the $\lambda_{R_e}$ measured from PSF-Convolved and PSF-Deconvolved IFU data
using MaNGA DR15 galaxies as the PSF-Free data, 
when the $\fpsf$ used for the deconvolution or $\lre$ correction is different from the $\fpsf$ of the convolved PSF. 
Upper and lower rows show the results when $\nsrc < 2$ and $\nsrc \ge 2$, respectively. First column shows $\lambda_{R_e}^{Conv}/\lambda_{R_e}^{Free}$ as a function of $\rm \sigma_{PSF}/R_{e}^{maj.}$. Second, third, and fourth columns shows $\lambda_{R_e}^{Deconv}/\lambda_{R_e}^{Free}$ as a function of $\rm \sigma_{PSF}/R_{e}^{maj.}$, but when $\fpsf$ used for the deconvolution or $\lre$ correction is 2.3, 2.6, and 2.9$\arcsec$, respectively. The median and standard deviation of $\Delta \lambda_{R_e}$ are noted in each panel.} \label{fig:spin_fwhm_nsersic}
\end{figure*}



\end{document}